\documentclass[11pt,a4paper]{article}
\pdfoutput=1 

\usepackage{amssymb,amsmath,amsfonts}
\usepackage[dvipsnames]{xcolor}
\usepackage{graphicx}
\usepackage{longtable}
\usepackage{verbatim}
\usepackage{color}
\usepackage[mathscr]{euscript}
\usepackage{framed}
\usepackage{mdframed}
\usepackage{amsmath}
\usepackage{amsfonts}
\usepackage{mathrsfs}
\usepackage{amssymb}
\usepackage{mathrsfs}  
\usepackage{subcaption}
\usepackage{caption}

\usepackage[left=26mm,top=20mm,right=26mm,bottom=20mm]{geometry}

\usepackage{color}
\usepackage{cite}
\usepackage{hyperref}
\hypersetup{colorlinks, citecolor=bluscuro, linkcolor=black, urlcolor=bluscuro}
\definecolor{rossos}{cmyk}{0,1,1,0.55}
\definecolor{bluscuro}{rgb}{0.15, 0.2, .85}
\definecolor{bluchiaro}{cmyk}{1,.3,0.,0.1}

\graphicspath{{./Figures/}}

\newcommand{\be}{\begin{equation}}
\newcommand{\ee}{\end{equation}}
\newcommand{\bea}{\begin{eqnarray}}
\newcommand{\eea}{\end{eqnarray}}
\newcommand{\beq}{\begin{equation}}
\newcommand{\eeq}{\end{equation}}

\def\beqa{\begin{eqnarray}}

\def\eeqa{\end{eqnarray}}

\def\lsim{\mathrel{\rlap{\lower4pt\hbox{\hskip0.5pt$\sim$}}
    \raise1pt\hbox{$<$}}}         %less than or approx. symbol
\def\gsim{\mathrel{\rlap{\lower4pt\hbox{\hskip0.5pt$\sim$}}
    \raise1pt\hbox{$>$}}}         %greater than or approx. symbol

\DeclareMathOperator\Mpc{Mpc^{-1}}

\usepackage[normalem]{ulem}

\newcommand{\arXiv}[2]{\href{http://arxiv.org/pdf/#1}{{\tt [#2/#1]}}}
\newcommand{\arXivold}[1]{\href{http://arxiv.org/pdf/#1}{{\tt [#1]}}}

\begin{document}

%\begin{titlepage}
%\begin{flushright}
%%DESY 16-xxx\\
%%CERN ...
%\end{flushright}
\vspace{0.1in}

%\vspace{1cm}
\begin{center}
{\Large\bf\color{black}
Primordial Black Holes from  $\alpha$-attractors
}\\
\bigskip\color{black}
\vspace{.5cm}
{ {\large Ioannis Dalianis${}^a$, Alex Kehagias${}^a$  and  George Tringas${}^{a,b}$ }
\vspace{0.3cm}
} \\[5mm]
{\it {$^a$\, Physics Division, National Technical University of Athens\\ 15780 Zografou Campus, Athens, Greece}}\\[2mm]

{\it {$^b$\, Physikalisches Institut der Universit\"at Bonn, \\ Nussallee 12, 53115 Bonn, Germany}}\\
\end{center}
%\bigskip
%\vskip1in 

\vskip.2in

\noindent
%\rule{15.8cm}{0.4pt}

\vspace{.3cm}
\noindent
\begin{center}
\large{ABSTRACT}
\end{center}
\vskip.15in 

%\begin{abstract}
\noindent

We consider primordial black hole (PBH) production in 
 inflationary  $\alpha$--attractors.  
We discuss  two classes of models, namely  models  with a minimal polynomial superpotential as well as  modulated  chaotic ones  that admit PBHs.   
We find that  a significant amplification of the curvature power spectrum ${\cal P_R}$ can be realized in this class of models with a moderate tuning of the potential parameters.  
We consistently examine the PBH formation during radiation and additionally during reheating eras 
where the background pressure is negligible.
It is shown that basic features of the curvature power spectrum 
are explicitly related with the postinflationary cosmic evolution and that the PBH mass and abundance expressions are accordingly modified. 
PBHs in the mass range $10^{-16}-10^{-14} \, M_{\odot}$ can form with  a cosmologically relevant abundance 
for a power spectrum peak  ${\cal P_R} \sim 10^{-2}$ and large  reheating temperature and, furthermore, for a moderate peak  ${\cal P_R} \sim 10^{-5}$ and 
 reheating temperature
$T_\text{rh}\sim 10^7$ GeV, 
characteristic of the position of the power spectrum peak.
Regarding the CMB observables, the $\alpha$--attractor models utilized here to generate PBH in the low-mass region  predict in general a  smaller $n_s$ and larger $r$ and $\alpha_s$ parameter values compared to the conventional inflationary $\alpha$--attractor  models.

%\end{abstract}
\bigskip

%\end{titlepage}
\vspace{7cm}
\noindent
\rule{5.8cm}{0.4pt}\\
\href{mailto:dalianis@mail.ntua.gr}{dalianis@mail.ntua.gr} \\
\href{mailto:kehagias@central.ntua.gr}{kehagias@central.ntua.gr}\\
\href{mailto:tringas@uni-bonn.de}{tringas@uni-bonn.de}

%\vskip.4in
%\hypersetup{colorlinks=true, citecolor=red, filecolor=green, linkcolor=red, urlcolor=green, citebordercolor={0 1 0}}

\pagebreak 
\tableofcontents
%\pagebreak

\section{Introduction}

The recent detection of gravitation waves due to the merging of two  massive ($\sim 30 M_\odot$) black holes  \cite{ligo}, has revived the idea that the dark matter in the universe, or some fraction of it, is composed of Primordial Black Holes (PBHs)  \cite{PBH1,PBH2,PBH3,PBH4, kam, Clesse:2016vqa, rep2, c0, c1, c2, revPBH}. 
PBHs are distinguished from stellar black holes since they are not remnants of the gravitational collapse of  massive stars. Instead, they are formed by the   collapse of primordial density perturbations which are of order unity upon horizon entry  \cite{s1,s2,s3}.  Therefore, the mass of such PBHs is not limited by  the usual mass  bound  ($\gsim 3M_\odot$) of astrophysical black holes but can attain much smaller masses and they  can be as light as $10^{-18}M_\odot$, but not lighter in order the evaporation rate in the late universe to be suppressed.

If PBHs are created in the early universe it is natural to contemplate upon the exciting  possibility  that they  comprise the elusive dark matter  of the universe \cite{c0}. Contrary to the stellar black holes, the PBHs evade the bounds on the baryonic matter abundance from the big bang nucleosynthesis.
An appealing and minimal assumption is to attribute the creation of PBHs  to  the inflationary phase itself.
Indeed,  recent studies indicate that large primordial density perturbations that lead to the PBH production can be triggered by the inflaton field stochastic fluctuations.
 This hypothesis can be realized in 
   single-field inflationary models  \cite{ Alabidi:2009bk,  Drees:2011hb, Drees:2011yz,  sone1,sone0,sone2, ssm1, Hertzberg:2017dkh,  sone3, Ozsoy:2018flq}, in double inflation, see e.g. \cite{Kawasaki:1997ju,  Kawasaki:2016pql, Kawaguchi:2007fz, Pi:2017gih},  as well as through some spectator field  \cite{stwo1,Carr:2016drx,gauge,cs}. 
   More interesting it is the fact that such a spectator field can be accommodated in the Standard Model and it can be identified with  the Higgs field  \cite{ssm2,ERR}.   PBHs are found to be formed with a  relic abundance  possibly large enough to  account for  a significant  fraction of the bulk dark matter in the universe.

In order for a model of inflation to seed  
the PBH formation, a mechanism is needed for the enhancement of the power spectrum of the curvature perturbation at small scales.  
The power spectrum has to be normalized  at the CMB scales by the amplitude of the temperature anisotropies, ${\cal P_R} \sim10^{-9}$, and  increased about seven orders of magnitude at much smaller scales,  ${\cal P_R} \sim 10^{-2}$, relevant to the PBH formation in a thermal radiation background. This striking change of the curvature power spectrum with respect to the scale inevitably has to be attributed to some peculiar feature of the inflationary potential. 
The power spectrum, according to the standard approximate analytic estimation, has a dependence ${\cal P_R}\sim H^2/\epsilon_1$, where $\epsilon_1$ is the first Hubble flow function and $H$ the Hubble parameter. 
At first sight, an enhancement in ${\cal P_R}$ could  be achieved if the $\epsilon_1$ decreases substantially during the inflationary evolution.  For $\epsilon_1\ll 1$ we have  for 
 the first slow-roll parameter $\epsilon_V$ of the inflaton potential  $\epsilon_V \simeq \epsilon_1 $,  which  implies that at some field value before the end of inflation the derivative of the potential nearly vanishes, $V' \simeq 0$. 
However, it is not obvious at all  that the $\epsilon_1$ and the other Hubble flow functions, $\epsilon_2$ and $\epsilon_3$, are also small in that field region since the acceleration of the inflaton becomes important  and the slow-roll approximation breaks down. The inflaton may overshoot the stationary point without any enhancement in the power spectrum.  A local plateau is necessary which implies that $V''$ has also to vanish. 
The general picture drawn from the attempts  so far converge to a potential that features an inflection point about a local plateau \cite{sone1}.  The inflaton field slowly rolls down the potential in the region that corresponds to the CMB scales, afterwards accelerates and  substantially decelerates in the region of the inflection point and, finally, slow-rolls again until the end of inflation. In between the initial and the final slow-roll phase there is a stage  known as ultra slow-roll (USR) phase \cite{Kinney}. It is during this stage that the power spectrum of the curvature power spectrum gets amplified and the PBH formation becomes possible.

The violation of the slow-roll phase means that the power spectrum has to be computed numerically. The approximate analytic expression ${\cal P_R}\sim H^2/\epsilon_1$,   gives wrong results because there is a ${\cal O}(1)$ change in Hubble flow functions.
The exact Mukhanov-Sasaki equation has to solved. The exact solution, obtained numerically,  reveals that in the regions of field space where the Hubble flow functions get larger than one the expression  ${\cal P_R}\sim H^2/\epsilon_1$ underestimates  several orders of magnitude the actual ${\cal P_R}$ value.  A careful look indicates that the power spectrum peak is generated when  the Hubble flow functions change rapidly. On the other hand, the $H^2/\epsilon_1$ value becomes  maximum when the velocity of the inflaton minimizes.  After that point, the slow-roll regime takes over and the analytic value for the power spectrum approximates well the numerical value.   The  difference in the amplitude between the numerical and the $H^2/\epsilon_1$ peak  is  found to be  hundreds or even thousands times large and it depends on the way    the  Hubble flow functions change.

Although the production of dark matter in the form of PBHs is rather appealing, its explicit implementation in  inflationary models seems to be particular delicate as it demands a considerable amount of fine tuning of the parameters of the potential.
Indeed, an inflection point, $V''=0$, together with a local plateau, $V'\approx 0$, appear only if the terms in the potential are carefully balanced. 
Nevertheless, it is notable 
that such a behavior can be implemented by several inflationary models.  After the first suggestion of an inflection point  potential constructed by a ratio of polynomials \cite{sone1}, a number of scenarios have been proposed in the literature.  Critical Higgs inflation \cite{ssm1}, simple polynomials or potentials with a radiative plateau a la Coleman-Weinberg where the inflaton is non-minimally coupled to the Ricci scalar \cite{sone0, sone2}, fibre inflation  \cite{sone3} and axion models  \cite{Ozsoy:2018flq} inspired by string theory, supergravity \cite{Gao:2018pvq} are examples of single field inflationary models that can accommodate the PBH formation.

\begin{figure}[!htbp]
  \centering
  \includegraphics[width=.7 \linewidth]{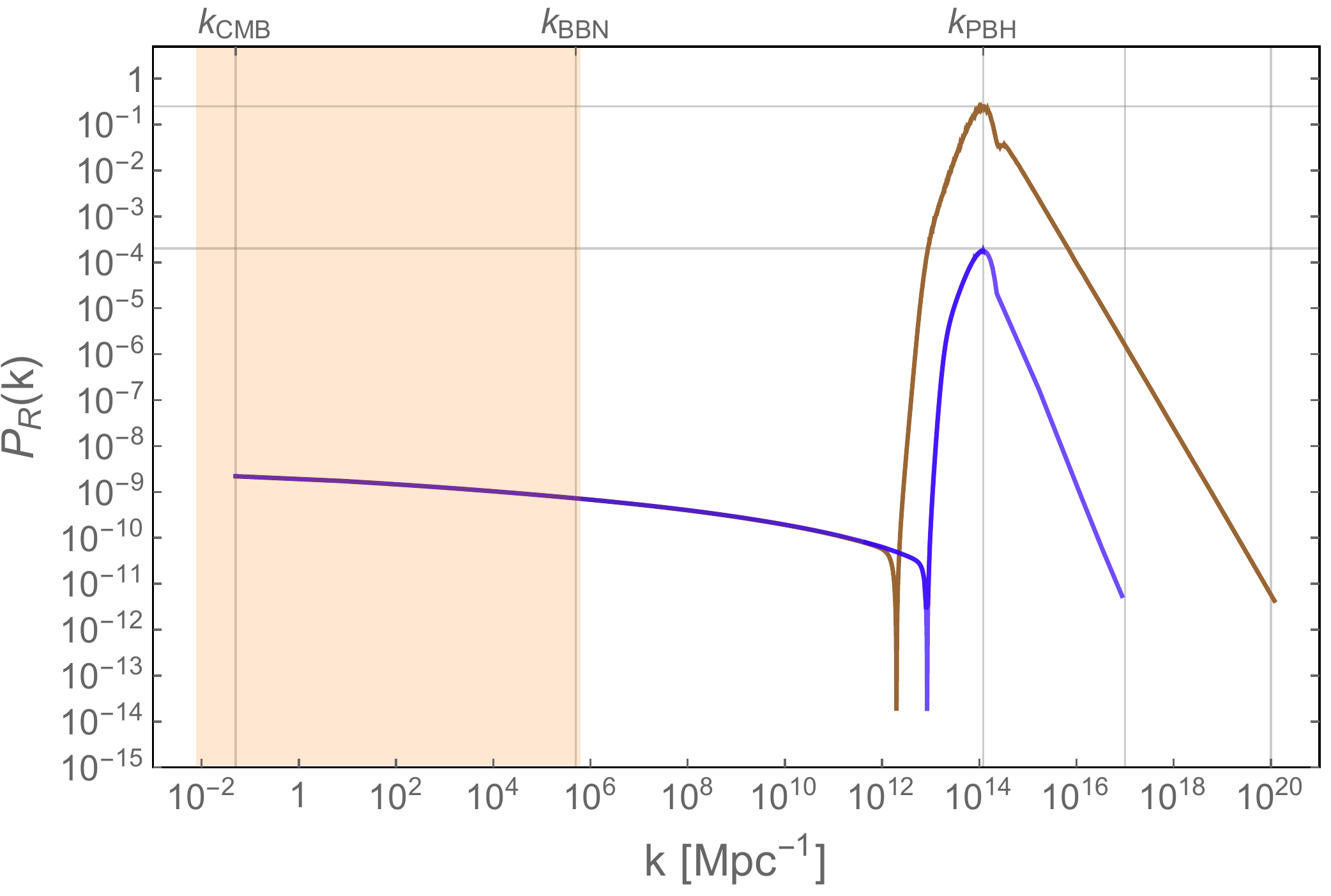}
 \caption{\label{psmdrd}~~  Two typical power spectra produced by our models (Model  $I_2$ family) for PBH formation with $\Omega_\text{PBH}/\Omega_\text{DM}  \sim 0.1$ during radiation domination (upper curve) and matter domination (lower curve). The shaded area  indicates the scales $k^{-1}$ that reenter horizon during the cosmic period  confirmed to be radiation dominated.}
\end{figure}

During the USR phase the inflaton velocity promptly decreases and one should check whether the quantum jumps influence the inflaton trajectory and the relic density of the PBH.  
This issue has been examined  in \cite{Biagetti:2018pjj} where it was found that the quantum diffusion has a significant impact on the PBH mass fraction making the classical standard predictions not trustable, unless extreme values for the density contrast threshold $\delta_c$ are adopted. In Ref. \cite{Ezquiaga:2018gbw} it has been argued that diffusion could induce an enhancement of the power spectrum.   Apart from the diffusion issue,  the amplitude of the peak can produce non-gaussianities  \cite{Franciolini:2018vbk} that lead to additional interesting constraints to the parameter space for the PBH production mechanism.

The amplitude of the power spectrum can be reduced if we depart from the simple assumption of a continuous radiation dominated era right after inflation. During a pressureless matter dominated era a sizable fraction of the total energy can collapse to PBH for ${\cal P_R} \sim 10^{-4}$ or even ${\cal P_R} \sim 10^{-5}$.  
A critical epoch is the transition of the universe from the supercooled inflaton dominated phase to hot thermalized plasma with temperature, $T_\text{rh}$, which may not be instantaneous. 
The only existing bound on the reheating duration is that  $T_\text{rh}>{\cal O}(1)$ MeV, in order to  proceed BBN, and a prolonged non-thermal phase may take place in the early universe. 
 In addition it is natural to assume that the early universe energy density is dominated for some period by late decaying scalar fields, which are natural in the context of supersymmetric and stringy scenarios. 
The radiation dominated (RD) era is actually observationally established to occur from the BBN epoch until the equality epoch.  At that cosmic epochs the horizon scale was respectively $k^{-1}_\text{eq}=(3.6\times 10^{-2})^{-1}$ Mpc and $k^{-1}_\text{BBN}\sim 10^{-5}$ Mpc. Scales $k>k_\text{BBN}$ may enter during a radiation dominated era or during an era where the equation of state deviates from 1/3.

In this paper, along the lines of the aforementioned works, we investigate the  generation of PBH in the context of the supergravity and superconformal theory considering in addition that the PBH may form during reheating or modulus dominated eras.  
We consider the general class of $\alpha$-attractor models which have the advantage to flatten the scalar potential for large values of the field. 
We identify the models in this class that exhibit an inflection point, in small field values where the potential cannot be flatten, and examine their parameter space for PBH production.  
We find that a broad set of choices of functions exists that can implement the double r\^ole of the inflaton, to generate the CMB anisotropies and the PBH formation. 
 The power spectrum gets enhanced after about $N=38$ e-folds of observable inflation and there, at small field values,  the general predictions for the spectral index  $n_s$  of the simple $a$-attractors do not apply.
   The particular  shape of the potential together with the multiple observational constraints 
require  particularly tuned parameters  
 but these parameters should be seen as moduli of a string theory compactification, and not necessarily as shortcomings of these scenarios. 
 Our models predict PBH in the low mass window since the corresponding scales that collapse to PBH reenter the horizon not long after inflation,  see Figure \ref{psmdrd}. At  that energy scales the reheating phase or a late decaying scalar fields can affect 
the PBH formation and we consistently  investigate the PBH formation rate taking into account aspherical or angular momentum effects.
 In addition, the models that we discuss and analyze predict distinctive values both for the CMB observables and for the PBH mass and abundance that can be tested by near future observational programs.  We mention that our analysis is general and applies to any inflationary model with a spike in its power spectrum.

The paper is organized as follows: In section 2 we review supergravity inflationary potentials in the 
$\alpha$-attractor setup. In section 3 we work out particular cases of supersymmetric $\alpha$-attractors.  In section 4 we present the computation of the power spectrum of the curvature perturbation.  In section 5 we give the expressions for the PBH mass and the abundance for three different cosmic eras, radiation domination, reheating and modulus condensate domination.  In section 6 we analyze the scenarios for the PBH formation in  superconformal attractor models,  and in section 7 we  present our results and discuss the observational implications. Finally we conclude in section 8.

\section{Superconformal Attractors}
\label{sec:intro}

We will consider here a  superconformal model of  ${\cal N}=1$ supergravity coupled to chiral multiplets $X^I$, $(I=  0,\cdots, n)$, where $X^0$ is the ``reference" superfield (conformon) in the general context of supersymmetric $\alpha$--attractor models \cite{KL1,KL2,CK,KLR2,KLR0,KLR,Kal, FKS}.  The key feature of such models  was firstly introduced in the context of general inflation potentials in supergravity \cite{Kallosh:2010ug}. 
The superconformal action is \cite{FvP}
\begin{eqnarray}
{\cal L}=[N(X,\overline X)]_D+[{\cal W}(X)]_F. \label{ac}
\end{eqnarray}
Here $N(X,\overline X)$ is a generic real function with Weyl weight $2$, which includes the conformon superfield, ${\cal W}(X)$ is the superpotential with Weyl weight $3$, and  the subscripts $D$ and $F$ denote F and D-terms as usual. The theory has extra symmetries and in particular, 
Weyl symmetry,  special conformal symmetry,  special supersymmetry and a $U(1)$ R-symmetry. Poincare supergravity is obtained after gauge fixing these symmetries.  
A first approach and ingredients of this construction can be found in the early paper \cite{Stewart:1994ts}.  

 The F-term potential for the scalars originates from the auxiliary fields of the chiral multiplets and reads
\begin{eqnarray}
V=G^{I\overline{J}}{\cal W}_I {\overline{{\cal W}}}_{\overline J},
\end{eqnarray}
where $G^{I\overline{J}}$ is the inverse of the 
K\"ahler metric $G_{I\overline{J}}=\partial_I\partial_{\overline{J}}N$ and ${\cal W}_I=\partial_I {\cal W}$. The bosonic part of the Lagrangian (\ref{ac}) turns  out to be
\begin{align}\label{la}
e^{-1}\mathcal{L}=-\frac{1}{6}N(X,\bar{X})R-G_{I\bar{J}}{\cal D}^{\mu}X^I{\cal D}_{\mu}\bar{X}^{\bar{J}}-G^{I\bar{J}}W_I\bar{W}_{\bar{J}} ,
\end{align}
where 
%$N$ stands for the K\"ahler potential, $W$ for the holomorphic superpotential, R the Ricci scalar, $G_{I\bar{J}}$ the K\"ahler metric and
 ${\cal D}_{\mu} X^I=\partial_\mu X^I-i A_\mu X^I$ and
$A_\mu=i(N_{\overline{I}} \partial_\mu \overline{X}^{\overline{I}}-N_I\partial_\mu X^I)/2N$. 
% 
%  is the  $U(1)$ R-symmetry 
% and  $A_\mu$ is the composite 
% gauge field of the local $U(1)$ R-symmetry.  
This is the K\"ahler  connection of the Poincar\'e supergravity after gauge fixing. 
Here we will consider the ``minimal" case of two chiral superfields $X^1=\Phi$ and $X^2=S$ in addition to  the compensator $X^0$, see also Ref. \cite{Roest:2015qya} where a formulation with one chiral superfield was suggested. 
The superfield $\Phi$ contains the inflaton whereas $S$ is the goldstino multiplet. 
This is a large  universality class of models  and we will consider the subclass described by superconformal models where the embedding K\"ahler potential exhibits an $SU(1,1)$ symmetry in the complex $X^0-X^1$  field space. In particular, we will consider the stabilizing along the inflationary trajectories embedding K\"ahler potential 
\begin{align}
N(X,\bar{X})=-|X^0|^2\left(1-\frac{|X^1|^2+|S|^2}{|X^0|^2}+3\zeta\frac{(S\bar{S})^2}{|X^0|^2(|X^0|^2-|X^1|^2)}\right)^\alpha, \label{NN}
\end{align}
where $\alpha$ is a numerical constant. 
Apparently, (\ref{NN}) is invariant under $SU(1,1)$ transformations of $X^0,X^1$. 
 This can be implemented in superconformal attractor theory by considering the superpotential
\begin{align}
W= S\, f\Big(\frac{X^1}{X^0}\Big) (X^0)^2\Bigg(1-\frac{(X^1)^2}{(X^0)^2}\Bigg)^{\frac{3\alpha-1}{2}}. \label{W}
\end{align}

Clearly, this choice of the superpotential breaks partially  $SU(1,1)$ to just $SO(1,1)$ for a constant   function $f(X^1/X^0)$ and $\alpha=1$, and completely if $f$ is not constant, or $\alpha\neq 1$.  
To recover Poincar\'e supergravity a gauge fixing that fixes the local conformal and local $U(1)$ symmetry should be imposed. Here we use the standard  D-gauge in which the compensator is fixed at $X^0=\overline{X} ^0=\sqrt{3}$. In this case, standard Poincar\'e supergravity is recovered with K\"ahler potential $K=-3\log (-N/3)$. In particular, we find that in this gauge the K\"ahler potential and the superpotential are given by 
\begin{eqnarray}
K=-3\alpha \log \left(1-\frac{|\Phi|^2+|S|^2}{3}+ \frac{\zeta\,|S|^4}{3-|\Phi|^2}\right), \label{ka}
\end{eqnarray}
and 
\begin{eqnarray}
{\cal W}=S\, f\Big(\Phi/\sqrt{3}\Big)\Big(3-\Phi^2\Big)^{\frac{3\alpha-1}{2}},
\end{eqnarray}
respectively. Along the trajectory $S=Im\Phi=0$, the effective Lagrangian 
for the field $\phi=Re \Phi$ turns out to be
\begin{eqnarray}
e^{-1}{\cal L}= \frac{1}{2}R-\frac{\alpha}{\left(1-\frac{\phi^2}{3}\right)^2}
\Big(\partial_\mu \phi\Big)^2-f^2\Big(\phi/\sqrt{3}\Big). \label{la1}
\end{eqnarray}
Defining 
\begin{eqnarray} \label{boost}
\phi=\sqrt{3}\tanh\frac{\varphi}{\sqrt{6\alpha}}, 
\end{eqnarray}
we may write (\ref{la1}) as 
\begin{equation} \label{la11}
e^{-1}{\cal L}= \frac{1}{2}R-\frac{1}{2}
\Big(\partial_\mu \varphi\Big)^2-f^2\Big(\tanh\frac{\varphi}{\sqrt{6\alpha}}\Big). 
\end{equation}
Note that the K\"ahler metric on the $S=0$ scalar submanifold corresponds to a $\Sigma=SU(1,1)/U(1)$ coset symmetric space of constant curvature $R_\Sigma=-2/3\alpha$ and therefore, one may recognize
 $\alpha$  as the square of the ``radius" of $\Sigma$ \cite{KLR,FKLP1,FK}.

The inflationary trajectory  $S=Im\Phi=0$  is always  stable for appropriate $\zeta$. Alternatively, one may assume that the goldstino superfield is nilpotent $S^2=0$ in which case the stability condition of the inflationary trajectory becomes milder. 

In this class of models inflation occurs for large fields $\varphi \gg 0$ (or $\phi\approx \sqrt{3}$). In this case, we can approximate the potential as
\begin{eqnarray}
V=f_1^2-4 f_1f'_1e^{-\sqrt{\frac{2}{3\alpha }}\varphi}+\mathcal{O}(e^{-2\sqrt{\frac{2}{3\alpha}}\varphi}), \label{pp}
\end{eqnarray}
where 
\begin{eqnarray}
f_1=f
|_{\varphi\to \infty}, ~~~f_1'= \partial_\varphi f |_{\varphi\to \infty}.
\end{eqnarray}
Similar expansion of the potential in Eq. (\ref{pp})  appeared recently in \cite{sone3}. 
It is  straightforward to verify that for the potential (\ref{pp})
the spectral index $n_s$ and the scalar-to-tensor ratio $r$ are given to leading order in the number of e-folds $N$ by  \cite{FKLP1,KLR} 
\begin{eqnarray} \label{nsr}
   1-n_s=\frac{2}{N},~~~r=\frac{12\alpha}{N^2}.  
   \end{eqnarray}   
In other words, in the universality class of superconformal models, inflation gives the correct CMB anisotropies and normalization if $f(1)\approx 10^{-5}M_\text{Pl}^2$.  
For $\alpha=1$ and $f(\phi)\sim c_1 \phi$ we get a monoparametric $\alpha$-attractor model that is equivalent to the Starobinsky model \cite{Starobinsky:1980te}.
Here we investigate  polyparametric models that generate  an inflection point in the window of small field values where extra e-folds take place. The small fields values are far from the asymptotically flat regions of $\alpha$-attractors and there a significant  deviation from  the relation (\ref{nsr}) is expected.  In the following sections we will show that the  relation (\ref{nsr}) is still valid in our $\alpha$-attractors  models but for $N<N_{0.05}$. In particular it is $N\sim 38$ characteristic of the PBH mass, where $N$ are the elapsed e-folds and $N_{0.05}$ the total number of e-folds of the observable inflation.

\subsection{Conditions for inflection points in superconformal attractors}\label{conditions}

Having ensured the correct power spectrum for the density perturbations, we are going to determine the necessary conditions the superpotential function $f(X)$ should satisfy for PBH production.  The potential of the scalar Lagrangian (\ref{la1}) is 
\begin{eqnarray}
V(\varphi)=f^2(\tanh\varphi/\sqrt{6\alpha}),
\end{eqnarray}. The conditions the potential should satisfy are:

\begin{enumerate}
 \item A global minimum at $\varphi=\varphi_0$ where the potential vanishes
\begin{eqnarray}
V(\varphi_0)=0.
\end{eqnarray}
This is the point where the inflaton will settle down without giving rise to  a cosmological constant. 
\item An inflection point $\varphi=\varphi_{\rm infl}$ satisfying 
\begin{eqnarray}
%V'(\varphi_{\rm infl})=
V'(\varphi_{\rm infl})\approx 0,~~~~ V''(\varphi_{\rm infl})=0,
\end{eqnarray}
 where the inflaton slows down and generates large amplification in the power spectrum. Both points have to lie into a specific interval before the potential becomes asymptotically flat.
\end{enumerate}
 
 \noindent
The above conditions can also be written as conditions for the function $f$ as
\begin{align}
f|_{\varphi_{0}}&=0 , \nonumber \\
\partial_{\varphi}f|_{\varphi_{{\rm infl}}}&\approx 0, \nonumber \\
\partial^2_{\varphi}f|_{\varphi_{{\rm infl}}}&=0. \label{fc}
\end{align}
However, the function $f$  is asymptotically constant for $\varphi\gg 0$ thus the global minimum and the inflection point have to lie into a specific interval before the potential becomes asymptotically flat. The interval where the potential is deformed (not stretched) depends
mildly on the choice of the arbitrary function.  Therefore,  a more proper way to define the interval where the extrema appear should be addressed.     

Since the function $f$ appearing in the superpotential depends on $\Phi$, it is more convenient to express the conditions (\ref{fc}) in terms of 
$\phi=Re\Phi|_{\theta=\bar\theta=0}$, which are then written 
\begin{eqnarray}
f|_{\phi_{0}}=0 ,~~~
\partial_{\phi}f|_{\phi_{{\rm infl}}}\approx 0, ~~~
\partial^2_{\phi}f|_{\phi_{{\rm infl}}}=0. \label{fci}
\end{eqnarray}
Note that $\phi$ has not canonical  kinetic term as can be seen from Eq.(\ref{la11}). In this case, positivity 
of the kinetic term of $\phi$ 
requires that $|\phi|<\sqrt{3}$. 
The first condition signifies that the largest root $\phi_0$  of $f$ will be the point where the inflaton will settle down if inflation occurs for $\phi$ close to  $\sqrt{3}$. 

So the conclusion here is that the  superpotential function $f(\Phi)$ should have  roots and  an inflection point with no root after the latter. 
 The two following conditions address again the inflection point of $V(\phi/\sqrt{3})$.

\section{Specific Models}
\label{sec:intro2}

The main feature of the cosmological attractor mechanism is that for a arbitrary potential function chosen, it is compatible with the CMB constraints \cite{Planck1,Planck2} due to the flat potential for $\varphi \gg 1$. Here we will employ a family of functions which shares the characteristics introduced in (\ref{conditions}) to make the attractors a candidate for generating PBHs. 

\subsection{Model $I$: Polynomial Superpotentials}\label{POL}

\begin{figure}[!htbp]
\begin{subfigure}{.5\textwidth}
  \centering
  \includegraphics[width=1.\linewidth]{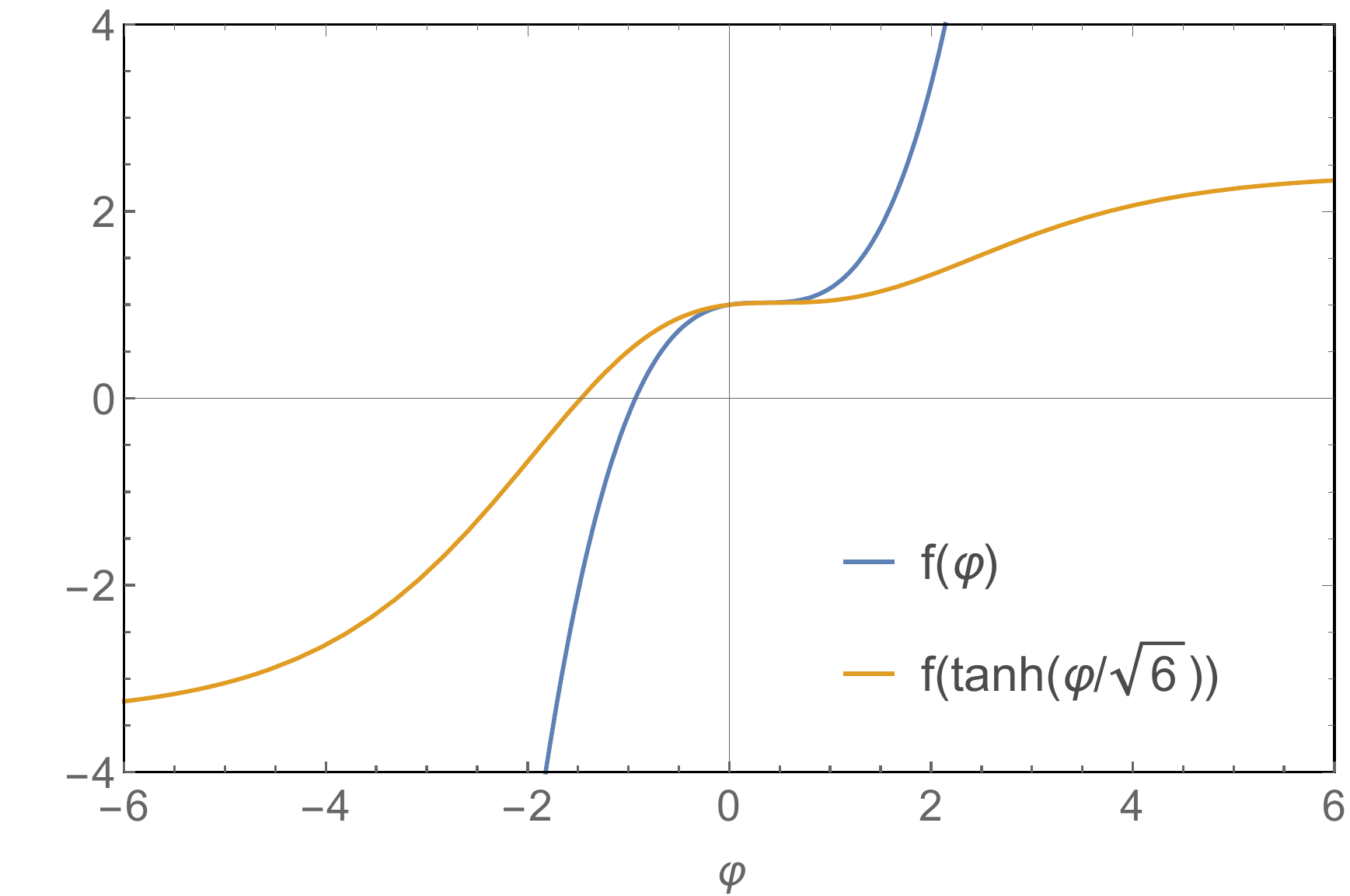}
  %\caption{1a}
  %\label{fig:sfig1}
\end{subfigure}%
\begin{subfigure}{.5\textwidth}
  \centering
  \includegraphics[width=1.\linewidth]{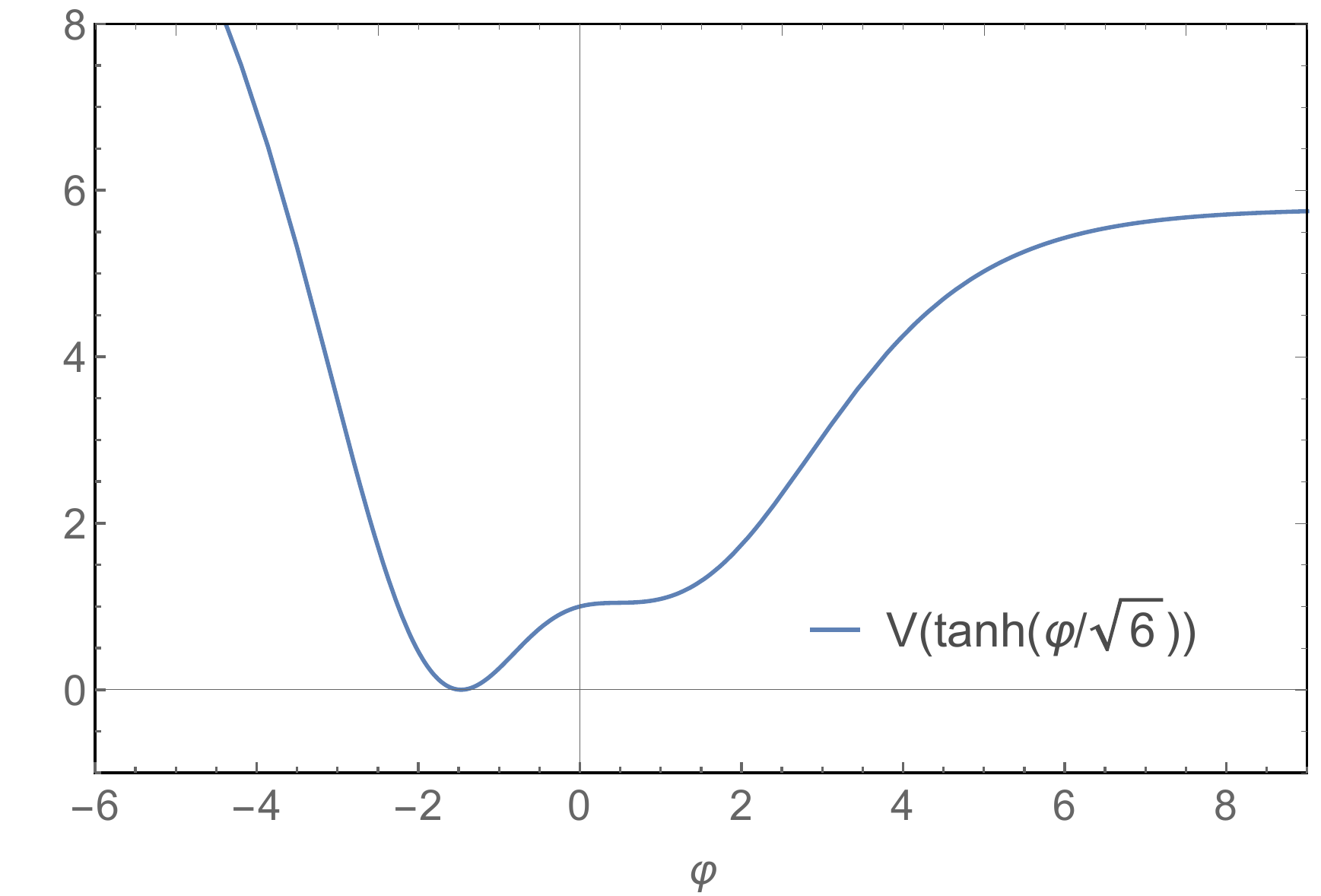}
 % \caption{1b}
  %\label{fig:sfig2}
\end{subfigure}
\caption{\label{function}~ {\it Left panel}: The function $f(\phi)$ given by Eq. (\ref{f1})  before and after the boost $\phi= \sqrt{3}\tanh(\varphi/\sqrt{6})$ with parameters $c_0=1,c_2=-c_3=-1/2, \xi=0.01$.  {\it Right panel}: The attractor potential $V=f^2$ .
% for the same set of parameters is presented.
}
\end{figure}

In general, an $n$-th degree polynomial has $n-2$ inflection points. Hence, the simplest polynomial with one  inflection point is the cubic 
 
\begin{equation}\label{f1}
f(\phi)=c_0+c_1\phi+c_2\phi^2+c_3\phi^3, 
\end{equation} 
 where $c_i$ are constant parameters.
There is always an  inflection point at 
\begin{eqnarray}
\phi_{\rm IP}=-\frac{c_2}{3 c_3}
 \end{eqnarray} 
 where the derivative of $f(\phi)$ is 
 \begin{eqnarray}
 f'(\phi_{\rm IP})=c_1-\frac{c_2^2}{3 c_3}.
 \end{eqnarray}
In order the function $f(\phi)$ to have a single root, it should be strictly monotonic, which happens for 
\begin{eqnarray}
 c_2^2\geq 3 c_1 c_3.
 \end{eqnarray} 
 In addition, the derivative of $f(\phi)$ at the inflection point should be small, that is
 \begin{eqnarray}
 c_2^2\approx 3 c_1 c_3,
 \end{eqnarray}
and therefore, we may write 
\begin{eqnarray}
 c_1=\frac{c_2^2}{3 c_3}+ \xi, ~~~~\xi \ll 1.
 \end{eqnarray} 
 In the following figures, we plot for illustrations the function $f(\phi)$ and  $V=f(\phi)^2$ for a boost $\phi= \sqrt{3}\tanh(\varphi/\sqrt{6})$. 
 It is clear that the inflection point in $f$ is transferred to the potential $V$, which exhibits a plateau for large values of $\varphi$. In addition, for a given value $a_3$, the parameter $a_2$ controls the relative height of the plateau, whereas $\delta$ controls the slope at the inflection point.

The potential so constructed  has therefore the following characteristics:

\begin{enumerate}
  \item  For large values of $\varphi$ slow-roll inflation is a good approximation due to the stretched plateau compatible with CMB observations.
  \item  The inflaton field can slow down due to the second plateau to generate PBHs, which for the above chosen parameters of the  potential is around the point $\varphi\approx 0.4$.
  \item It provides a vacuum for reheating at $\varphi\approx -1$ with zero cosmological constant.
\end{enumerate}

\begin{figure}[!htbp]
\begin{subfigure}{.5\textwidth}
  \centering
  \includegraphics[width=1.\linewidth]{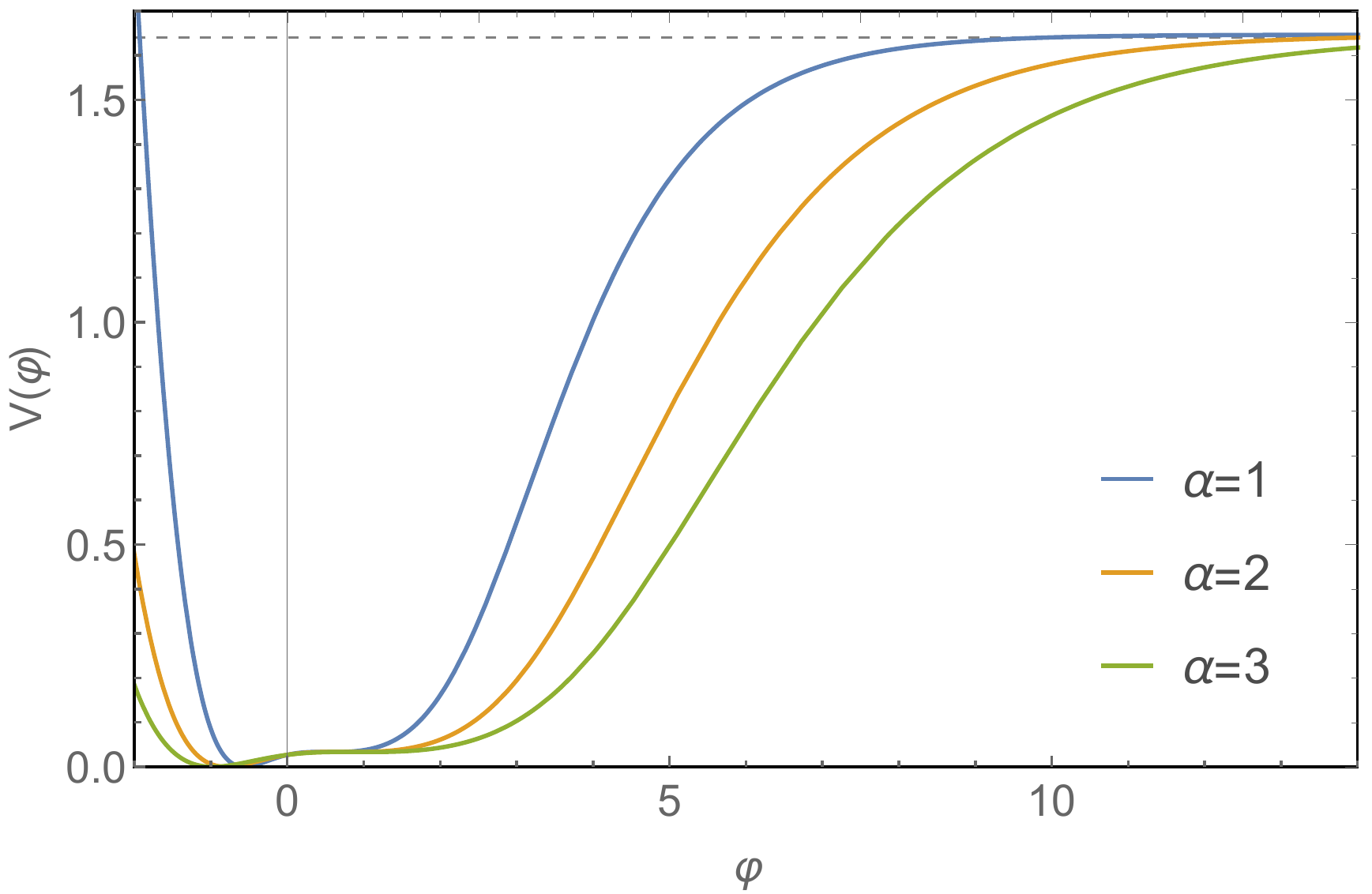}
\end{subfigure}%
\begin{subfigure}{.5\textwidth}
  \centering
  \includegraphics[width=1.\linewidth]{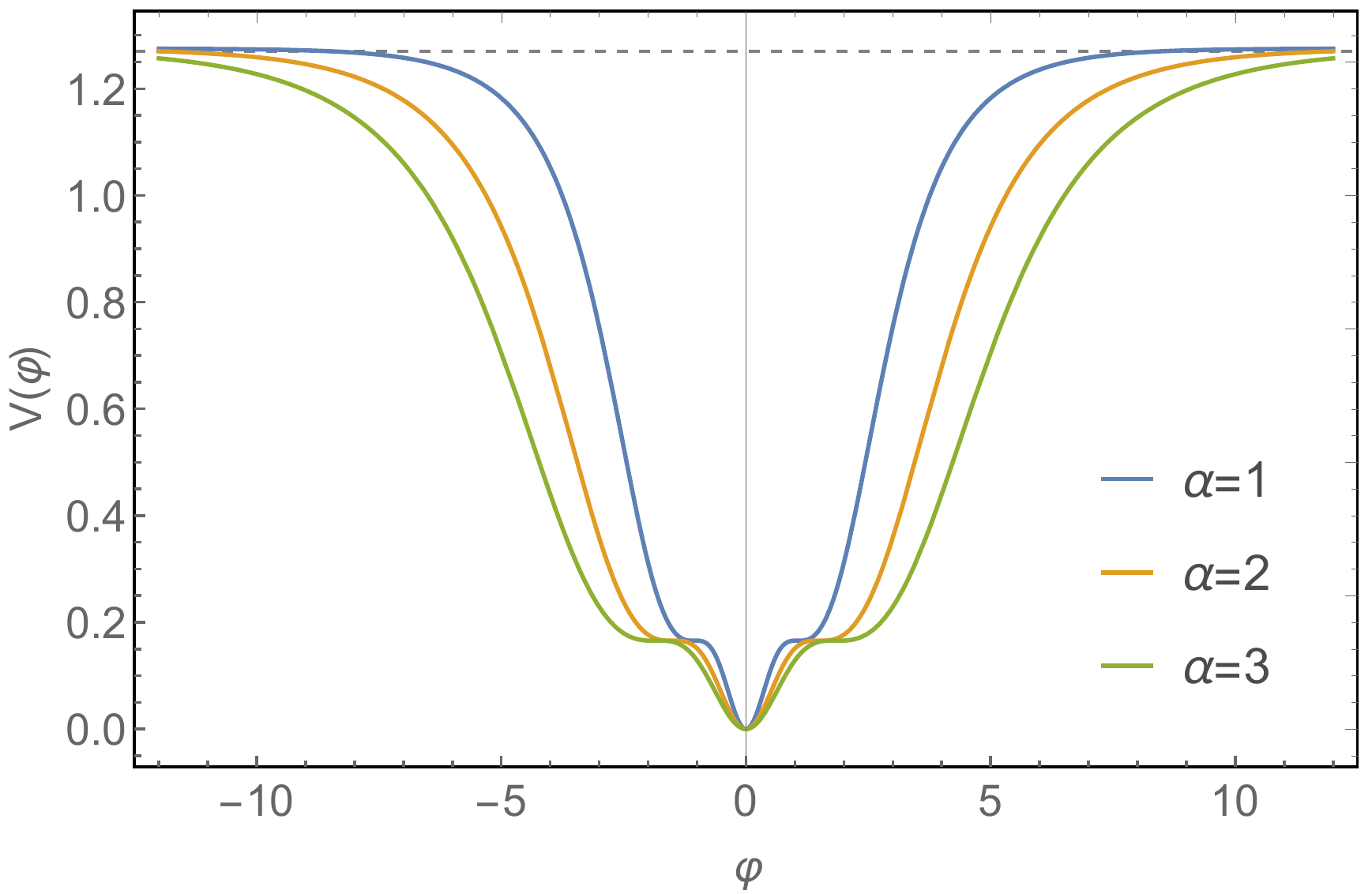}% change to SinPot if you want the old plot
 % \caption{1b}
  %\label{fig:sfig2}
\end{subfigure}
%\captionof{figure}
\caption{\label{potentials} {The inflationary potentials with inflection point for the Model $I_\alpha$ (polynomial).  and the Model $II$ (chaotic modulated), on the {\it left} and {\it right panel} respectively.
The potential parameters are given in Tables 1 and 2. The  different curves correspond to different $\alpha$ values. The orange and green curve depict the change in the potential shape when the value of the $\alpha$ parameter is allowed to vary. 
}}
\end{figure}

Based on the above consideration, the supergravity action is determined by the superpotential 
\begin{align}
W= S\Big((X^0)^2-(X^1)^2\Big)\Bigg(c_0+c_1\Big(\frac{X^1}{X^0}\Big)+c_2\Big(\frac{X^1}{X^0}\Big)^2+c_3\Big(\frac{X^1}{X^0}\Big)^3\Bigg),
\end{align}
and we call the corresponding models $I_\alpha$ depending on the $\alpha$ parameter. 
After gauge fixing the local conformal symmetry in the D-gauge $X^0=\overline X ^0=\sqrt{3}M_\text{Pl}$ and stabilizing the goldstino superfield $S$ at $S=0$, the   inflationary trajectory is determined by the canonically normalized inflaton potential 

\begin{equation}\label{dims}
V=|f(\varphi)|^2=V_0 \left\{c_0+
%\Big(\frac{c_2^2}{3a_3}+\delta\Big)
c_1\tanh\left(\frac{\varphi}{\sqrt{6\alpha}}\right)+c_2\tanh^2\left(\frac{\varphi}{\sqrt{6\alpha}}\right)+c_3\tanh^3\left(\frac{\varphi}{\sqrt{6\alpha}}\right)\right\}^2.
\end{equation}
This is the potential of the Model $I$.
 
For fixed $V_0$  the CMB normalization at the pivot scale $k_\text{cmb}=0.05 \Mpc$ gives a constraint  for the parameters $c_i$.
%in order the power spectrum ${\cal P_R} (k)$ to be compatible with the observed one 
Further constraints are obtained by the number of e-folds and the mass window where the PBHs form.  
For the Model $I$, three set of values for the parameters that can realize the PBH formation are  listed in Table \ref{tab1}. The potential for the Model $I_1$, where the index 1 stands for $\alpha=1$, is depicted with the blue curve  in the Figure \ref{potentials}.  One can see that the potential possesses the required characteristics: an asymptotically flat plateau for large  $\varphi$ values, a second flat plateau about the inflection point  that  decelerates the inflaton and a global minimum where reheating takes place. 
This potential produces  a strong amplification of the curvature power spectrum and thus a significant PBH abundance, see Figure \ref{FigRad}.
The PBH production will be  discussed in detail in the following sections.

\begin{center}
\begin{tabular}{|c||c|c|c|c|c|c|c|} 
\hline
$\alpha$ & $c_3$ & $c_2$ & $c_1$ & $c_0$ & $\varphi_*$ & $ V_0$ & $N_{0.05}$\\ [0.5ex] 
 \hline \hline
 1  & 2.20313  &  -1.426    & 0.3 & 0.16401 & 7.0328 & $2.1\times10^{-10}$ & 55.4   \\ 
\hline
\end{tabular}
\captionof{table}{\label{tab1} A set of  values for the parameters of the Model $I$. The  $\varphi_*$ is the value of the  inflaton field when the scale $k_\text{cmb}=0.05\Mpc$ exits the horizon $N_{0.05}$ e-folds before the end of inflation. The $V_0$ is an overall constant written in Planck units and  fixed by the CMB normalization. } 
\end{center}

\begin{figure}[!htbp]
\begin{subfigure}{.5\textwidth}
  \centering
  \includegraphics[width=1.\linewidth]{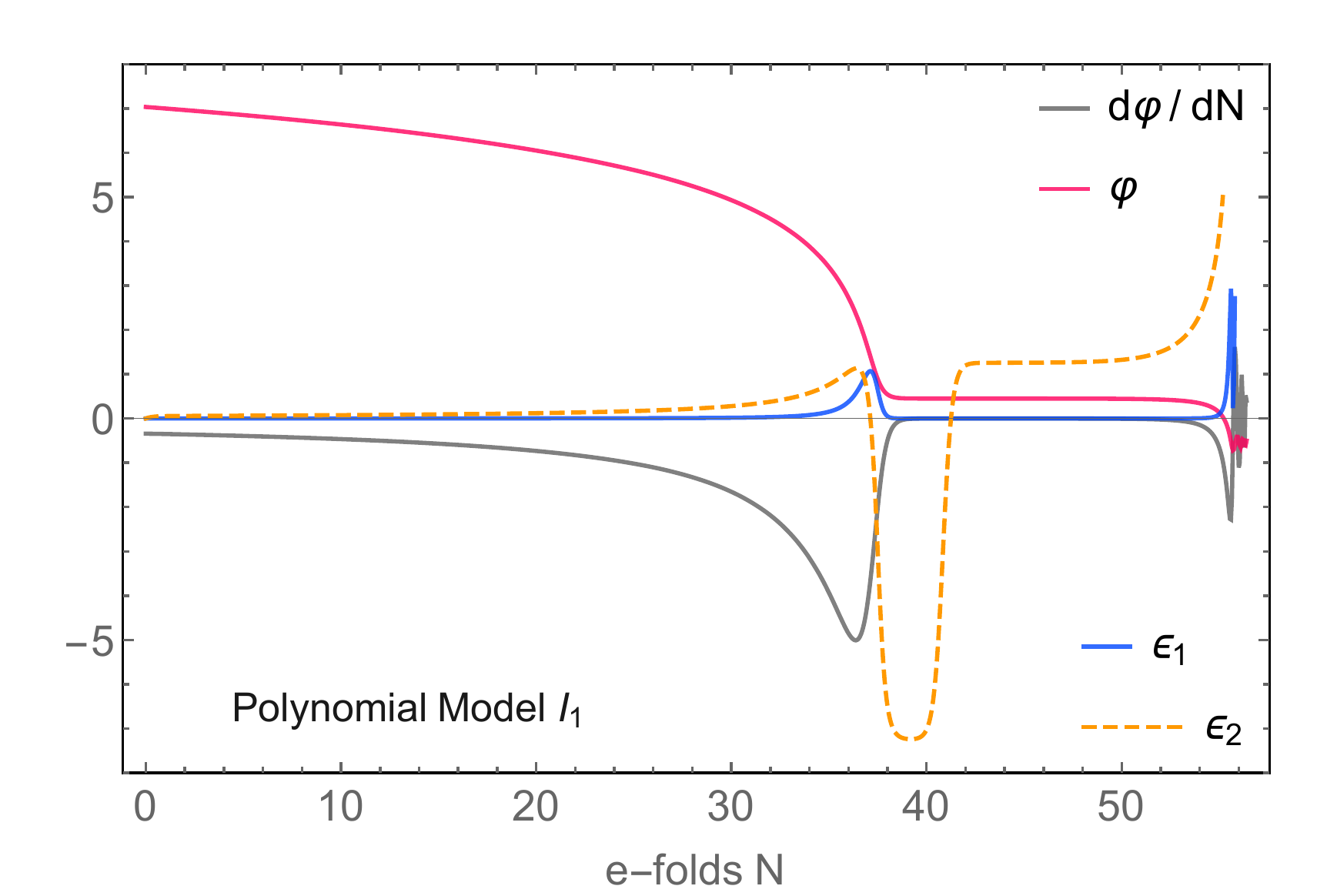}
  %\caption{1a}
  %\label{fig:sfig1}
\end{subfigure}%
\begin{subfigure}{.5\textwidth}
  \centering
  \includegraphics[width=1.05\linewidth]{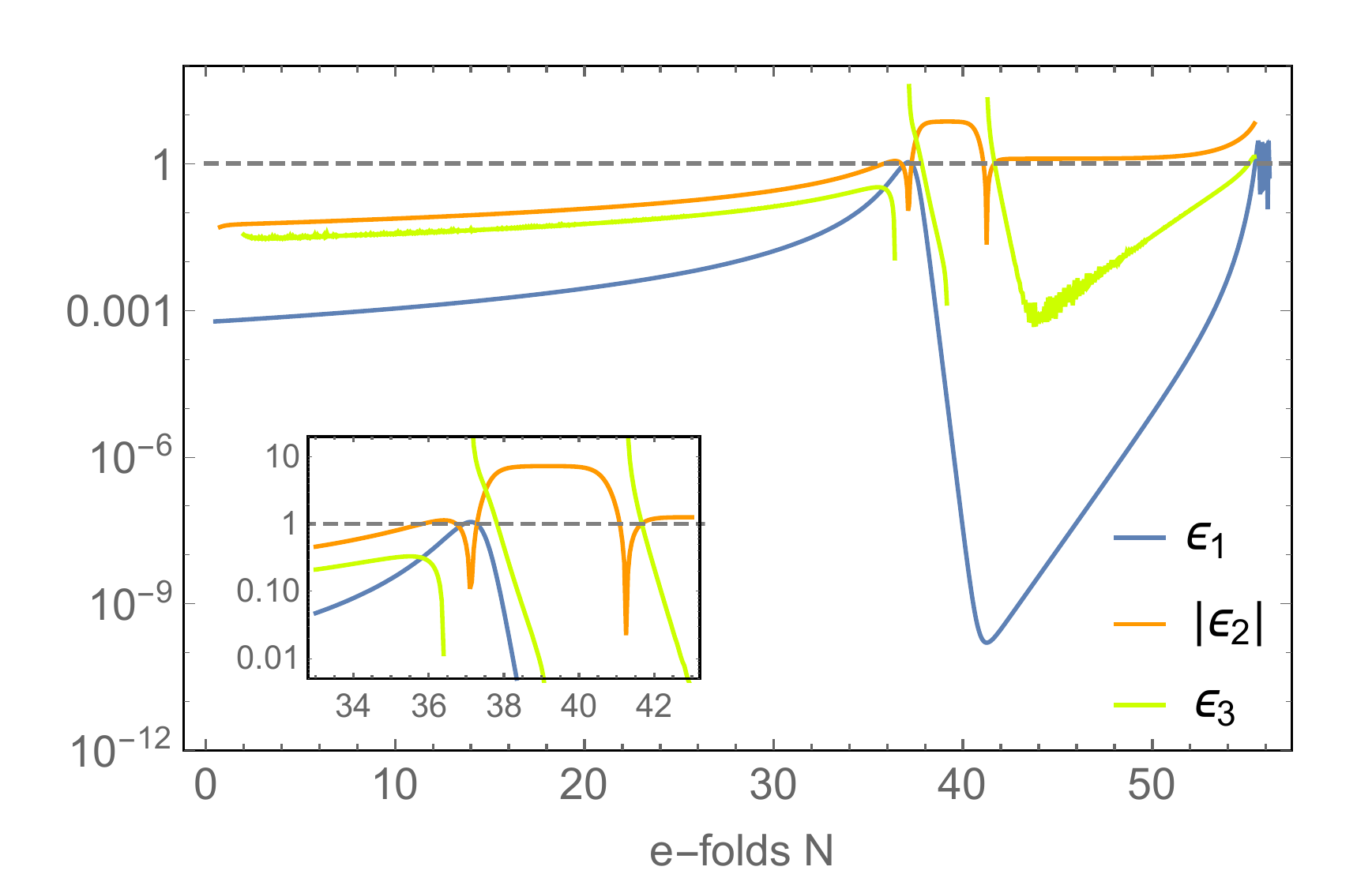}
 % \caption{1b}
  %\label{fig:sfig2}
\end{subfigure}
\caption{\label{TOT} {\it Left panel}: It is shown the evolution of the inflaton field, the inflaton velocity, and the Hubble flow parameters $\epsilon_1$  and $\epsilon_2$  with respect to the number of e-folds $N$ for the polynomial superpotential Model $I_1$. The inflaton field value and velocity (magnified $10^6 $ times) are depicted in Planck units. {\it Right panel}: The $\epsilon_1$, $\epsilon_2$ and $\epsilon_3$ against the number of  e-folds $N$. In the magnified plot one can see that the $\epsilon_1$ parameter is violated and becomes slightly larger than one.
}
\end{figure}

The amplification of the  power spectrum for the curvature perturbation can be understood by examining the inflaton dynamics. Initially, the inflaton finds itself in the asymptotically flat plateau and slowly rolls down the slope. In that region the Hubble flow parameters, defined in Eq. (\ref{eek}), are $\epsilon_1, \epsilon_2, \epsilon_3 \ll 1$ until the gradient of the potential starts to increase. 
Before the inflaton reaches the second plateau 
the kinetic energy becomes maximum  the $|\epsilon_2|, |\epsilon_3|$ become larger than one and for a moment the  first Hubble flow parameters is $\epsilon_1>1$.  For the Model $I_1$ this happens about 36 e-folds after the $k_\text{cmb}^{-1}$ scale exited the horizon, see Eq. (\ref{efolds}) for the definition of the number of e-folds.
 When the inflaton reaches the second plateau 
  the acceleration  becomes comparable to the gradient of the potential. There the slow-roll approximation is not valid. 
The equation of motion in this region becomes \cite{Kinney, Martin:2012pe, Motohashi:2014ppa, Germani, Dimopoulos}

\begin{equation}
\ddot{\varphi}+3H\dot{\varphi}= -V'\simeq 0\,,
\end{equation}
and this phase is known as ultra slow-roll (USR) phase. 
It is during that phase that at least ${\cal O}(1)$ changes in the slow-roll parameters happen and  the power spectrum amplitude gets amplified \cite{Motohashi:2017kbs}. 
The inflaton escapes the USR phase when is reaches the local maximum with the minimum velocity  at $N\approx 41$ e-folds  and,  afterwards, inflation continues till the end for about 16 more e-folds. The aforementioned evolution of the inflaton field is depicted in  Figure \ref{TOT}.

\subsection{Model $II$: Modulated Chaotic Inflation Potentials}
\label{SIN}

It has been demonstrated in the previous section  that polynomial functions with an almost critical  inflection point and single root  maintains these characteristics after the boost $\phi\rightarrow \sqrt{3} \tanh(\varphi/\sqrt{6\alpha})$.  
In addition, appropriate adjustment of the  parameters  can successfully produce a large peak in the scalar power spectrum at specific PBHs mass window. \par

Beside the obvious polynomial functions which possess this behavior, sinusoidal functions appear periodical inflection points and roots.  In this section we study  a class of models with potentials inspired from the natural modulated potentials \cite{Kallosh:2014vja} and the axion monodromy models \cite{SW1,SW2,Easther:2013kla,Flauger:2009ab,Kobayashi:2014ooa}.

\begin{figure}[!htbp]
\begin{subfigure}{.5\textwidth}
  \centering
  \includegraphics[width=1.\linewidth]{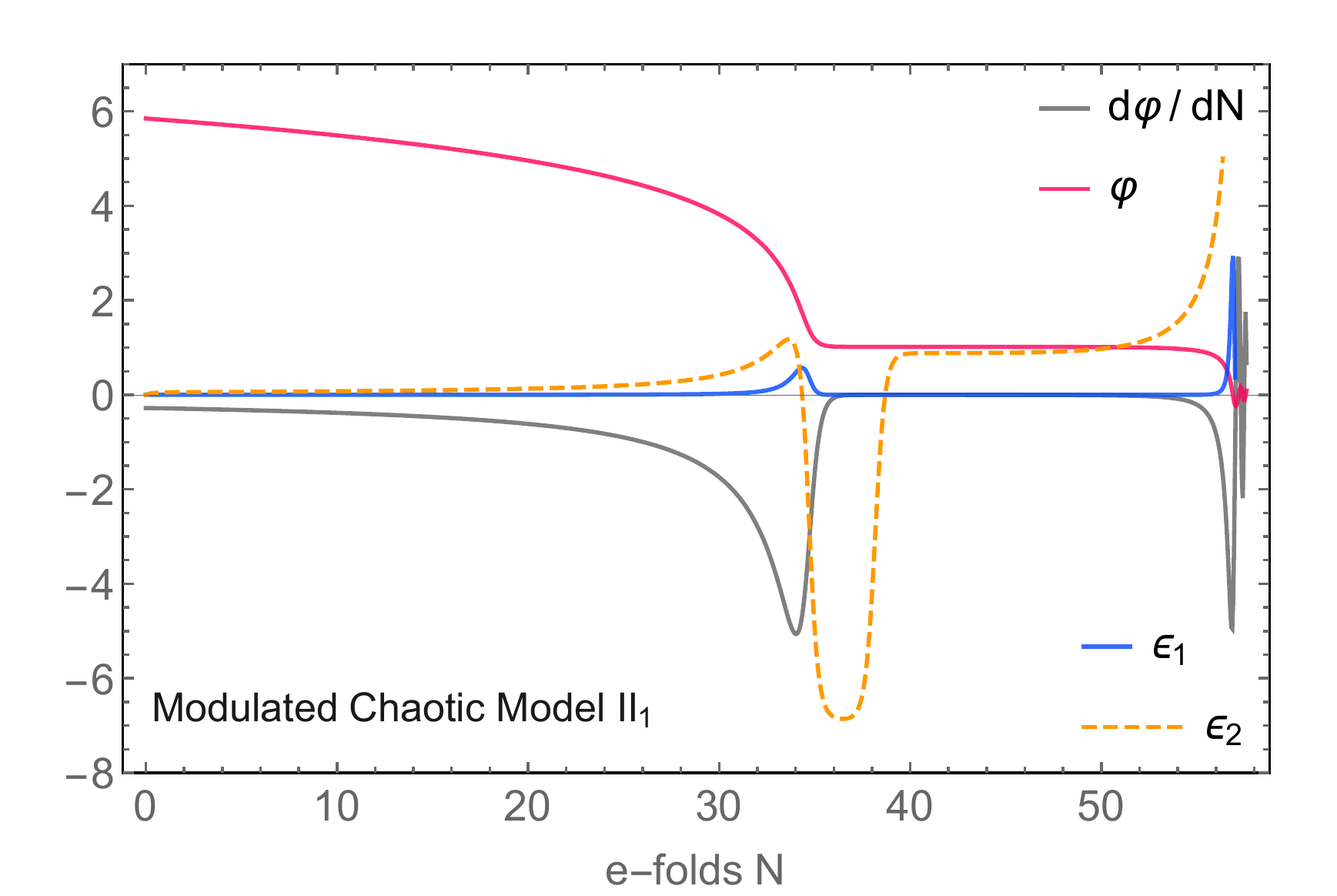}
  %\caption{1a}
  %\label{fig:sfig1}
\end{subfigure}%
\begin{subfigure}{.5\textwidth}
  \centering
  \includegraphics[width=1.05\linewidth]{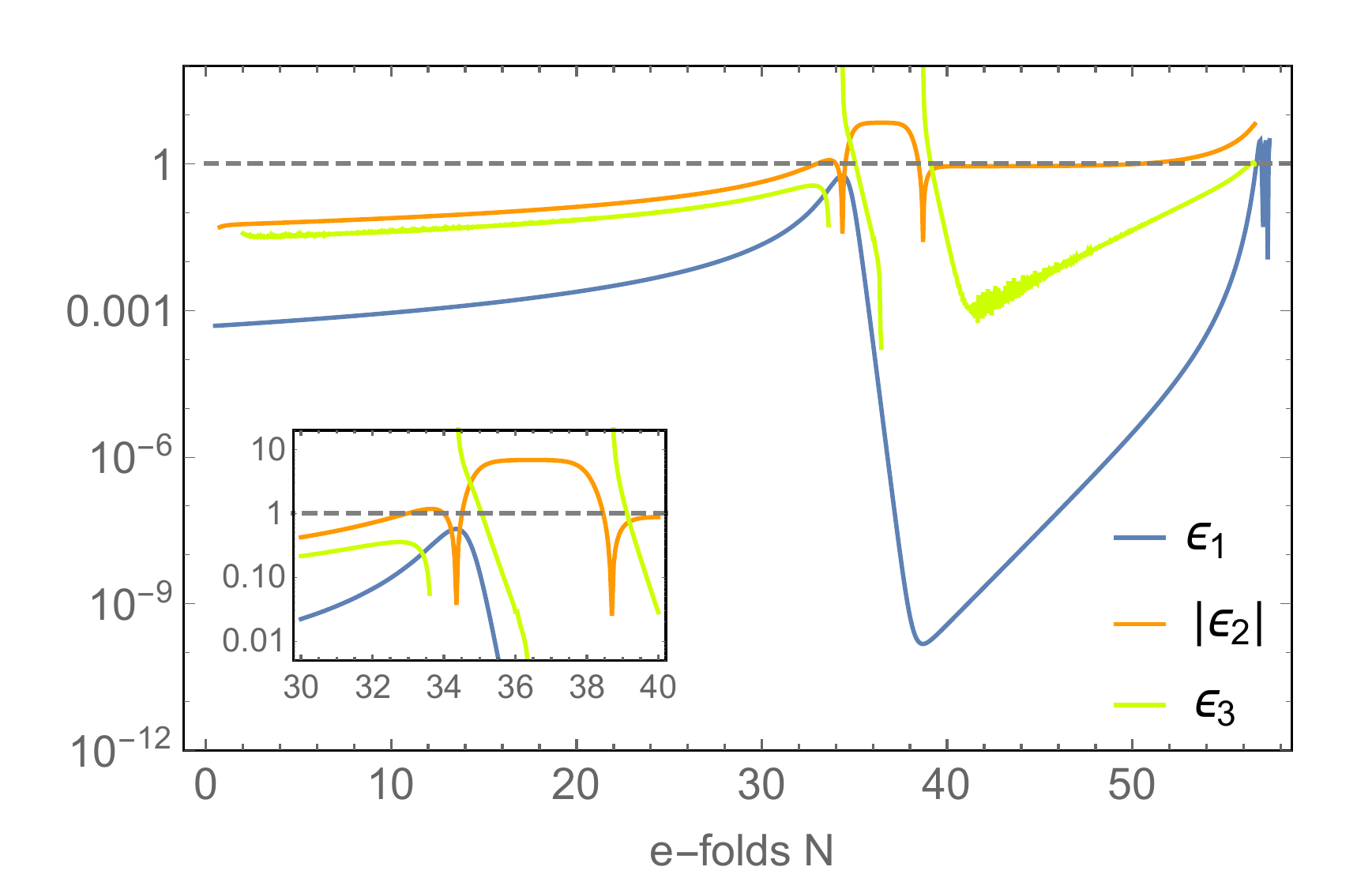}
 % \caption{1b}
  %\label{fig:sfig2}
\end{subfigure}
\caption{\label{SRpar} {\it Left panel}: As in Figure \ref{TOT},  the evolution of the inflaton field, the inflaton velocity, and the Hubble flow parameters $\epsilon_1$  and $\epsilon_2$  with respect to the number of e-folds $N$  for the chaotic modulated Model $II$. The inflaton field value and velocity (magnified $10^6 $ times) are depicted in Planck units.  {\it  Right panel}:  The  $\epsilon_1$,  $\epsilon_2$ and $\epsilon_3$ parameters against the e-folds $N$.  In the magnified plot one sees that the $\epsilon_1$ parameter,  contrary  to the polynomial model, remains less than one.}
\end{figure}

We will study here the  particular case of a modulated chaotic superpotential  \cite{Kallosh:2014vja} of the form (\ref{W}) where $f$ is given in the D-gauge by ($X^1=\Phi$)
\begin{eqnarray}
f(\Phi/\sqrt{3})=g(\Phi/\sqrt{3})+A \lambda  \sin\frac{\Phi}{\sqrt{3}f_\phi},
\end{eqnarray}
where $A,f_\phi,\lambda$ are numerical constants. In others words, the superpotential depends on a generic function $g(\Phi)$ with a sinusoidal modulation.  We will work out the simple case of 
 $\alpha=1$ (other values of $\alpha$ can be analyzed similarly) and a linear function $g(\Phi)=\lambda \Phi$ so that after stabilizing the imaginary part of the complex $\Phi|$, its real part $\phi$ will specify $f(\phi)$ to be
 
 \begin{eqnarray} \label{ssin}
  f(\phi)=\lambda\Big(\phi+A\sin\frac{\phi}{f_\phi}\Big).
  \end{eqnarray} 
Hence, the potential for the canonical normalized field $\varphi$ turns out to be
\begin{eqnarray}\label{chaomod}
V(\varphi)=V_0 \Big[\tanh(\varphi/\sqrt{6})+A\sin\left(\tanh(\varphi/\sqrt{6})/f_\phi\right)\Big]^2, ~~~~V_0=\lambda^2.
\end{eqnarray}
This is the potential of the model that we call  Model $II$.
The position and the amplitude of the power spectrum peak depends critically on the values of the parameters $V_0, \,A$ and $f_\phi$.
The parameters $A$ and $f_\phi$ determine the  plateau  about the inflection point as well as the energy scale. The parameter $\lambda$ is fixed by the CMB scalar power spectrum as usual. 
The values that we assign to the parameters of the Model $II$  are listed in the Table \ref{tab2}. 

\begin{center}
\begin{tabular}{|c||c|c|c|c|c|} 
\hline
  $\alpha$ & $A$ & $f_{\phi}$ & $\varphi_*$  & $V_0$ & $N_{0.05}$ \\ [0.5ex] 
 \hline\hline
  1 & 0.130383 & 0.129576  & 5.85 &  $2\times 10^{-10}$  & 56.2   \\  % 59.65 
 \hline
\end{tabular}
\captionof{table}{\label{tab2} A set of values for the parameters of the Model $II_1$ 
}
\end{center}
For those parameters the inflaton potential has the form illustrated in Figure \ref{potentials}. The dynamics along the inflaton trajectory are very similar to the those described for the polynomial Model $I$.  However, we note that here the first Hubble flow parameter remains less than one, see Figure \ref{SRpar}.

Let us also comment that the potentials of our models, at large field values,  share  similar features with the potentials of Ref. \cite{sone3}.
However,  the potentials and aspects of the phenomenology differ since 
the models correspond to different microscopic configurations and described by distinct K\"ahler and superpotentials.

\begin{figure}[!htbp]
\begin{subfigure}{.5\textwidth}
  \centering
  \includegraphics[width=1.\linewidth]{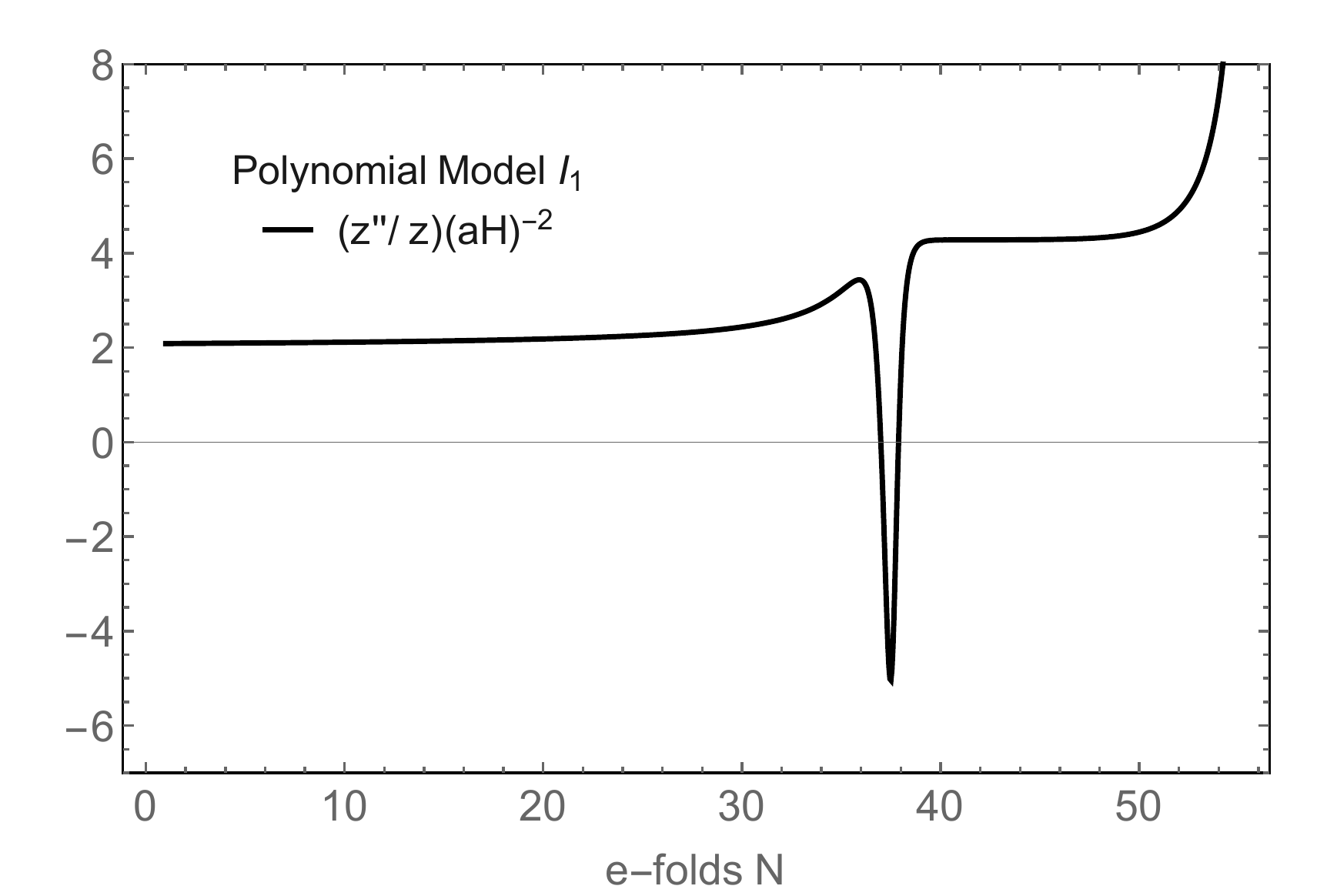}
  %\caption{1a}
  %\label{fig:sfig1}
\end{subfigure}%
\begin{subfigure}{.5\textwidth}
  \centering
  \includegraphics[width=1.\linewidth]{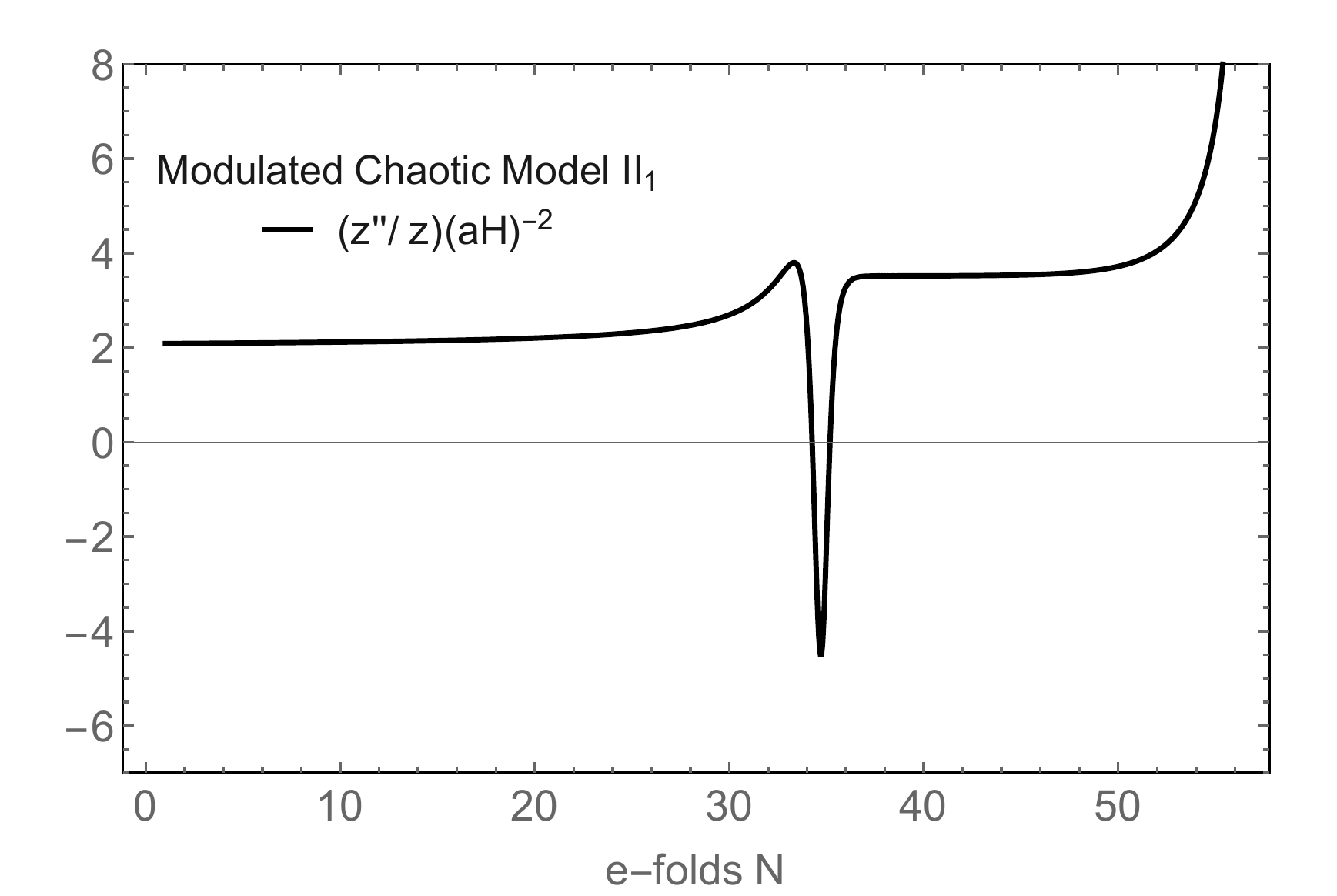}
 % \caption{1b}
  %\label{fig:sfig2}
\end{subfigure}
\caption{ The spike in the $(z''/z)(aH)^{-2}$ evolution caused by the USR phase, for the Models $I_1$ ({\it left panel}) and $II_1$ ({\it right panel}).
\label{ZZ}}
\end{figure}

\section{The Power Spectrum of the Curvature Perturbation}
\label{sec:intro3}

Fluctuations of the scalar field $\phi$ in a homogeneous background induce scalar perturbations in the metric. A gauge invariant quantity is the comoving curvature perturbation ${\cal R}$,
and in the comoving gauge we have  $\delta \phi=0$ and $g_{ij}=a^2 \left[(1-2{\cal R}) \delta_{ij} +h_{ij}\right]$.
The variance in the fluctuations are measured by the power spectrum ${\cal P_ R}(k)$.
In the framework of the inflationary universe the fluctuations naturally
emerge from quantum zero-point fluctuations \cite{Starobinsky:1979ty, Mukhanov:1981xt}. 
Expanding the inflaton-gravity action
to second order in ${\cal R}$ one obtains
\begin{equation}
S_{(2)}= \frac12 \int {\rm d}^4x \sqrt{-g} a^3 \frac{\dot{\phi}^2}{H^2} \left[\dot{{\cal R}}^2 -\frac{(\partial_i {\cal R})^2}{a^2}\right]\,.
\end{equation}
After the variable redefinition $v=z{\cal R}$ where $z^2=a^2\dot{\phi}^2/H^2=2a^2\epsilon_1$ and switching to conformal time $\tau$ 
(defined by $d\tau=dt/a$), the action is recast into
\begin{equation}
S_{(2)} =\frac12 \int {\rm d}\tau {\rm d}^3x \left[(v')^2 -(\partial_i v)^2 +\frac{z''}{z}v^2 \right]\,.
\end{equation}
The evolution of the Fourier modes $v_k$ of $v(x)$ are described by the Mukhanov-Sasaki equation
\begin{equation} \label{MS}
v''_k +\left(k^2-\frac{z''}{z}\right) v_k =0 ,
\end{equation}
where $z''/z$ is  expressed in terms of the Hubble flow functions  
\begin{equation} \label{eek}
\epsilon_1 \equiv -\frac{\dot{H}}{H^2}, \quad
 \epsilon_2 \equiv -\frac{\dot{\epsilon}_1}{H\epsilon_1}, \quad 
 \epsilon_3 \equiv -\frac{{\dot{\epsilon}}_2}{H\epsilon_2}, 
\end{equation}
as 
\begin{equation}
\frac{z''}{z} =(aH)^2 \left[2-\epsilon_1 +\frac32 \epsilon_2 - \frac12\epsilon_1 \epsilon_2 +\frac14 \epsilon^2_2+\frac{1}{2} \epsilon_2 \epsilon_3 \right]. 
\end{equation}
What is critical for our calculations is the super-Hubble evolution of the curvature perturbation, that is for $k^2 \ll z''/z$.  In the large-scale limit the  solution of Eq. (\ref{MS}) is a linear combination of $z$ and $z\int d\tau/z^2$ and since ${\cal R}=v_k/z$ one finds 
\begin{equation} \label{Rsr}
{\cal R} =C_1 +C_2 \int \frac{dt}{a^3 \epsilon_1} \,,
\end{equation}
where $C_1$ and $C_2$ are integration constants. In the conventional single-field inflationary scenarios based on the slow-roll analysis the second term corresponds to a decaying mode and ${\cal R}$ soon becomes a constant after the Hubble exit. 
The power on a given scale  of ${\cal R}$ is obtained  once the  solution $v_k$ of the Mukhanov-Sasaki equation is known and estimated at a time well after it exits the horizon and its value freezes out,
\begin{equation}
 \left. {\cal P_ R} =\frac{k^3}{2\pi^2}\frac{|v_k|^2}{z^2} \right|_{k\ll aH } \,.  \label{ppp}
\end{equation}
The initial conditions for the modes $v_k$ are set by the Bunch-Davies vacuum.
Deep inside the Hubble horizon, $k \gg aH$, the evolution of the $z''/z$ is unimportant because $k^2 \gg z''/z$. There, 
 all modes have time independent 
frequencies and the Eq. (\ref{MS}) reads $v''_k+k^2v_k=0$ that gives  the Minkowski initial condition $v_k =\frac{1}{\sqrt{2k}} e^{-ik\tau}$ for the Mukhanov-Sasaki equation.
In de Sitter space the Eq. (\ref{MS}) is simplified, since it is $z''/z=2/\tau^2$, and one can solve it explicitly. In such a case the power spectrum for ${\cal R}$ in scales larger than the Hubble radius is found to be
\begin{equation}\label{formP}
%\left. 
{\cal P_ R} = \frac{H^2}{8\pi^2 \epsilon_1} \,.
% \right|_{k=aH}\,.
\end{equation}  
The above analytic result  is valid beyond the slow-roll approximation, in the sense that it is derived without neglecting the acceleration and the kinetic energy of the inflaton, and is a very good approximation as long as the Hubble flow slow-roll parameters are much less than one during the inflationary phase. 
If this is not the case the numeric solution of the exact Mukhanov-Sasaki equation has to be pursued.

\begin{figure}[!htbp]
\begin{subfigure}{.5\textwidth}
  \centering
  \includegraphics[width=1.\linewidth]{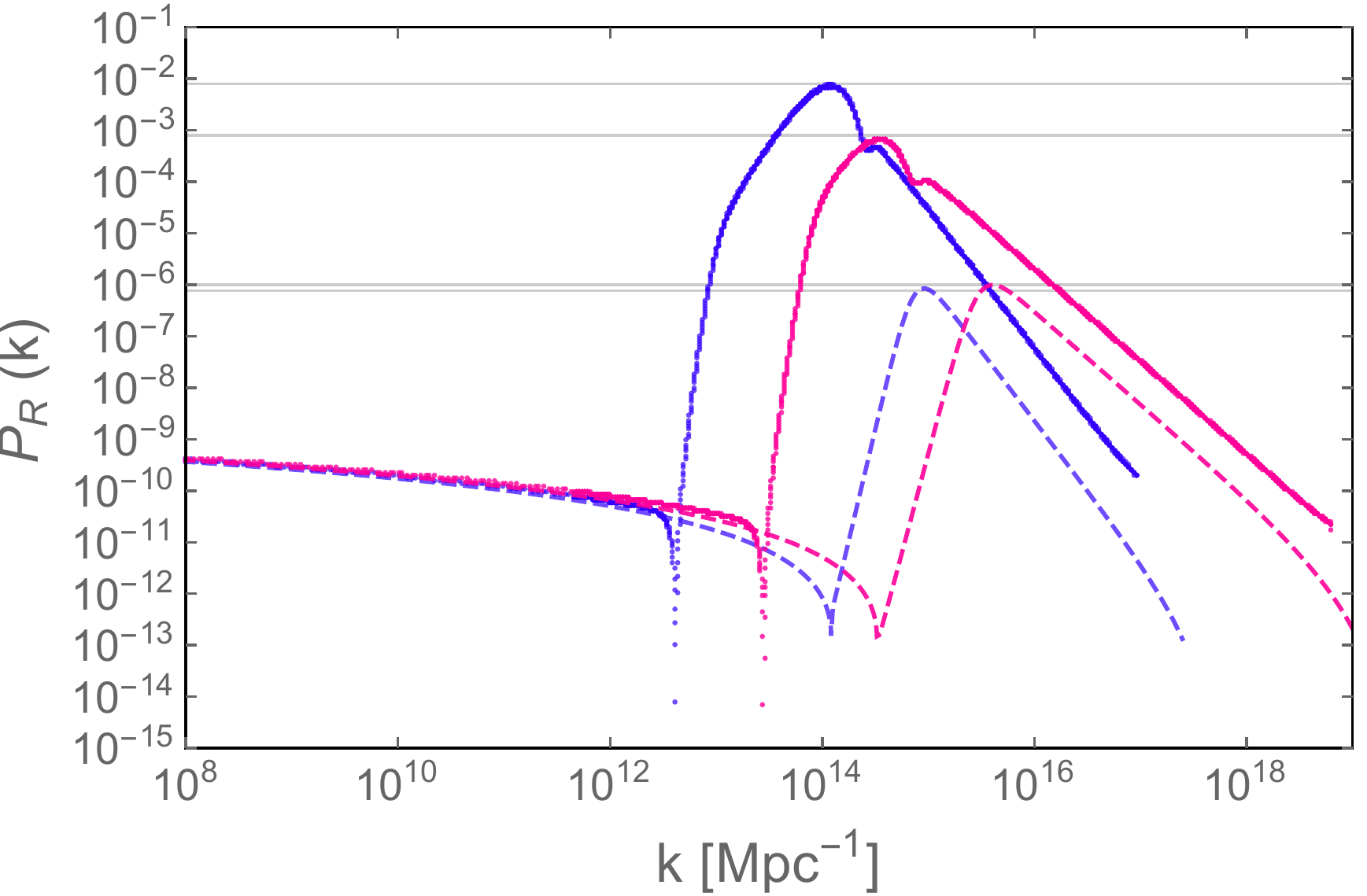}
  %\caption{1a}
  %\label{fig:sfig1}
\end{subfigure}%
\begin{subfigure}{.5\textwidth}
  \centering
  \includegraphics[width=1.\linewidth]{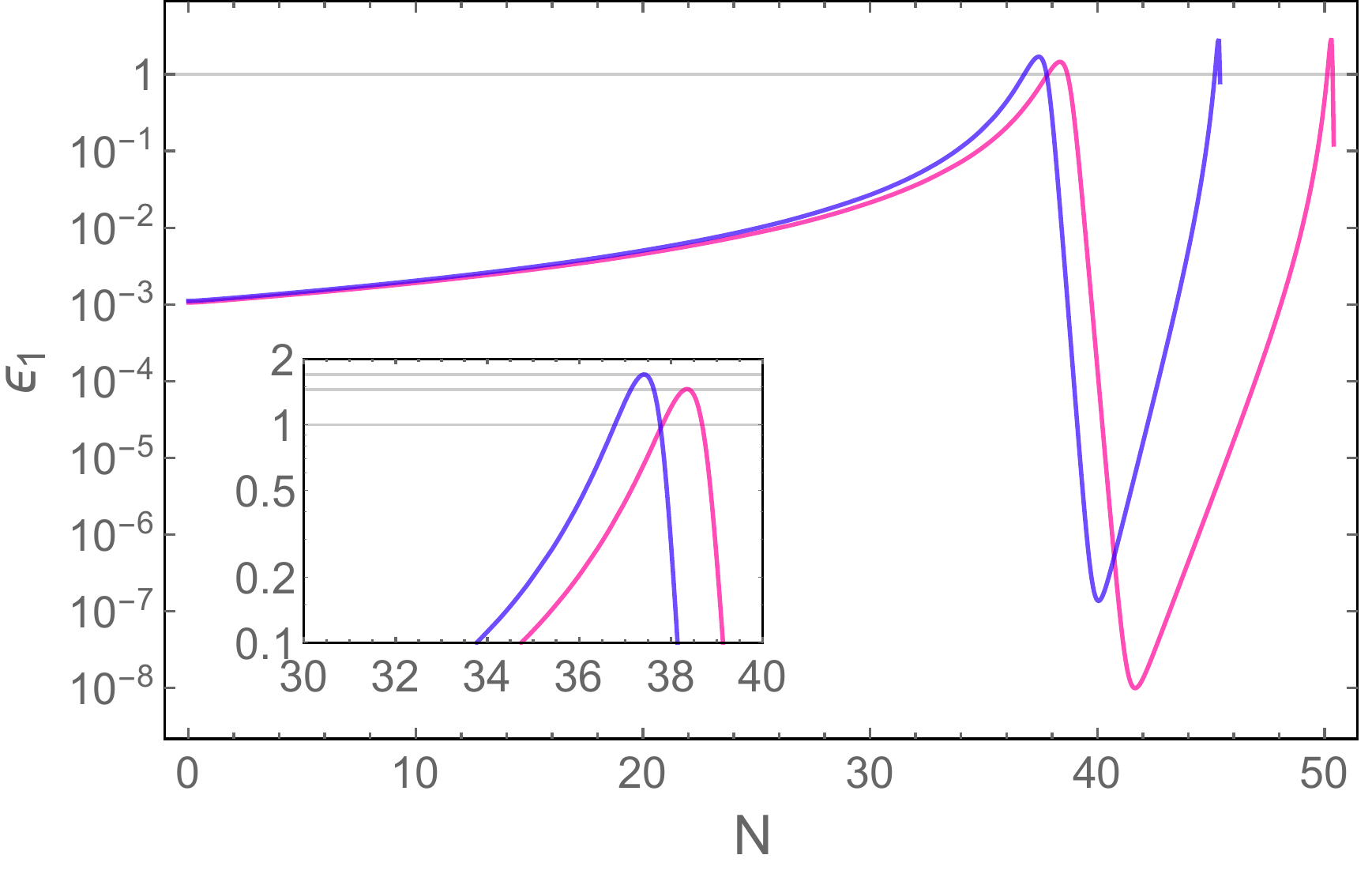}
 % \caption{1b}
  %\label{fig:sfig2}
\end{subfigure}
\caption{ \label{enhanc} 
%In the left plot 
{\it Left panel}: The large enhancement of the power spectrum that comes with the violation of the first slow-roll parameter $\epsilon_1$, shown in the {\it right panel}. 
Although the $\epsilon_1$ parameter becomes 13 times smaller for the model depicted in red than the model depicted in blue, hence the analytic  PS (dashed lines) is larger,  surprisingly enough the actual PS (dense dotted lines) enhancement is 10 times greater for the model in blue. We attribute this enhancement to the fact that the violation of the $\epsilon_1$ parameter is stronger for the  model in blue.  The models  of this figure belong to the Model $I_2$ family with parameters $(c_0, c_1, c_2,  c_3, V_0)=(0.0438, 0.3041, -1.405, 2.09824, 5.5 \times 10^{-10})$ and $(0.0439,  0.272, -1.41711, 2.29189, 4.4 \times 10^{-10})$ depicted in red and blue respectively. 
% MD_Attr1_a=2_new, MD_Attr4_a=2_FIN
%(0.0439, 0.272, -1.41711, 2.29189) and (0.194959,  1.20795, -6.29336, 10.1782)
 }
\end{figure} 

Indeed, during the USR phase there is an ${\cal O}(1)$ change in $\epsilon_1$ and it is $\epsilon_2 \sim -6$, hence the slow-roll approximation breaks down. 
The evolution of $(z''/z) (aH)^{-2}$ with respect to the number of e-folds for inflationary models that feature a USR phase  is depicted in Figure \ref{ZZ}.  We see that the $(z''/z) (aH)^{-2}$ instantaneously vanishes and hence the solution (\ref{Rsr}), obtained for $k^2\ll z''/z$, is not valid for the  inflationary models that feature an inflection point plateau. The $\epsilon_1$ decreases rapidly during the USR due to the decrease of the $\dot{\phi}$, and the second term in Eq. (\ref{Rsr}) is not a decaying mode. 
In Ref. \cite{Motohashi:2014ppa}, after examining particular solutions of the equation of motion for the inflaton field with constant-roll, i.e. for $\kappa_ \text{c} \equiv V'/H\dot{\phi}$ constant, it has been shown that for $-3/2<\kappa_\text{c} <0$ the second solution of Eq. (\ref{Rsr}) corresponds to a growing mode.  This is because the $\epsilon_1 $ decays as $a^{-2(3+\kappa_\text{c})}$ which is faster than $a^{-3}$ for $\kappa_\text{c}>-3/2$, thus the ${\cal R} \simeq \int dt/(a^3 \epsilon_1)$ grows and a super-Hubble evolution of the curvature perturbation takes place. In our scenarios it is $\kappa_c\simeq 0$ in the region about the inflection point and there the curvature perturbation grows fast.  
 Only after the $\epsilon_1$ stops decreasing the  conventional solution ${\cal R} \simeq C_1$ becomes valid again, that is, once  the USR phase is over, the conservation of the  curvature perturbation at super-Hubble scales is recovered.

\subsection{The computation  of the curvature power spectrum}

We define the {\it elapsed} number of e-folds
\begin{equation} \label{efolds}
N\equiv \ln\frac{a}{a_\text{cmb}}=\ln a\, ,
\end{equation}
where we set $a_\text{cmb}=1$ at the moment that the CMB scale $k_\text{cmb}=0.05$ Mpc$^{-1}$ exited the Hubble horizon.  This should not be confused with the usual definition where the e-folds number is given by the logarithm $\ln (a_\text{end}/a)$.  
We also denote $N_{0.05}$ the total the number of e-folds till the end of inflation, i.e. $N_{0.05}=\ln a_\text{end}$ . 
The computation of the solutions to the  Mukhanov-Sasaki equation requires a   systematic numerical approach, that we outline below.
\begin{itemize}
\item We calculate the evolution of the coupled inflaton-metric system at the background level.

\item We solve numerically the Mukhanov-Sasaki equation and find the evolution of the real and the imaginary part of the solution $v_k$.
A numerical iteration is carried out  for more than 2500 modes $k$ that range from $k_\text{cmb}=0.05$ Mpc$^{-1}$ to $k_\text{end}=k_\text{cmb}  (H_\text{end}/H_\text{cmb})\,e^{N_{0.05}}$.
 We apply the Bunch-Davies initial conditions for each mode five e-folds before it crosses the Hubble horizon.  
\item We calculate the power spectrum of each $v_k$  well after the scale exits the Hubble horizon and its value freezes out. 
The region around the peak, which is the region of interest, is shaped by 2000 points.   %that are  produced after solving the Mukhanov-Sasaki equation  and after estimating the $u_k$ values at the end of inflation.
%In total, more than 2500
The points produced numerically construct the 
${\cal P_ R}(k)$ that allows the estimation of the variance $\sigma^2(M_k)$ of the density perturbations and in turn the fraction of the mass that collapses to form PBHs.

\end{itemize}
The scalar power spectrum is normalized at the CMB scale $k_\text{cmb}={0.05}$, where the amplitude $A_s$ of ${\cal P_R}$ is measured to be   $\ln(10^{10} A_s) = 3.089$ \cite{Planck1, Planck2}. 

Special attention should be paid to the case where the slow parameter $\epsilon_1$ becomes larger than one.  The inflaton exits the first slow-roll phase with large velocity and, before it enters the USR regime, the inflationary expansion may break for a moment.  For the models $I_\alpha$, values of  $\epsilon_1 \sim 1.5$ are found for an interval about less than one e-fold. Although the duration is short it may lead to a miscalculation of the spectrum at that scale, for the relevant  $k$ modes cross the horizon three times: they exit, reenter  %for $\epsilon_1>1$, 
and finally exit again.    
 When  such a case is encountered we subtract from the calculation of the power spectrum, ${\cal P_R}(k)$, the points that correspond to solutions of the Mukhanov-Sasaki equation for modes that enter the Hubble horizon, that is in the interval where $\epsilon_1>1$, and keep the final contribution to ${\cal P_R}(k)$ for the same modes during the $\epsilon_1<1$ stage. 

\subsection{Remarks on the amplification of the  power spectrum }

In the initial slow-roll regime the power spectrum is very well approximated by Eq.(\ref{ppp})  until the moment that the Hubble flow functions start to grow considerably. In the region about the inflection point the $\epsilon_1$, $\epsilon_2$ and $\epsilon_3$ are rapidly changing and a peak at the curvature power spectrum is produced. The effective frequency squared $ \omega^2_k\equiv k^2-z''/z$ of the Mukhanov-Sasaki equation is accordingly changing, see Figure \ref{ZZ}, and the strong increase in the ${\cal P_R}$ can be qualitatively understood as a momentarily violation of the adiabaticity condition.    
It is known from the WKB theory that if the frequency $\omega_k$ is varying slowly with time,
then the solutions to the equation (\ref{MS}) do not grow and approximate well the solutions one would obtain assuming that $\omega_k$  is constant. On the other hand, if the $\omega_k$ is changing rapidly, then the WKB analysis breaks down. This behavior 
is quantified by the dimensionless ratio $r_k\equiv {\dot{\omega}_k}/{\omega^2_k}$
which gives the often-called adiabaticity condition. For $|r_k| \ll 1$ we have the adiabatic regime, while for $|r_k|\gg 1$ adiabaticity is violated and a significant amplification of the power spectrum is expected.  This happens for modes that exit the Hubble radius before the non-attractor phase. After horizon exit it is $|r_k|\ll 1$ and the $|r_k|$ is decreasing as $(aH)^{-1}$ until the moment the inflaton reaches the inflection point region. There, the $z''/z$ instantaneously vanishes, it  becomes negative and,  in less than about one e-fold, positive again. The instantaneous divergence of $r_k$ suggests that a resonance-like change in the ${\cal P_R}$ takes place.

An interesting remark is that the inflationary phase is momentarily violated for the polynomial superpotential models, see Figure \ref{TOT}.  The violation is found  to be stronger for $\alpha>1$ values.
Large $\alpha$ values increase the potential energy difference between the CMB and the inflection point plateau. In the intermediate region the kinetic energy of the inflaton is large enough to interrupt for a moment the accelerating expansion.  
It is observed that the stronger the violation of the slow-roll phase is the more  the power spectrum amplitude is amplified with respect to the approximate   $H^2/{(2\pi^2 \epsilon_1)}$ value. The amplification is found to be of order $10^3$, see the Figure \ref{enhanc}.
Hence, large ${\cal P_R}$ values can be also achieved for not too small $\epsilon_1$ or, equivalently, for not too small $\dot{\varphi}$ values.

\section{Primordial Black Hole production}
\label{sec:PBHP}

In this section we  discuss the basic formalism for the PBH formation relevant to the
 models investigated in this work. 
Here, PBH are black holes which are thought to  form in the early universe before the big-bang nucleosynthesis epoch due to large density fluctuations caused by an inflaton that features an inflection point plateau.  Large perturbations of scale $k^{-1}$ create overdense regions that collapse to form black holes  with mass $M_k$ after the horizon reentry. 
 Notably, the PBH formation process depends critically on the background pressure. In the following subsections we discuss the PBH formation during radiation (RD) and matter domination (MD) eras separately.

For a scale $k^{-1}_\text{}$,  which exits the Hubble horizon
 $\Delta N_{k\text{}}$  e-folds  before the end of inflation
 % and the energy density contained collapses and forms a black hole.
there is the relation %between the wavenumber and the e-folds, % $\Delta N_k$ are related to the scale $k^{-1}$ as 
\begin{equation} \label{Dnk}
\Delta N_k= \ln\left(\frac{k_\text{end}}{k}\right)-\ln\left(\frac{H_\text{end}}{H_k}\right)\,.
\end{equation}  
%where  $\Delta N_k=2\tilde{N}_k$.  
Let us assume that  $H_\text{end}\simeq H_k $ which a very good approximation for scales $k^{-1}$ that exit  the Hubble horizon during or after the USR inflationary phase, so the second term in Eq. (\ref{Dnk}) can be neglected.   
In our study, the energy density contained in the horizon of scale $k^{-1}$ may collapse and form a black hole. In order to estimate the probability of such an event it is crucial to specify the epoch of horizon reentry.  
After the end of inflation the Hubble horizon, $H^{-1}$, grows faster than the background expansion and the scales gradually reenter the horizon. The rate that the $H^{-1}$ increases depends on the background energy density and for barotropic parameter $w=p/\rho$, it is $H^{-1} \propto a^{\frac32(1+w)}$.  At the same time the physical  scale $ak^{-1}_\text{}$ grows as $a(t)$ and reenters the horizon when 
\begin{equation}
\left( \frac{a_{k,\text{re}_\text{}}}{a_\text{end}} \right)^{\frac12(1+3w)} =\,e^{\,\Delta N_{k\text{}}}\,,
\end{equation}
where $a_{k,\text{re}_\text{}}$ is the  scale factor at horizon reentry. It is useful to define\footnote{Throughout the text the e-folds that take place during postinflationary expansion are denoted with a tilde, $\tilde{N}$, to make a clear distinction  with the e-folds that take place during inflation. } $\tilde{N}_{\text{}k}\equiv \ln (a_{\text{re},k}/a_\text{end})$, the e-folds that take place after the end of inflation until reentry, hence 
\begin{equation} \label{Ntild}
 \tilde{N}_{k\text{}}=\frac{2}{(1+3w)}\, \Delta N_{k_\text{}}\,,
\end{equation}
where $w>-1/3$. After inflation, there is a transitional era between the supercool inflationary phase and the thermalized radiation dominated phase, required for the BBN, that  is known as reheating. 
The reheating stage can be short or prolonged, depending on the inflaton decay rate, $\Gamma_\text{inf}$. In general, it is characterized by an equation of state that is close to zero, $w=w_\text{rh} \simeq 0$, i.e. the inflaton condensate gravitates as pressureless matter.    Reheating ends at $a_\text{rh}$ with the complete inflaton decay and the energy density stored in the inflaton oscillations transforming into radiation with temperature $T_\text{rh}$. The duration of the reheating phase $\tilde{N}_\text{rh}\equiv \ln (a_\text{rh}/a_\text{end})=\left[3(1+w_\text{rh}) \right]^{-1}\ln \left( {\rho_\text{end}}/{\rho_\text{rh}}\right)$ determines whether the scale $k^{-1}_\text{}$ reenters the horizon during reheating or radiation domination and, in turn,  this determines the PBH formation process.  If it is 
\begin{equation} \label{mrc}
\begin{split} %\nonumber
&\tilde{N}_\text{rh}  < \tilde{N}_{\text{}k}  \quad \Longrightarrow \quad  k^{-1}\,\text{reenters after reheating} \\
& \tilde{N}_\text{rh} > \tilde{N}_{\text{}k} \quad \Longrightarrow \quad  k^{-1}\,\text{reenters during reheating}.
\end{split}
\end{equation}  
%PBH form during a matter domination era. 
For $\tilde{N}_\text{rh} > \tilde{N}_{\text{}k}$  part of the inflaton condensate energy density collapses and forms black holes.
 On the other hand, for $\tilde{N}_\text{rh} > \tilde{N}_{k,\text{}}$ the PBH likely form during a radiation era and it is the thermal plasma that collapses.  But this might not be the case.  In supersymmetric  and stringy scenarios the universe is well possible to enter an era that the energy density is dominated by moduli fields, that we collectively label $X$.  Such a phase is often required due to constraints on the LSP abundance \cite{Dalianis:2018afb} or other stable hidden sector particles which might supplement the dark matter relic density. Thus, it is possible that $\tilde{N}_\text{rh}<\tilde{N}_k$ but the $k^{-1}$ scale enters during a cosmic era where the background pressure is almost zero.

\begin{figure}[!htbp]
\begin{subfigure}{.5\textwidth}
  \centering
  \includegraphics[width=1.\linewidth]{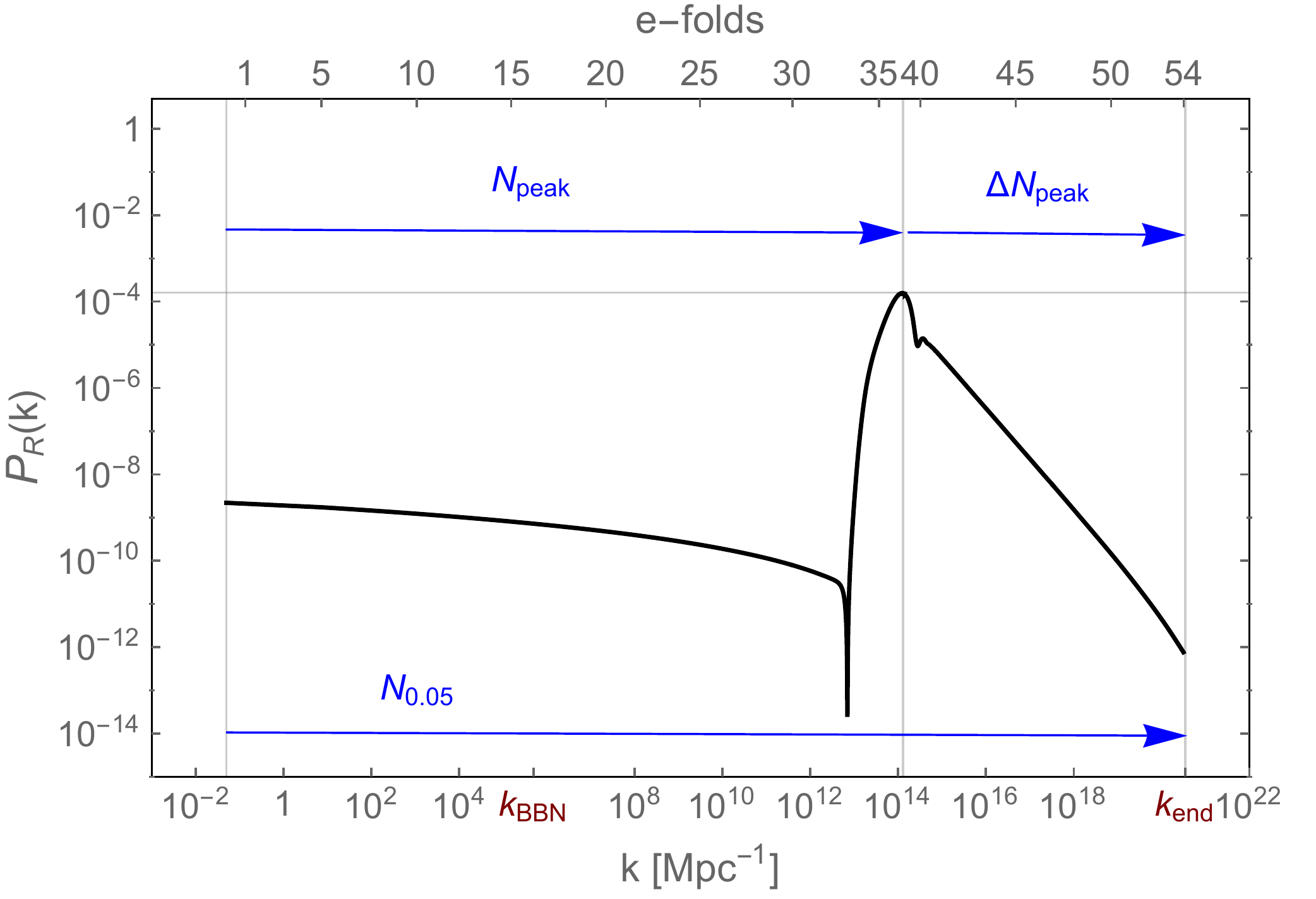}
  %\caption{1a}
  %\label{fig:sfig1}
\end{subfigure}%
\begin{subfigure}{.5\textwidth}
  \centering
  \includegraphics[width=1.\linewidth]{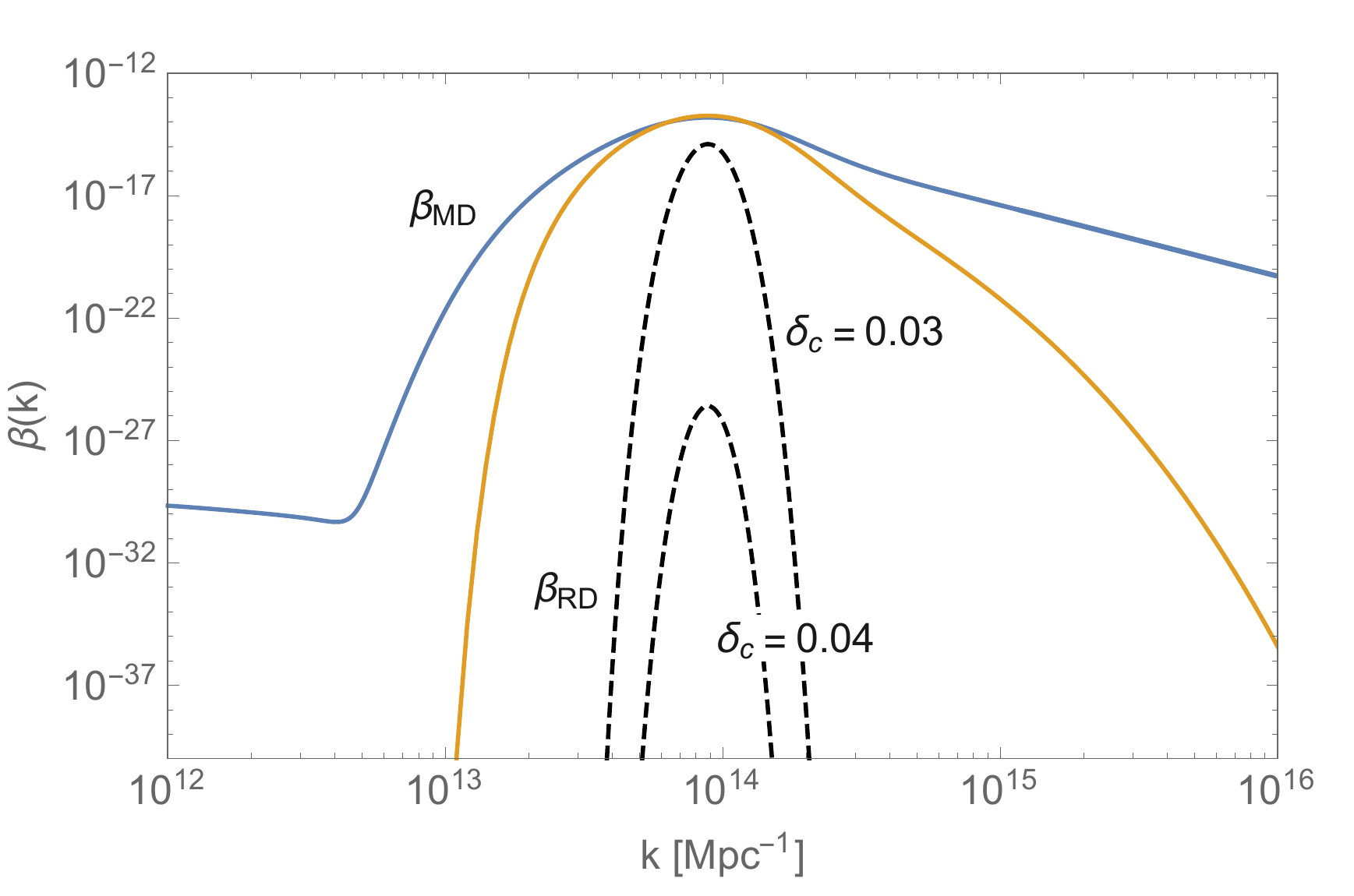}
 % \caption{1b}
  %\label{fig:sfig2}
\end{subfigure}
\caption{\label{PSbeta} {\it Left panel}: 
The power spectrum of the curvature perturbation (here Model $I_5$, example $\#$ 5). 
{\it Right panel}: 
The corresponding mass fraction $\beta(k)$ that collapses into a black hole for matter and radiation domination background for the power spectrum on the left. The upper solid curve gives the $\beta_\text{MD}$  when spin effects are neglected and the lower when spin is important.  Extreme $\delta_c$ values for $\beta_\text{RD}$, depicted with dashed curves, were chosen to make the distributions visible.}
\end{figure}

The characteristic scale is that of the power spectrum peak, $k^{-1}_\text{peak}$, and in our scenarios we find that 
\begin{equation}
5 \lesssim  \Delta N_\text{peak}  \lesssim 20\,,
\end{equation}
see Figure \ref{PSbeta}. If the inflaton condensate decays promptly, it is $w=1/3$ and  the scale $k^{-1}_\text{peak}$ enters the horizon  after $5 \lesssim \tilde{N}_{\text{peak}} \lesssim 20$ e-folds, whereas if prolonged matter domination era follows inflation, it is $w=0$ and  $10 \lesssim \tilde{N}_{\text{peak}} \lesssim 40$.   In our inflationary context, where $\rho_\text{end}>10^{-14} M^4_\text{Pl}$, 
it is possible a cosmic MD era to last more than 40 e-folds after inflation. Hence these  values for the $\tilde{N}_\text{peak}$ are in accordance with the requirement of a radiation dominated universe at the BBN epoch.

We note that the (\ref{mrc}) condition can be written in terms of the temperature  as well. If $T_k$ is the temperature of PBH formation then the hierarchy between $T_k$ and $T_\text{rh}$ determines the PBH formation process. In the following subsection we examine the $T_\text{rh}>T_k$ scenario,  where $T_\text{rh} =\left(\pi^2 g_{*}(T_\text{rh})/90 \right)^{-1/4} \left(\Gamma_\text{inf} M_\text{Pl}\right)^{1/2}$ is the temperature of the radiation due to the inflaton decay, and later the  $T_\text{rh}<T_k$ scenario.

\subsection{PBH production during radiation dominated era}

Let us assume that $T_\text{rh}>T_k$, hence PBH form during a radiation dominated era. The theory for the PBH formation that we follow is based on the traditional Press-Schechter formalism \cite{Press}.  Large perturbations of scale $k^{-1}$ create overdense regions that gravitationally dominate over the radiation pressure and collapse to form black holes  with mass $M_k$ after the horizon reentry. We assume that curvature perturbations are described by   Gaussian statistics in order to estimate the PBH formation probability and  connect the collapse threshold to the power spectrum.  For spherically symmetric regions the PBH form with rate $\beta$,

\begin{equation} \label{brad}
\beta_\text{RD}(M_k)=\int_{\delta_c}d\delta\frac{1}{\sqrt{2\pi\sigma^2(k)}}e^{-\frac{\delta^2}{2\sigma^2(k)}}\,
\simeq \, \frac{1}{2}\text{erfc}\left(\frac{\delta_c}{\sqrt{2}\sigma(k)}\right) 
\, \simeq \,  \frac{1}{\sqrt{2\pi}} \frac{\sigma(k)}{\delta_c} e^{-\frac{\delta^2_c}{2\sigma^2(k)}}  \,.
\end{equation}
%It might be useful to add the subscript RD at this formation rate, i.e.   $\beta=\beta_{RD}$, because it is characteristic of the RD era, however we will   
The parameter $\delta_c$ is the threshold density perturbation and erfc$(x)$ is the complementary error function.  For $\delta>\delta_c$ density perturbations overcome internal pressure and collapse.  The integration in Eq. (\ref{brad}) takes place for $\delta_c<\delta <\delta_\text{max}$ and for  $(\delta_\text{max}-\delta_c)/\sigma\gg 1$ the right part of (\ref{brad}) is a good approximation.
The PBH abundance is particularly sensitive to the threshold $\delta_c$ due to the exponential dependence.
A characteristic value for $\delta_c$ is 1/3 \cite{PBH3} for black hole formation during the radiation dominated era, that is the equation of state parameter value. However, different values for $\delta_c$ are cited in the literature, see e.g. \cite{Niemeyer:1997mt,  Shibata:1999zs, Musco:2008hv,  Musco:2012au, Harada:2013epa}, and its actual value seems to be rather uncertain.  It has been recently suggested that the morphology of the power spectrum plays also a r\^ole \cite{Germani:2018jgr}. 
The $\delta_c$ is a  key quantity for the calculation of the PBH abundance because of the exponential sensitivity of the $\beta_\text{RD}(M_k)$ on $\delta_c$. For example, it has been shown that the PBH abundance can increase during the QCD phase transition due to the softening of the equation of state \cite{Byrnes:2018clq}.    
In the comoving gauge Ref. \cite{Harada:2013epa} finds that 
\begin{equation} \label{dc}
\delta_c = \frac{3(1+w)}{5+3w}\sin^2 \frac{\pi \, \sqrt{w}}{1+3w}\,.
\end{equation}
For $w=1/3$ it is $\delta_c=0.41$.
In this work we adopt different  $\delta_c$ values and demonstrate how our results are modified by this choice. Also, in the comoving gauge the curvature perturbation  ${\cal R}$ is related to the density perturbation $\delta$, assuming a nearly scale-invariant curvature power spectrum for a few e-folds around horizon crossing  %\cite{Drees:2011hb, Young:2014ana} 
as
\begin{equation}
\delta(k, t) = \frac{2(1+w)}{5+3w} \left(\frac{k}{aH} \right)^2 {\cal R}(k, t)\,.
\end{equation}
%{\cal R}_c =\frac{9}{2 \sqrt{2}} \delta_c
Hence, the relation between curvature perturbation and density threshold for a radiation dominated universe is $\delta_c\,=\,{4}/{9}\, {\cal R}_c$ at the moment the perturbations reenter the horizon.
%\begin{equation}
%\delta_c\,=\,\frac{4}{9}\, {\cal R}_c
%\end{equation}
The variance of the density perturbations $\sigma(k)$ smoothed on a scale $k$ for radiation domination is given by \cite{Young:2014ana}

\begin{equation}
\sigma^2(k)= \left( \frac{4}{9} \right)^2  \int \frac{dq}{q}W^2(qk^{-1})(qk^{-1})^4{\cal P_R}(q)\,,
\end{equation}
where ${\cal P_R}(q)$ is the power spectrum of the curvature perturbations calculated numerically. $W(z)$ represents the Fourier transformed function of the Gaussian window,  $W(z)=e^{-z^2/2}$. 
In order to estimate the mass spectrum of the PBHs  
the horizon scale  at the time of reentry of the perturbation mode $k$ has to be related to the mass of  PBH formed. 
 During the radiation era the wavenumber scales like $k_\text{} \propto g^{1/2}_* g^{-2/3}_s  S^{2/3} a^{-1}$ and the 
 Hubble horizon as $H \propto g^{1/2}_* g^{-2/3}_s  S^{2/3} a^{-2}$ where $S$ denotes the entropy, and $g_*$, $g_s$ count the total number of the effectively massless degrees of freedom for the energy and entropy densities respectively. The horizon mass $M_H=4\pi \rho H^{-3} /3 $  scales like $\propto H^{-1}$. Hence,  assuming conservation of the entropy between the reentry moment and the epoch of radiation-matter equality, which is a good approximation unless significant entropy production takes place in between,
   the relation between the PBH mass $M_k$ and the comoving wavenumber $k$ is given by %\cite{}

\begin{align}\label{mas}
M_k\equiv M(k)=\gamma \rho \frac{4\pi H_k^{-3}}{3} \Big|_{k=a H}&\simeq \frac{\gamma M_\text{eq}}{\sqrt{2}}\left(\frac{g_{\text{eq}}}{g_{}(T_k)}\right)^{\frac{1}{6}}\left(\frac{k}{k_\text{eq}}\right)^{-2} \\ 
&\simeq 2.4 \times 10^{-16} M_{\odot}\Big(\frac{\gamma}{0.2}\Big)\left(\frac{g(T_{k})}{106.75}\right)^{-\frac{1}{6}}\left(\frac{k}{10^{14}\,\text{Mpc}^{-1}}\right)^{-2} \,. \nonumber
\end{align}

The factor $\gamma$ gives the fraction of the horizon mass $M_H$ that collapses to form PBHs. Its value depends on the details of the gravitational collapse and an analytical estimation  \cite{PBH3}  gives $\gamma=0.2$. However, the $\gamma$ value varies in the  literature and the choice of it alters the mass window that PBHs are produced.
 In Eq. (\ref{mas}) we assumed that the degrees of freedom  $g_*$, $g_{*s}$ are approximately equal and labled $g$. 
For the estimation of PBH mass we have considered the number of degrees of freedom at the temperature of the PBH formation $T_k$ in the radiation epoch to be $g_*(T_{k})=106.75$, assuming that particles are those of the Standard Model. 
Here, since  we consider that the supersymmetric $\alpha$-attractors realize the inflationary phase,  supersymmetric degrees of freedom may participate in the thermal equilibrium. In case the MSSM is thermalized then it is  $g_*(T_k)=228.75$. 
 At radiation-matter equality  the comoving wavenumber is $k_{\text{eq}}=0.07 \,\Omega_\text{m} h^2 \Mpc$ and $M_\text{eq}$ denotes the corresponding horizon mass.  

 The mass of each PBH formed  is given by the  Eq. (\ref{mas}).
 The other critical quantity is the PBH abundance.   
 In order to estimate the PBH abundance we take into account that PBHs behave as non-relativistic matter, hence $\rho_\text{PBH}/\rho$ grows inversely proportional to the temperature of the thermal plasma until the radiation-matter equality. We define $f_\text{PBH}$ the present ratio of the PBH abundance with mass $M_k$ over the total dark matter (DM) abundance
 \begin{equation}
\left. f_\text{PBH}(M_k)\equiv \frac{\Omega_{\text{PBH}}(M_k)}{\Omega_{\text{DM}}} 
\simeq  \frac{\Omega_\text{m}}{\Omega_\text{DM}} \frac{\rho_\text{PBH}(M_k)}{\rho_\text{rad}} \right|_\text{eq}\,.
 \end{equation}
 In the above expression we introduced the total matter relic density parameter $\Omega_\text{m}$ and took into account that $\Omega_\text{m}=\Omega_\text{rad}$ at equality epoch. Hence, given that at the moment of horizon reentry of the perturbation mode $k$, the energy density that collapses to PBH is $\gamma \beta_\text{}(M_k)\rho$, along with the fact that it redshifts slower than radiation, one finds that at the equality epoch it is 
% \begin{equation} 
 %\left .
  $f_\text{PBH}(M_k) =
      \gamma \, \beta_\text{RD}(M_k)\,  (T_{k}/T_\text{eq})({\Omega_\text{m} }/{\Omega_\text{DM} })$. 
      %\,  \right|_\text{eq} \,.
 %\end{equation} 
Assuming entropy conservation, thus the temperature scales like $T\propto g_s^{-1/3} a^{-1}$, that $H^2 \propto g_*T^4 $ and  $M_k\propto \gamma H^{-1}$  we get
 \begin{equation}
  f_\text{PBH}(M_k) \,= \, 2^{-3/4}\,
      \frac{\Omega_\text{m} h^2}{\Omega_\text{DM} h^2 } \,\gamma^{3/2} \, \beta_\text{RD} (M_k)\,\left( \frac{g_{\text{eq}}}{g(T_{k})}   \right)^{1/4}   \left(  \frac{M_k}{M_\text{eq}}\right)^{-1/2} \,.        
 \end{equation}
 In the above expression we utilized the formula for the  PBH mass $M_k$, Eq. (\ref{mas}),  formed at the moment that $T=T_{k}$. 
 We took again the effective degrees of freedom $g_*$ and $g_s$ approximately equal and labeled $g$.  The factor $2^{-3/4}$ comes from the fact that at equality epoch the energy density is equally partitioned between matter and radiation.
 We also ignored as negligible mass accretion and evaporation effects.
  Plugging in numbers the above expression  is recast into a more  informative form,
 
% &=\frac{\Omega_{\text{PBH}}(M)}{\Omega_{\text{DM}}}=\frac{\rho_{\text{PBH}}(M)}{\rho_m}\Bigg|_{\text{eq}}\frac{\Omega_mh^2}{\Omega_{\text{DM}}h^2} \\ % &
\begin{equation}
f_\text{PBH} (M_k)\, =\,\left(\frac{\beta_\text{RD}(M_k)}{8\times 10^{-15}}\right) \, 
\left(   \frac{\Omega_{\text{DM}}h^2}{0.12}   \right)^{-1}  
 \Big(\frac{\gamma}{0.2}\Big)^{\frac{3}{2}} 
\left(\frac{g(T_{k})}{106.75}\right)^{-\frac{1}{4}}  
\left(\frac{M_k}{M_{\odot}}\right)^{-1/2}\,.
\end{equation}
This is the fractional  relic density  of PBH with mass $M_k$ with respect to the overall dark matter density. %\cite{Inomata}
In order to see whether the total relic density of the PBH saturates the observed dark matter relic density, $\Omega_{\text{DM}}h^2\sim 0.12$ \cite{Planck1}, we have to integrate over the masses $M_k$ for $k$ in the RD era, 
\begin{equation} \label{ftot}
f_\text{PBH,tot} = \int_M\, d \ln M\, f_\text{PBH} (M)\,= 2 \int_k\, d \ln k\, f_\text{PBH} (M_k)\,.
\end{equation}

Here, we choose the parameters in order the  PBH to account for a significant part of the dark matter in the universe and hence, test our models against observations, see Table 3.

\subsection{PBH production during matter dominated era} 

After inflation the transition to thermalized plasma, in most of the models, is not instantaneous but a reheating stage follows.   Furthermore, 
it is well possible that  in the era before the BBN and after infation the universe energy density is dominated by a pressureless fluid other than the inflaton, e.g a modulus field.  Actually, this is rather common and often natural in the context of supersymmetric extensions of the Standard Model of particle physics and in stringy set-ups where late decaying scalar fields are generic.
 In both of these scenarios, reheating or modulus domination,  the PBH formation process gets modified.  
Such scenarios have been examined in the past in Ref. \cite{Khlopov:1980mg, Polnarev:1986bi}.
 We note that, in our context,  black hole formation by the collapse of a scalar condensate is generally expected to take place, see also e.g. \cite{Helfer:2016ljl} for a relevant analysis.
 
 According to the formula (\ref{dc}) in the limit $w\rightarrow 0$  we have $\delta_c \rightarrow 0$, which means that even minute perturbations would collapse to a black hole.  In  this case  the Eq. (\ref{brad}) is not the correct expression to estimate the PBH formation rate. 
Ref. \cite{Harada:2016mhb} examined the PBH production in a matter dominated universe and found that it is generally larger, though non-spherical effects are present and decrease the efficiency of the collapse in a pressureless background. For small $\sigma$ the PBH production rate  $\beta$ tends to be proportional to $\sigma^5$, 
\begin{equation} \label{bmat}
\beta_\text{MD}(M_k)\, = \, 0.056 \, \sigma^5(k)\,.\end{equation}  
This expression has to be contrasted with the gaussian expression (\ref{brad}). It was derived with semi-analytical calculations while analytically a lower and an upper bound were found, $0.014 \, \sigma^5 <\beta< 0.128 \, \sigma^5$ . The PBH production rate is further modified when the collapsing region has spin. The angular momentum suppresses  the formation rate which now  reads \cite{Harada:2017fjm},
\begin{equation} \label{spin}
\beta_\text{MD}(M_k)=2\times 10^{-6}f_q(q_c) {\cal I}^6 \sigma(k)^2 e^{-0.147\frac{ \, {\cal I}^{4/3}}{ \sigma(k)^{2/3}}}\,.
\end{equation}
Benchmark values are $q_c=\sqrt{2}$, ${\cal I}=1$, $f_q \sim 1$. According to \cite{Harada:2017fjm} this expression applies for $\sigma(k) \lesssim 0.005$  whereas Eq.  (\ref{bmat}) applies for $0.005 \lesssim \sigma(k) \lesssim 0.2$.  Depending on the amplitude of the power spectrum, we will examine  examples that fall in the validity regime of Eq. (\ref{bmat}) and examples that the Eq. (\ref{spin}) gives  major corrections to the final result,  see Figures \ref{FigReh} and \ref{FigMod}.
Also, the variance of the density perturbations $\sigma(k)$ smoothed on a scale $k$ for matter domination era is given by 
\begin{equation}
\sigma^2(k)= \left( \frac{2}{5} \right)^2  \int \frac{dq}{q}W^2(qk^{-1})(qk^{-1})^4{\cal P_R}(q)\,.
\end{equation}

During matter domination the relation of the perturbation mode $k$  to the PBH mass formed is different than Eq. (\ref{mas}). 
The horizon mass $M_H=4\pi\rho H^{-3}/3$ scales like $H^{-1}$. 
Right after inflation it is $M_\text{end}=4\pi M^2_\text{Pl}/H_\text{end} $ and grows like $a^{3/2}$ %or  $e^{\tilde{3\tilde{N}/2}}$   
during a pressureless reheating stage,  while during a RD era it grows like $a^2$. % $e^{\tilde{2N}}$. 
Hence, the relation between the PBH mass $M_k$ and the comoving scale $k^{-1}$  %that enters horizon  as $\propto a^{1/2}$ 
reads
\begin{align}\label{masMD}
M_k=\gamma \rho \frac{4\pi H^{-3}}{3} \Big|_{k=a H}&\simeq 
\gamma M_\text{end} \left(\frac{k}{k_\text{end}} \right)^{-3}\,.
\end{align}
Accordingly,  in terms of the $M_\text{rh}$, $M_\text{eq}$ and for $k>k_\text{rh}$ we obtain
\begin{align} \label{masMD2} 
 \nonumber
M_k & =\gamma M_{\text{rh}} \, \left(\frac{k}{k_\text{rh}}\right)^{-3} \\
& \simeq  \frac{\gamma M_\text{eq}}{\sqrt{2}}\left(\frac{g_{\text{eq}}}{g(T_\text{rh})}\right)^{\frac{1}{6}}\left(\frac{k_\text{rh}}{k_\text{eq}}\right)^{-2} \left(\frac{k}{k_\text{rh}}\right)^{-3} \\ 
&\simeq 2.4 \times 10^{-16} M_{\odot}\Big(\frac{\gamma}{0.2}\Big)\Big(\frac{g(T_\text{rh})}{106.75}\Big)^{-\frac{1}{6}}\left(\frac{k_\text{rh}}{10^{14}\,\text{Mpc}^{-1}}\right)^{-2} e^{-\frac{3}{2}(\tilde{N}_\text{rh}- \tilde{N}_k)}  \,, \nonumber
\end{align}
where $k^{-1}_\text{rh}$ is the comoving horizon scale at the epoch of reheating, $M_\text{rh}$ is the horizon mass for $k=k_\text{rh}$  and $\tilde{N}_\text{rh}$,  $\tilde{N}_k$ are respectively the e-folds after infation that the scales $k^{-1}_\text{rh}$ and $k^{-1}$ reenter the horizon, i.e.  $\tilde{N}_k\equiv \ln(a_k/a_\text{end})=2 \ln(k_\text{end}/k)$ with $ \tilde{N}_k<\tilde{N}_\text{rh}$.
 Again here, we took $g_* \cong g_\text{s}\equiv g$ and we comment that if the MSSM is thermalized at $T_\text{rh}$ then it is  $g_*(T_k)=228.75$.  
The Eq. (\ref{masMD2}) for $k_\text{end}$ gives the horizon mass at the end of inflation and has to coincide with the Eq. (\ref{masMD}). 
In the present analysis we assume a {\it one-to-one} correspondence between the scale of perturbation and the mass of PBHs, and we make the rough approximation of  {\it instantaneous collapse} during the MD era.

Turning to the abundance, one should take into account that the energy density of the PBHs will  redshift slower than radiation only after $T_\text{rh}$ hence, $f_\text{PBH}(M_k) = \gamma \, \beta_\text{MD}(M_k)\,  (T_\text{rh}/T_\text{eq})({\Omega_\text{m} }/{\Omega_\text{DM} })$. 
Assuming entropy conservation after reheating we get for $k>k_\text{rh}$
 \begin{align}\label{fmd} 
  f^{(\text{MD})}_\text{PBH}(M_k) & = \, 2^{-3/4}\,
      \frac{\Omega_\text{m} }{\Omega_\text{DM} } \,\gamma^{} \, \beta_\text{MD}(M_k)\, \left( \frac{g_{\text{eq}}}{g(T_\text{rh})}   \right)^{1/4}
       \,      \left(  \frac{M_\text{rh}}{M_\text{eq}}\right)^{-1/2}  \\
        &= \, 2^{-3/4}\,    \frac{\Omega_\text{m} }{\Omega_\text{DM} } \,\gamma^{3/2} \, \beta_\text{MD}(M_k)\, \left( \frac{g_{\text{eq}}}{g(T_\text{rh})}   \right)^{1/4}
       \,      \left(  \frac{M_k}{M_\text{eq}}\right)^{-1/2}   \, e^{-\frac{3}{4}(\tilde{N}_\text{rh}- \tilde{N}_k)} \,  . 
 \end{align}
Equivalently, the abundance of PBH  formed during an early matter domination era are modified with respect to the abundance of the RD era as 
 \begin{equation} \label{mod}
  f^{}_\text{PBH}(M_k) 
\, \longrightarrow \, \frac{T_\text{rh}}{T_k} \, f^{}_\text{PBH}(M_k)\,,
 \end{equation} 
for  $T_\text{rh}<T_k$ and $M_k$ given by Eq. (\ref{masMD2}).  
Here $T_k$ does not have the standard physical meaning since universe is not thermalized. 
The total abundance of PBH produced during pressureless reheating is numerically obtained by integrating in the momentum space,
\begin{equation} \label{ftotmd}
f_\text{PBH,tot} = \int_M\, d \ln M\, f_\text{PBH} (M)\,= \,3  \int^{k_\text{end}}_{k_\text{rh}}\, d \ln k\, f_\text{PBH} (M_k)\,.
\end{equation}

% Eq. (\ref{ftot}), and after taking into account the MD PBH abundance (\ref{fmd}). 

 In the case  PBH form during a modulus dominated era, $T_\text{rh}$ has to be replaced by the decay temperature of the modulus field. We will come back to the scenario of  modulus domination in subsection  6.2.

  \begin{figure}[!htbp]
\begin{subfigure}{.5\textwidth}
  \centering
  \includegraphics[width=1.\linewidth]{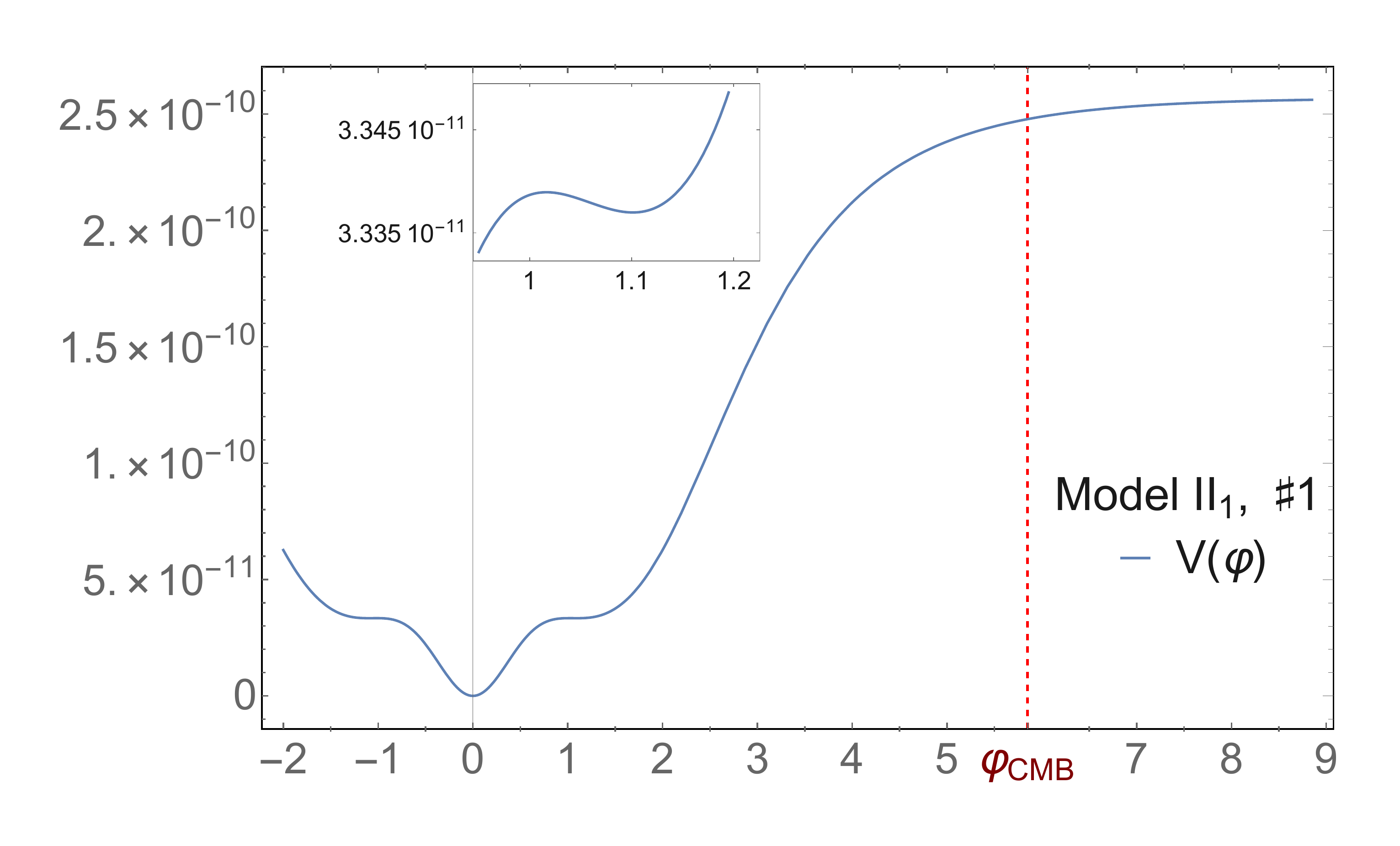}
  %\caption{1a}
  %\label{fig:sfig1}
\end{subfigure}%
\begin{subfigure}{.5\textwidth}
  \centering
  \includegraphics[width=1.\linewidth]{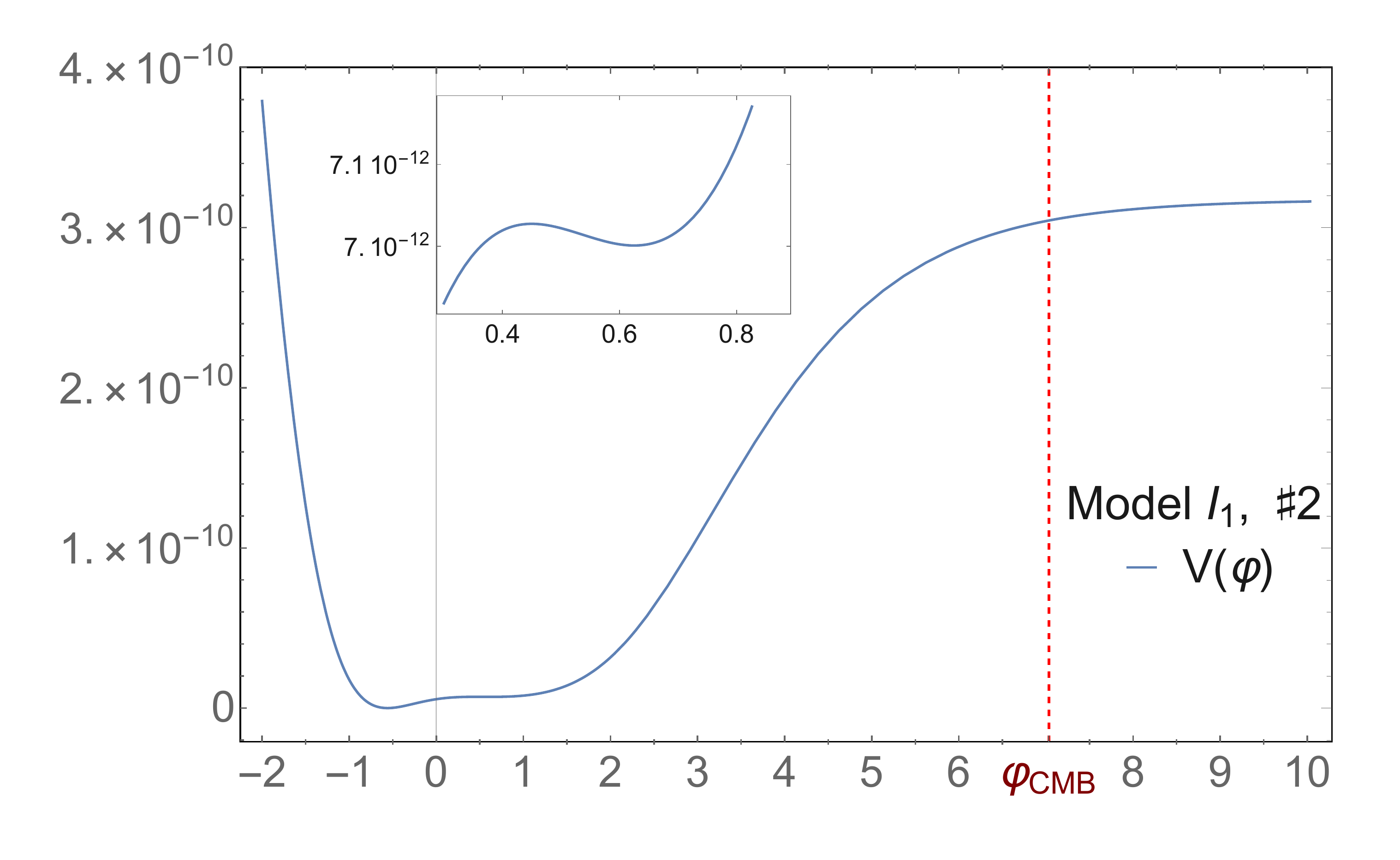}
 % \caption{1b}
  %\label{fig:sfig2}
\end{subfigure}
\begin{subfigure}{.5\textwidth}
  \centering
  \includegraphics[width=1.\linewidth]{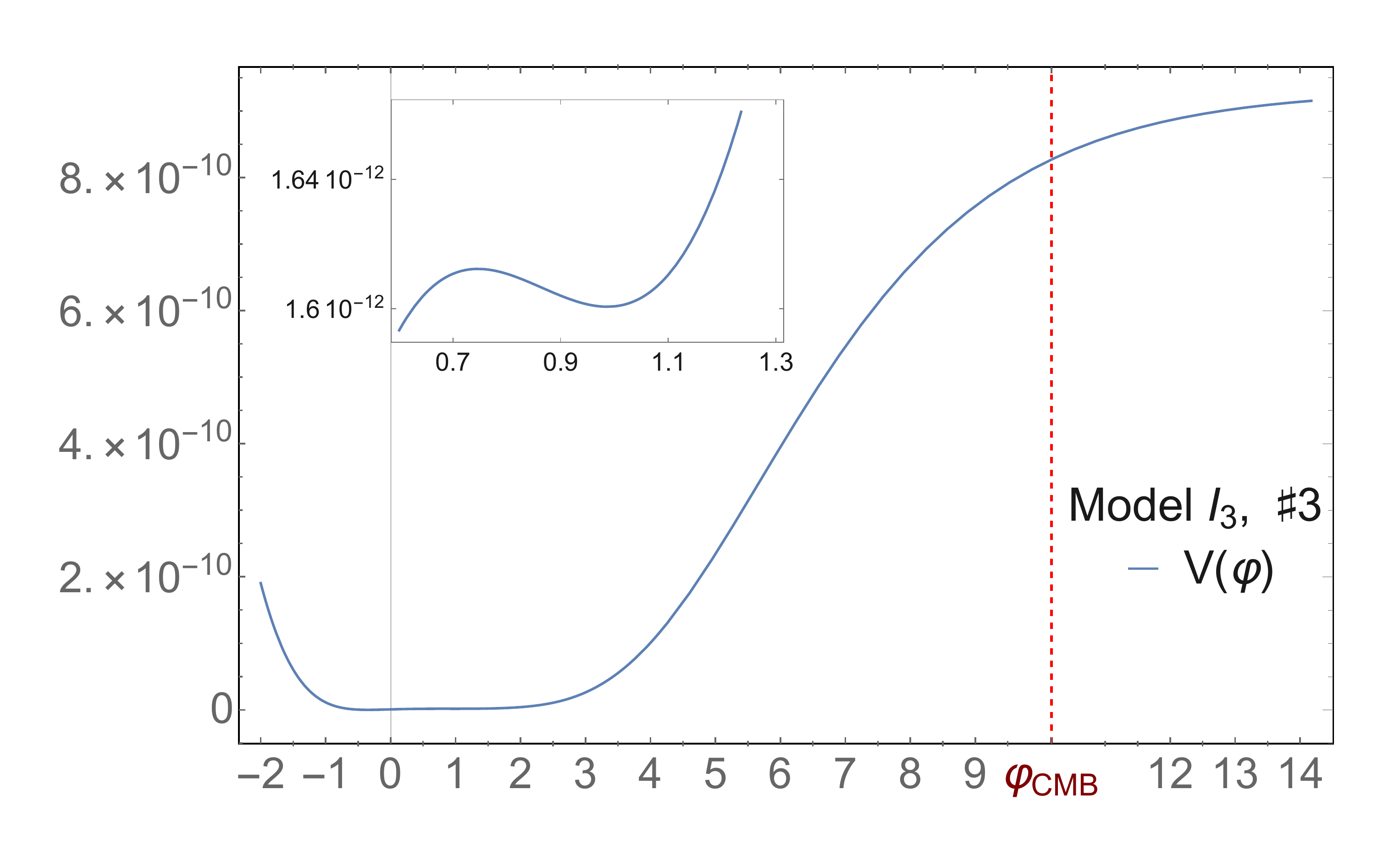}
  %\caption{1a}
  %\label{fig:sfig1}
\end{subfigure}%
\begin{subfigure}{.5\textwidth}
  \centering
  \includegraphics[width=1.\linewidth]{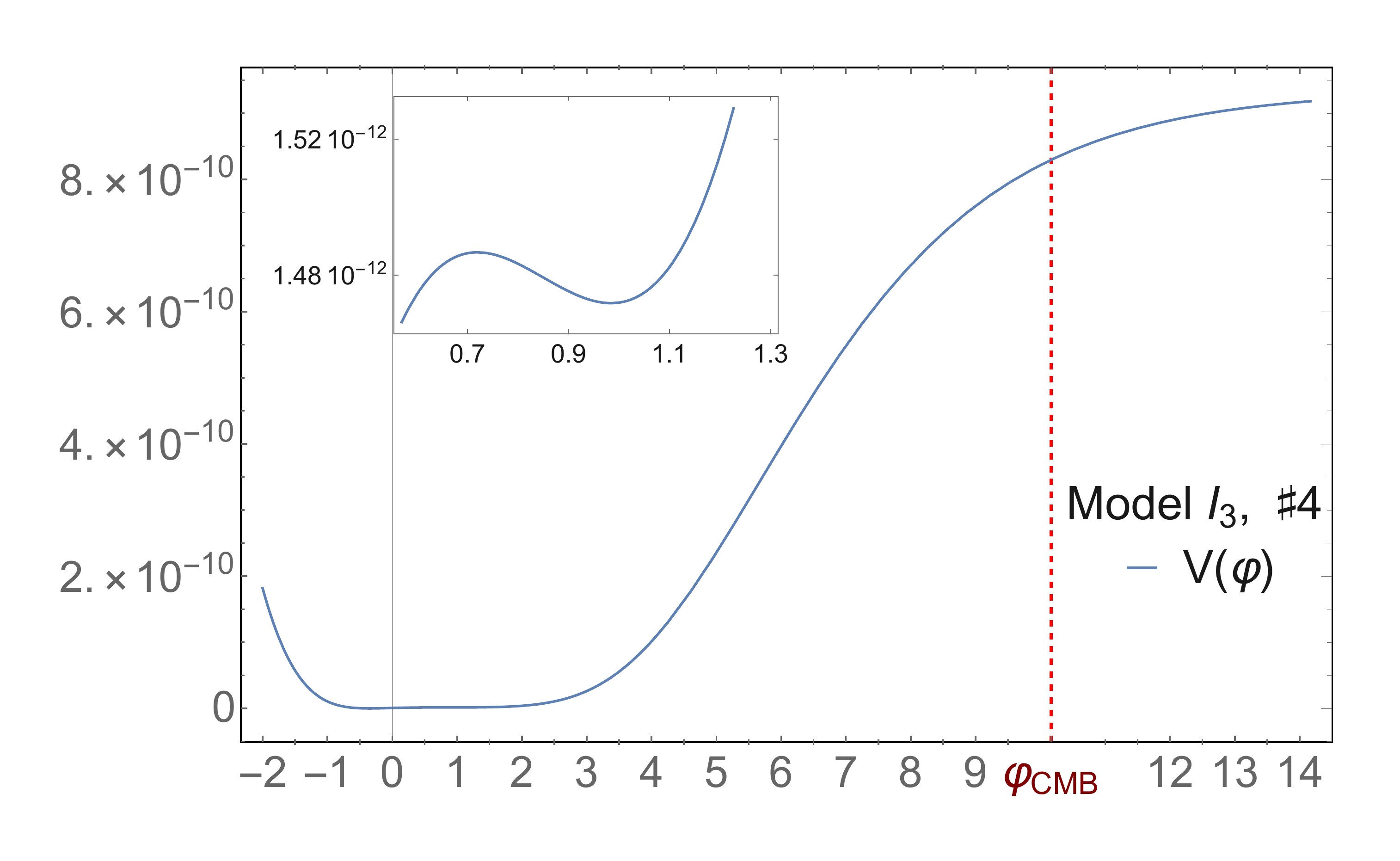}
 % \caption{1b}
  %\label{fig:sfig2}
\end{subfigure}
\begin{subfigure}{.5\textwidth}
  \centering
  \includegraphics[width=1.\linewidth]{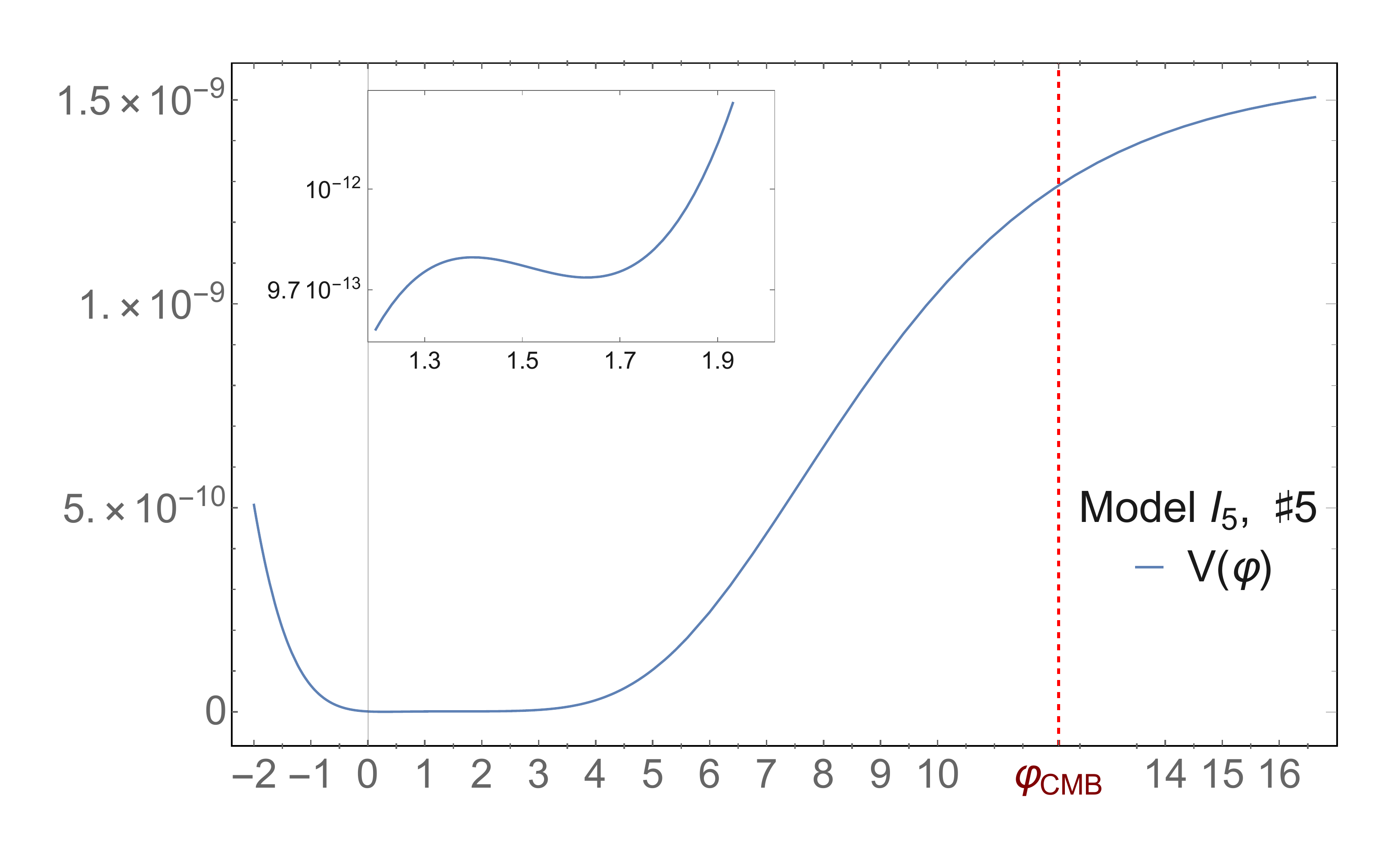}
  %\caption{1a}
  %\label{fig:sfig1}
\end{subfigure}%
\begin{subfigure}{.5\textwidth}
  \centering
  \includegraphics[width=1.\linewidth]{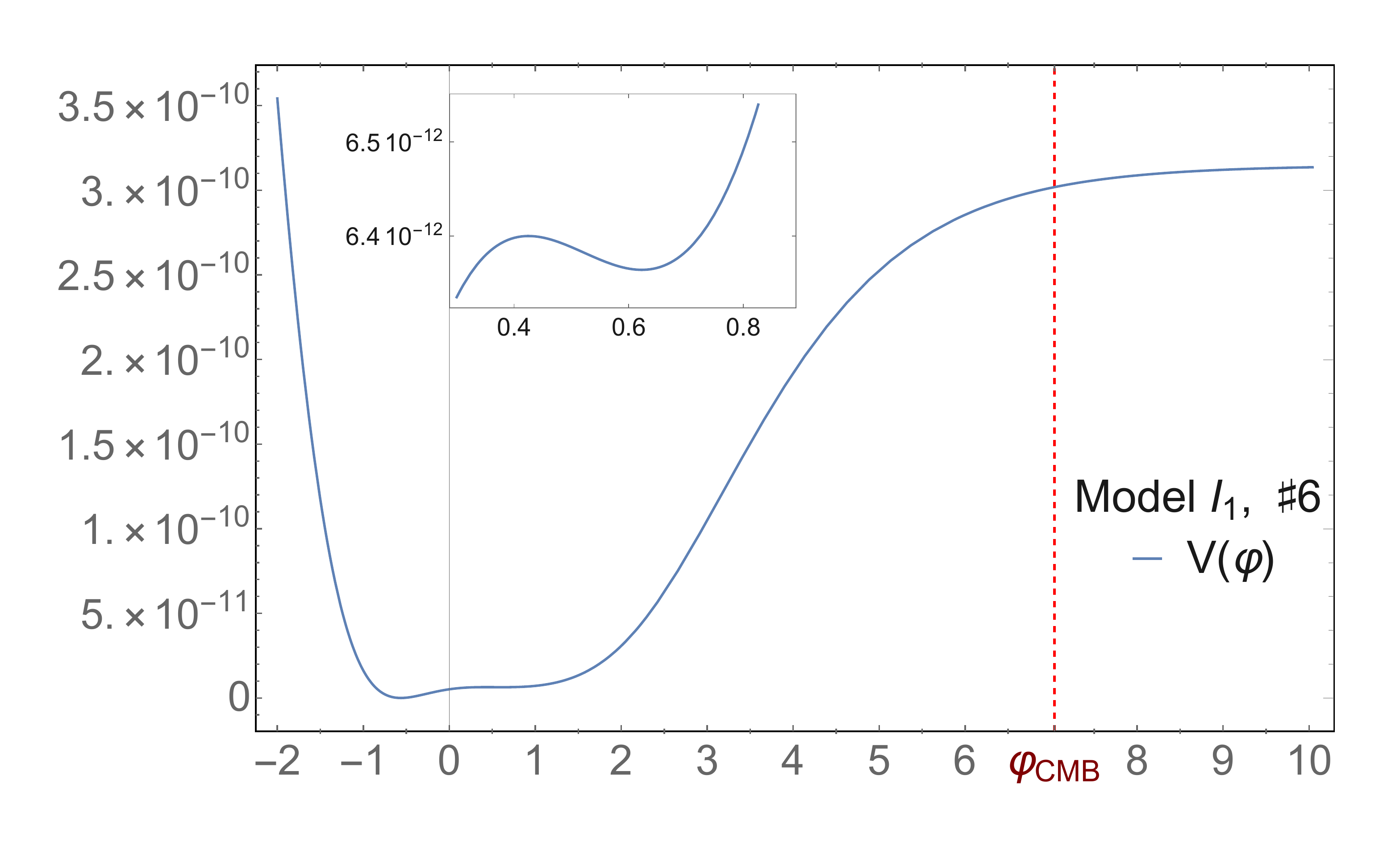}
 % \caption{1b}
  %\label{fig:sfig2}
\end{subfigure}
\caption{\label{6V} {The six potentials from superconformal attractors that we utilized to  examine the PBH formation during different cosmic eras in the early universe. The inner plots zoom in the plateau about the inflection point.  The values for the parameters for each potential are listed in Tables {\ref{tabII}} and \ref{tabI}.}}
\end{figure}

\section{PBH formation during radiation, reheating and modulus dominated eras in superconformal attractors}

In the following we analyze the PBH formation for superconformal attractor inflationary models, however our analysis is general and should apply to any inflationary model with a lagre peak in the curvature power spectrum.  We also expect that our results, presented here and in section 7, should be quantitatively similar with those of any inflationary potential that yields $1-n_s \simeq 2/N$ and $r\simeq 12/N^2$ in the large $N$ expansion. From this perspective the  superconformal attractor models account for an appealing framework to demonstrate  our analysis and results.

A characteristic of the inflationary models that produce both the CMB anisotropies and small scale overdensities that  form PBH is that the e-folds number $N_{0.05}$ is fully specified. It is 
\begin{equation} \label{sep}
N_{0.05} = N_\text{peak} +\Delta N_\text{peak}
\end{equation}
The $N_\text{peak}$ is bounded from above by the requirement not to have a too light PBH mass that would evaporate during the universe lifetime. Also $N_\text{peak}$ is bounded from below  
by the observed $n_s$ value.  
Roughly speaking, the approximate relation (\ref{nsr})  gives $N_\text{peak} \gtrsim 34$ for $n_s \gtrsim 0.942$ and  $N_\text{peak} \lesssim 39$ for $M_\text{PBH} \gtrsim 10^{-18} M_{\odot}$. 
The $\Delta N_\text{peak}$ is much less constrained. It has a minimum size implied by the duration of the reheating stage.  Also,  if $\Delta N_\text{peak} \rightarrow 0$ the range of momenta $k$  that collapse shrinks.
In our models, the  peak in the power spectrum that seeds the PBH formation is  generated about $N_\text{peak} \simeq 38 $ e-folds after the CMB scale $k^{-1}_{0.05}$ exited the Hubble horizon.  Afterwards  about $\Delta N_\text{peak}\sim 10-20$ e-folds of expansion follow until the end of inflation. In total these models require $N_{0.05} \gtrsim 45$ e-folds of inflation. The $N_{0.05}$ is related to the 
post-inflationary cosmic expansion rate \cite{Liddle:2003as} via the expression 
\begin{equation} \label{Nrh}
N_{0.05} \simeq 57.6 +\frac14 \ln \epsilon_* +\frac14 \ln \frac{V_*}{\rho_\text{end}} -\frac{1-3w_\text{}}{4} \tilde{N}_\text{dark} \,,
\end{equation}
where $\tilde{N}_\text{dark}$ are the e-folds that take place in the observationally "dark" era after the end of inflation until the BBN epoch,  where the universe is known to be thermalized, 
and $w$ denotes the average value for the equation of state during that era.
In our models it is $\ln (\epsilon_* V_*/\rho_\text{end})^{1/4} = \pm{\cal O}(1)$, where $\epsilon_* \sim 3/(4N^2_\text{peak})$. 
Hence the above expression constrains the $N_{0.05}$ to be $N_{0.05} \lesssim 58$ e-folds for $w \leq 1/3$.

\subsection{The reheating stage}

Let us now define a criterion that tells us whether PBH form during radiation era or during reheating with $w_\text{rh}\simeq 0$. %matter dominated era. 
  The pre-BBN e-folds of the adiabatically expanding universe can be split into  $\tilde{N}_\text{dark}=\tilde{N}_\text{rh} + \tilde{N}_\text{RD}+ \tilde{N}_X$, which are the e-folds that take place during reheating, radiation domination and modulus domination respectively, namely $\tilde{N}_\text{rh}=\ln(a_\text{rh}/a_\text{end})$,  $\tilde{N}_X=\ln(a^\text{dom}_X/a^\text{dec}_X)$ \cite{Dalianis:2018afb}. 
Then, the last term of the Eq. (\ref{Nrh}) reads $(\tilde{N}_\text{rh}+\tilde{N}_X)/4$. Leaving aside the modulus domination scenario for a while, i.e. $\tilde{N}_X=0$,  the Eq. (\ref{sep}), (\ref{Ntild}) and (\ref{Nrh}) yield the critical $\Delta N_\text{peak}$ value for PBH formation after inflation, 
\begin{equation}
\Delta N^\text{(cr)}_\text{peak} \, \equiv \, \frac{\tilde{N}_\text{rh}}{2}\,  \simeq \,  \frac{2}{3} \left(57 -N_\text{peak} \right)\,,
\end{equation}
where for simplicity we assumed that $\ln (\epsilon_* V_*/\rho_\text{end})^{1/4} \simeq -0.3$. 
Taking that $N_\text{peak} \simeq 38$, which gives the optimum value for $n_s$, we find that $\Delta N^\text{(cr)}_\text{peak} \simeq 12.7 $. Hence, for $\tilde{N}_X=0$,  the PBH formation condition  for the superconformal inflationary models reads,
\begin{equation} \label{mrc2}
\begin{split} %\nonumber
&\Delta N_\text{peak} > 12.7  \quad \text{or} \quad N_{0.05} > 50.7  \quad \Longrightarrow \quad  \text{PBH form during RD} \\
& \Delta N_\text{peak} < 12.7   \quad \text{or} \quad N_{0.05} <50.7  \quad \Longrightarrow  \quad  \text{PBH form during reheating}
\end{split}
\end{equation}  
We assume that the greatest part of the reheating phase corresponds to a MD era with $w_\text{rh} \simeq 0$. %, an assumption that is plausible. 
If it is  $\Delta N^\text{}_\text{peak} =\Delta N^\text{(cr)}_\text{peak} \simeq 12.7$ the $k^{-1}_\text{peak}$ reenters the horizon at the moment of the transition form MD to RD.  
We note that the condition (\ref{mrc2}) follows simply from the fact that the $N_{0.05}$ is directly related to the post-inflationary expansion history. 
 If the reheating is instantaneous it is $\tilde{N}_\text{rh} \rightarrow 0$ and PBH form during RD era regardless the position of the peak. In this case it is $N_{0.05} \simeq 57$.

The critical value, $\Delta N^\text{(cr)}_\text{peak} \simeq 12.7$, implies that PBH can form during a matter dominated era only if a prolonged reheating stage follows inflation with $\tilde{N}_\text{rh}\gtrsim 25$. At the end of reheating the energy density is $\rho_\text{rh} =e^{-3\tilde{N}_\text{rh}}\rho_\text{end}$, 
where $\rho_\text{end}=3H^2_\text{end} M^2_\text{Pl} \sim 10^{-12} M^4_\text{Pl}$ in our models, while the reheating temperature is 
\begin{align}
T_\text{rh}\, \sim   10^{15} \times e^{-3 \tilde{N}_\text{rh}/4}\,\, \text{GeV}. 
\end{align}
 Consequently, for $\tilde{N}_\text{rh}=2 \Delta N^\text{(cr)}_\text{peak} $, there is a critical reheating temperature  
 \begin{equation} \label{Tcr}
 T^\text{}_\text{cr} \sim  10^7 \, \text{GeV},
 \end{equation}
 that the peak placed $\Delta N^\text{(cr)}_\text{peak} \simeq 12.7$ e-folds before inflation end reenters the horizon.
 If $T_\text{rh}>T^\text{}_\text{cr}$  a PBH with mass $M_{k_\text{peak}}$
 forms during RD era while  if $T_\text{rh}<T^\text{}_\text{cr}$  it forms during the reheating stage. 
%We mention that the $T^\text{(cr)}_\text{rh}$ is the temperature that a particular 
Actually, one can find that the critical temperature (\ref{Tcr}) is the temperature that the scale $k^{-1}_\text{peak}$ reenters the horizon for any $ \Delta N_\text{peak} >  \Delta N^\text{(cr)}_\text{peak}$ value, 
%if the reheating has completed regardless the $\Delta N_\text{peak}$ value, 
that is 
\begin{equation}
T_\text{peak}=T_\text{cr}\,\quad \quad \quad \text{for}\,\quad \quad \quad   T_\text{rh}>T_\text{cr}.
\end{equation}
This  is expected since the scale $k^{-1}_\text{peak}$ is fixed  in our models at about  $10^{14}$ Mpc, and hence the ratio of temperatures in a RD era without entropy production has to be also fixed for particular scales, e.g. $T_\text{peak}/T_\text{BBN}$ has to be equal to  $k_\text{peak}/k_\text{BBN}$.

When a scale $k^{-1}$ reenters the horizon during MD era and PBHs form, the energy density of the produced PBHs does not grow relatively to the background in the course of the MD expansion. Only after the complete reheating of the universe does the PBH energy density start to increase with respect to the ambient radiation. Therefore, in the computation of the relic PBH abundance the delayed thermalization  has to be considered.  According to the expression (\ref{fmd}) we have to estimate the exponential $e^{-\frac34 \left(\tilde{N}_\text{rh} - \tilde{N}_k \right)}$. The exponential is actually equal to the ratio $T_\text{rh}/T_k$, where we call $T_k$  the temperature that the universe would have had if the energy density at the moment of $k^{-1}$ reentry, $\rho_k=e^{-3\tilde{N}_k} \rho_\text{end}$, had been transformed into thermal radiation. 
It is $H_\text{end}\simeq H_k $ for scales $k^{-1}$ that exit  the Hubble horizon during or after the USR inflationary  phase, so the second term in Eq. (\ref{Dnk}) can be neglected. Hence,  the  Eq. (\ref{fmd}) reads in terms of the wavenumber $k$,
\begin{align}\label{fmdre} 
  f^{(\text{MD})}_\text{PBH}(M_k)  = 2^{-3/4}\, \frac{\Omega_\text{m} }{\Omega_\text{DM} } \,\gamma^{3/2} \, \beta_\text{MD}(M_k)\, \left( \frac{g_{*\text{eq}}}{g_*(T_\text{rh})}   \right)^{1/4}
       \,      \left(  \frac{M_k}{M_\text{eq}}\right)^{-1/2}    \, e^{-\frac34 \left(  \tilde{N}_\text{rh}- 2\ln (k_\text{end}/k)   \, \right)  } \,.
 \end{align}
 The $\tilde{N}_\text{rh}$ is generally given by the Eq. (\ref{Nrh}) and for our models, under the assumption that $w_\text{rh} \simeq 0$,  it is $\tilde{N}_\text{rh}\simeq 4\,(57-N_{0.05})$. 
Therefore, when the reheating stage is taken into account, the formula (\ref{ftot}) for the  relic PBH abundance is {\it generalized} as 
 \begin{equation} \label{gen}
  f_\text{PBH,tot}\, =
  \begin{cases} 
 \, 3 \,\int_k^{} d \ln k\,    f^\text{(MD)}_\text{PBH} (M_k)  
       \,, \,\, \quad\quad\quad\quad\quad\quad\quad\quad\quad\quad\quad \text{for} \quad\quad k >k_\text{rh}  \\      \\ 
       
    \,2  \,  \int_k  d \ln k\,    f^\text{(RD)}_\text{PBH} (M_k) 
        \,, \,\, \quad\quad\quad\quad\quad\quad\quad\quad\quad\quad\quad \text{for} \quad\quad k <k_\text{rh}
  \end{cases}
   \end{equation}
 where  $k_\text{rh}=e^{-\frac{\tilde{N}_\text{rh}}{2}}\, k_\text{end} $ is the momentum that corresponds to the horizon scale at the time of the reheating completion. In the above formula,  the $f^\text{(RD)}_\text{PBH}$ is written in terms of the PBH  production rate in a RD era given by Eq. (\ref{brad}),  whereas  the  $ f^\text{(MD)}_\text{PBH} $ is written in terms of the production rate in a MD era given by Eq. (\ref{bmat}). 
 Apparently,  the formula (\ref{gen}) is general and applies beyond the framework of superconformal attractors. It holds for any $N_{0.05}$, $N_\text{peak}$ and $\Delta N_\text{peak}$, 
 not only for the particular values that our models admit.  

 Until now we have implicitly assumed 
 an instantaneous transition from the matter domination era to the thermalized radiation era. However, the inflaton does not decay instantaneously. Apart from the possible preheating effects (that are strong right after inflation) at the end of its lifetime inflaton gradually looses energy towards lighter degrees of freedom with energy density $\rho_\text{rad}= (\Gamma_\text{inf}/4H)\rho_\text{inf}$. This implies that the equation of state  deviates from zero and pressure gradually appears.  The effective equation of state for this combination of fluids is 
\begin{equation}
w_\text{}=\frac{p_\text{inf}+p_\text{rad}}{\rho_\text{inf}+\rho_\text{rad}} \simeq \frac{w_\text{rad}}{4} \frac{\Gamma_\text{inf}}{H}\,,
\end{equation}
 where the right hand approximation holds for  $\Gamma_\text{inf} \ll H$and $w_\text{rad}=1/3$. The effective equation of state may have interesting implications for the PBH  formation mechanism. Ref. \cite{Carr:2018nkm} took into account the gradually produced radiation and specified under what conditions the transition between matter and radiation PBH production  occurs.  They found that the effective $\tilde{N}_\text{rh}$ number for the PBH formation in MD is reduced about $10\%$  for variance $\sigma <0.05$ and, also, this reduction is  independent of the decay rate $\Gamma$.  This result  modifies the $f_\text{PBH,tot}$ for $k_\text{peak}=k_\text{rh}$ and can be neglected for  $k_\text{peak} \gg k_\text{rh}$.  

In this work, we take into account the generic presence of a reheating phase and we use the formula (\ref{gen}) for the computation of the PBH abundance during  RD or  MD eras.

 \subsection{Modulus domination}

The  requirement $T_\text{rh}<T_\text{cr} \sim 10^7$ GeV for PBH formation in MD era means that a slow reheating takes place. This can be realized if the inflaton decay rate is particularly suppressed.  
Such a scenario is not impossible, nevertheless in a  supersymmetric and stringy framework the presence of extra late decaying scalar fields, that we collectively call them moduli $X$ fields, is natural. 
 These fields can dominate the early universe energy density and realize a prolonged MD domination era before BBN. In such a case the condition  for PBH formation during MD era reads $\Delta N_\text{peak}<\Delta N^\text{(cr)}_\text{peak}=  ({\tilde{N}_\text{rh}+ \tilde{N}_X})/ 2$, or in terms of the modulus decay temperature,  
\begin{align} \label{X1}
T_X^\text{dec} \lesssim  T^\text{}_\text{cr} \,,
\end{align} 
 for a modulus domination phase that follows {\it continuously} right after inflaton  decay. The $T_X^\text{dec}$ stands for the modulus decay temperature. 
 
 However, it is well possible that a modulus $X$ field dominates the energy density of the universe several e-folds after the complete reheating of the universe, i.e. $T_X^\text{dom}<T_\text{rh}$. 
 In such a case the condition (\ref{mrc2}) for PBH formation during reheating/RD era  does not apply.  
 A PBH of scale $k^{-1}$ can form during a modulus dominated era if the $\Delta N_k$ %te-folds  before the end of inflation  $\Delta N_k$ that the perturbations at the scale $k^{-1}$ were generated   
 satisfies
\begin{align} \label{X2}
 \frac{\tilde{N}_\text{rh}}{2} +\Delta \tilde{N}_\text{RD} \, <\,  \Delta N^\text{}_k\text{} \,  < \, \frac{\tilde{N}_\text{rh} + \tilde{N}_X}{2} +\Delta \tilde{N}_\text{RD}\,.
 \end{align} 
 where $\tilde{N}_X=[3(1+w_X)]^{-1}\ln(\rho^\text{dom}_X/\rho^\text{dec}_X)$ are the e-folds that take place during the  modulus domination until the moment it decays, and $\Delta \tilde{N}_\text{RD} = \ln(\rho_\text{rh}/\rho^\text{dom}_X)^{1/4}$ are the e-folds of radiation domination that take place between reheating and the epoch of the modulus domination. We assume that during modulus domination the equation of state is that of pressureless matter, i.e. $w_X=0$. 
 Considering the momenta $k$, the condition (\ref{X2}) is recast into
 \begin{align}
 k_X^\text{dec} < k < k_X^\text{dom} \quad \Longrightarrow \quad  \text{PBH form during MD}
 \end{align}
  where  $k_X^\text{dom}=e^{-{\tilde{N}_\text{rh}}/{2}-\Delta \tilde{N}_\text{RD}} k_\text{end}$ and $k_X^\text{dec}=e^{-(\tilde{N}_\text{rh}+\tilde{N}_X)/2-\Delta \tilde{N}_\text{RD}} k_\text{end}$.
 Thus, the PBH abundance (\ref{fmd}) during modulus domination  reads  here,
  \begin{align}\label{fmdre2}  \nonumber
  f^{(\text{MD})}_\text{PBH}(M_k) \, & = \, 2^{-3/4}\,
      \frac{\Omega_\text{m} }{\Omega_\text{DM} } \,\gamma^{} \, \beta_\text{MD}(M_k)\, \left( \frac{g_{\text{eq}}}{g(T^\text{dec}_X)}   \right)^{1/4}
       \,      \left(  \frac{M(T_X^\text{dec})}{M_\text{eq}}\right)^{-1/2}  \\ 
  & =\, 2^{-3/4} \frac{\Omega_\text{m} }{\Omega_\text{DM} } \,\gamma^{3/2} \, \beta_\text{MD}(M_k)\, \left( \frac{g_{\text{eq}}}{g(T^\text{dec}_X)}   \right)^{1/4}
       \,      \left(  \frac{M_k}{M_\text{eq}}\right)^{-1/2}    e^{-\frac34 \left(  \tilde{N}_X   -2\ln (k^\text{dom}_X/k)   \, \right)}\,,
 \end{align}
 where $M(T_X^\text{dec})$ is the horizon mass at the time of the modulus decay. 
 Collectively, the fractional  PBH abundance, when a modulus domination era with vanishing pressure takes place in the post-reheated universe,  reads
   \begin{equation} \label{genMod}
  f_\text{PBH,tot} =
  \begin{cases} 
   %\, \int_k^{} d \ln k\,    f^\text{(RD)}_\text{PBH} (M_k) \times\frac{1}{D_X}
   \, \left[\text{Eq}.(\ref{gen}) \right]\times\frac{1}{\Delta_X}
       \,,  \quad\quad\quad\quad\quad\quad\quad\quad\quad\quad  \text{for} \quad\quad\quad    k >k^\text{dom}_X  \\  \\
       
 \, 3\int_k^{} d \ln k\,    f^\text{(MD)}_\text{PBH} (M_k) 
      \quad\quad\quad  \quad\quad\quad \quad\,   \quad \text{for} \quad\quad\quad k^\text{dec}_X<k <k^\text{dom}_X  \\      \\ 
       
    \,2  \int_k  d \ln k\,    f^\text{(RD)}_\text{PBH} (M_k)\,, 
       \quad\quad\quad\quad\quad\quad\quad\,\,\,\, \text{for} \quad\quad\quad k <k^\text{dec}_X
  \end{cases}
   \end{equation}
where 
% the $M_k$ is given by Eq. (\ref{masMD2}) after the replacement  $T_\text{rh} \rightarrow T_X^\text{dec}$ and
 the $  f^\text{(MD)}_\text{PBH} (M_k)$ is given by Eq. (\ref{fmdre2}).
The $\Delta_X\simeq T_X^\text{dom}/T_X^\text{dec}$   is the dilution factor due to the modulus low entropy production. 
In the limit $\tilde{N}_\text{rh} \rightarrow 0$ the e-folds that take place during modulus domination can be specified for a given power spectrum by the equation, $\tilde{N}_X/4\simeq 57-N_{0.05}\simeq  19-\Delta N_\text{peak}$.  

We are mostly interested in the epoch that the scale $k^{-1}_\text{peak}$ reenters the horizon. 
If $k_\text{peak} \gg k_X^\text{dom}$ the net effect of the modulus domination is a dilution of the PBH abundance  $\Delta_X$ times.  In the interesting case that  $k^\text{dec}_X<k_\text{peak} <k^\text{dom}_X $ the PBH form in a pressureless background and the production rate and abundance  might be enhanced, particularly if $k_\text{peak} \sim k_X^\text{dec}$. 
We note that $k^{-1}_\text{peak}$ can enter during the modulus domination era  for any  $N_{0.05}$ value smaller than 57. Finally, in the case that $k_\text{peak} \ll k_X^\text{dec}$ the modulus domination era has minor effects on PBH formation that takes place mostly at $T_\text{peak}$. 

The modulus domination scenario is particularly attractive in the framework of the supersymmetric theories, see also Ref. \cite{Georg:2017mqk} for a relevant work.  PBH can account only for a fraction of the total DM density, and thus an extra dark matter component should exist. Supersymmetric theories provide DM candidates that can supplement the dark matter density. However, the increase of the lower bounds on the sparticle masses by LHC searches is at odds with the standard  thermal dark matter scenario and dilution effects are often required, see e.g.   \cite{Dalianis:2018afb} for a recent analysis.
From this perspective,  the  PBH formation during a modulus domination era is a realistic and a very interesting scenario.

\section{Observational signatures and constraints for PBH from superconformal attractors}

The models explored in this work predict PBHs in the small mass window, see Tables {\ref{tabRad}}, \ref{tabMD}  and \ref{tabX} and very characteristic values for the scalar tilt, $n_s$ and the scalar tilt running, $\alpha_s$.

\begin{center}
\begin{tabular}{|c|c||c|c|c|c|c|} 
\hline
 $\#$&  $\boldsymbol{\alpha}$ & $\boldsymbol{A}$ & $\boldsymbol{f_{\phi}}$ & $\boldsymbol{\varphi_*}$  & $\boldsymbol{V_0}$ & $\boldsymbol{N_{0.05}}$ \\ [0.5ex] 
 \hline\hline
 \textbf{1}& 1 & 0.130383& 0.129576  & 5.85 &  $2\times 10^{-10}$  & 56.6   \\  % 59.65 
 \hline
  \textbf{7} & 1 & 0.128114& 0.126331  & 5.7725 &  $1.7\times 10^{-10}$  & 52.5 
  \\ 
 \hline
\end{tabular}
\captionof{table}{\label{tabII} A set of values for the parameters of the Model $II_1$. The  $\varphi_*$ (or $\varphi_\text{CMB}$) is the value of the  inflaton field 
$N_{0.05}$ e-folds before the end of inflation. 
}
\end{center}

\begin{center}
\begin{tabular}{|c||c||c|c|c|c|c|c|c|} 
 \hline
$\#$ 
& $\boldsymbol{\alpha}$ 
& $\boldsymbol{c_3}$ 
& $\boldsymbol{c_2}$ 
& $\boldsymbol{c_1}$ & $\boldsymbol{c_0}$ 
& $\boldsymbol{\varphi_*}$  
 & $\boldsymbol{V_0}$ & $\boldsymbol{N_{0.05}}$ \\ [0.5ex] 
 \hline \hline
 \textbf{2} &  1  & 2.20313  &  -1.426    & 0.3 & 0.16401 & 7.0328 & $2.1\times10^{-10}$  &  55.4 \\ 
  \hline
 \textbf{3} & 3  &   2.4843 & -1.5 &  0.2964 & 0.0355 & 10.173  &  
 $5.4 \times 10^{-10}$  &  50.8 \\ 
   \hline
    \textbf{4} &  3  &2.5259 & -1.5002 & 0.2902 & 0.0355 & 10.166 & $5.1\times10^{-10}$ & 50.2 \\ 
\hline
\textbf{5} &   5   &2.11917 & -1.7147 & 0.46 & -0.0198 & 12.128 & 
$2.2\times10^{-9}$ & 54  \\ 
\hline
\textbf{6} &  1  & 2.25336 & -1.42226 & 0.2892 & 0.16401 & 7.04 & 
$1.9\times10^{-10}$ & 53  \\ 
\hline
\end{tabular}
\captionof{table}{\label{tabI} Five set of  values for the parameters of the Models $I_\alpha$, %The  $\varphi_*$ is the inflaton field value $N_{0.05}$ e-folds before the end of inflation and  $V_0$ is written in Planck units,
 as in Table \ref{tabII}. 
} 
\end{center}

\begin{figure}[!htbp]
\begin{subfigure}{.5\textwidth}
  \centering
  \includegraphics[width=1.\linewidth]{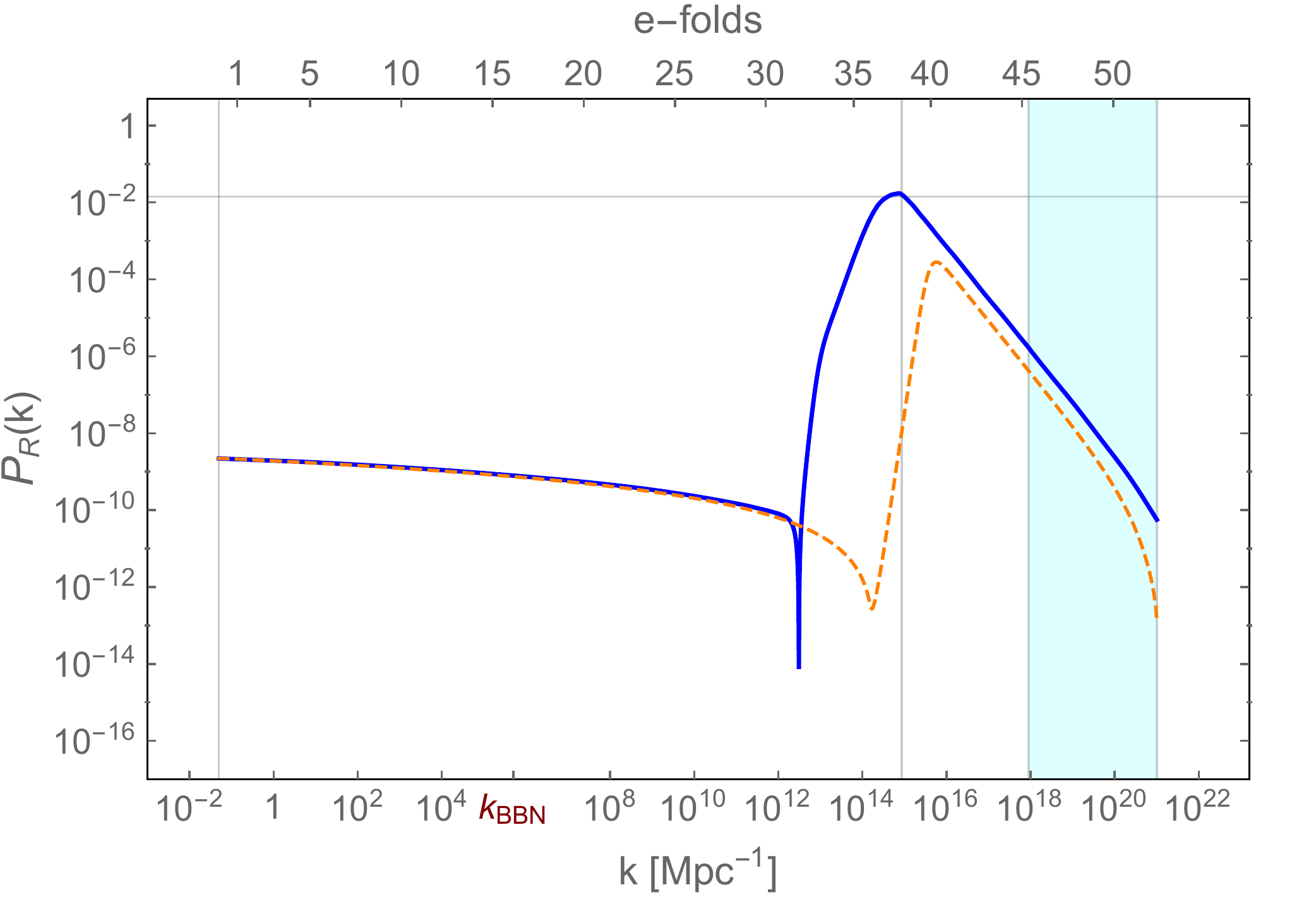}
\end{subfigure}%
\begin{subfigure}{.5\textwidth}
  \centering
  \includegraphics[width=1.\linewidth]{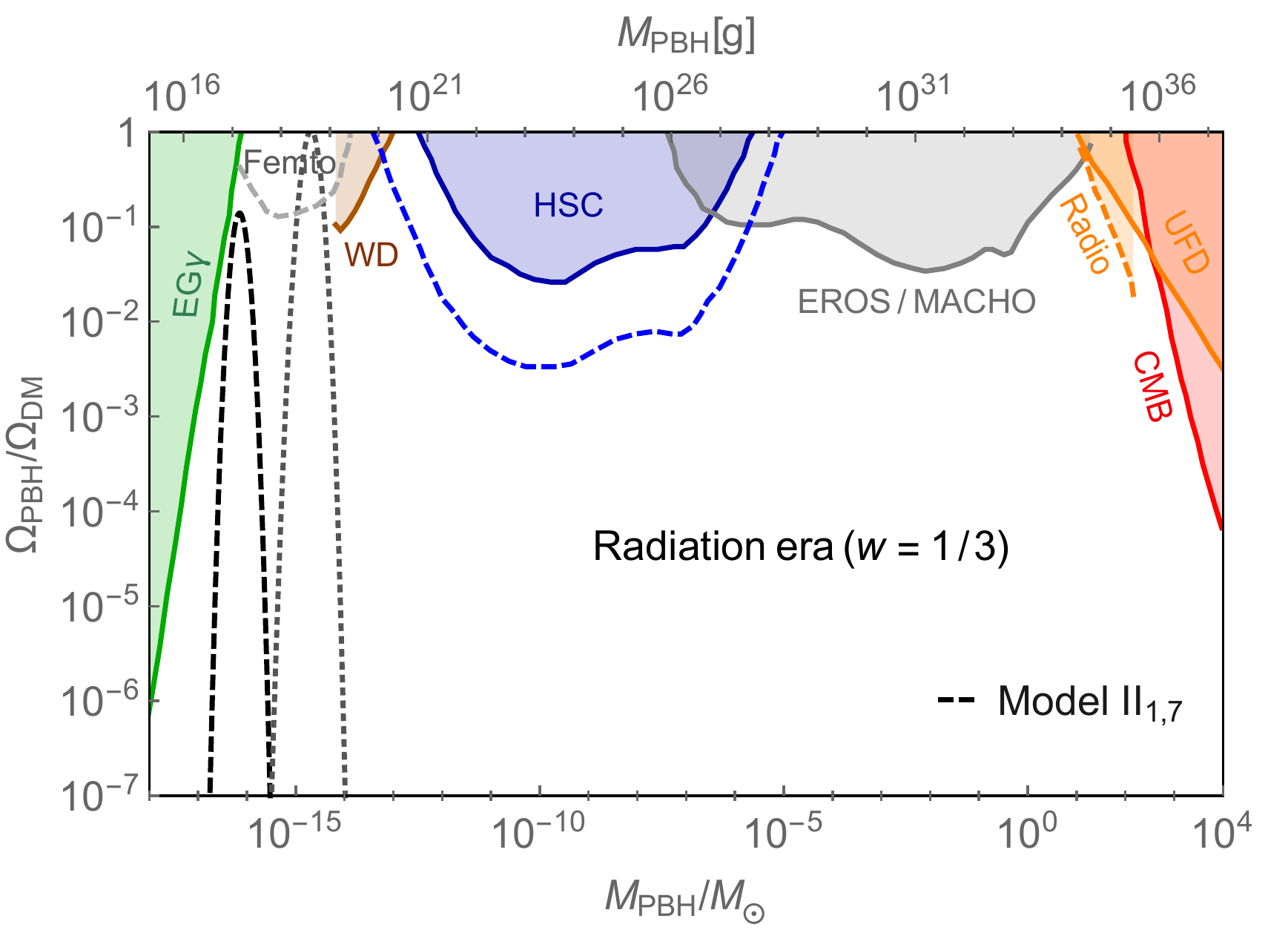}
\end{subfigure}
\\
\\
\begin{subfigure}{.5\textwidth}
  \centering
  \includegraphics[width=1.\linewidth]{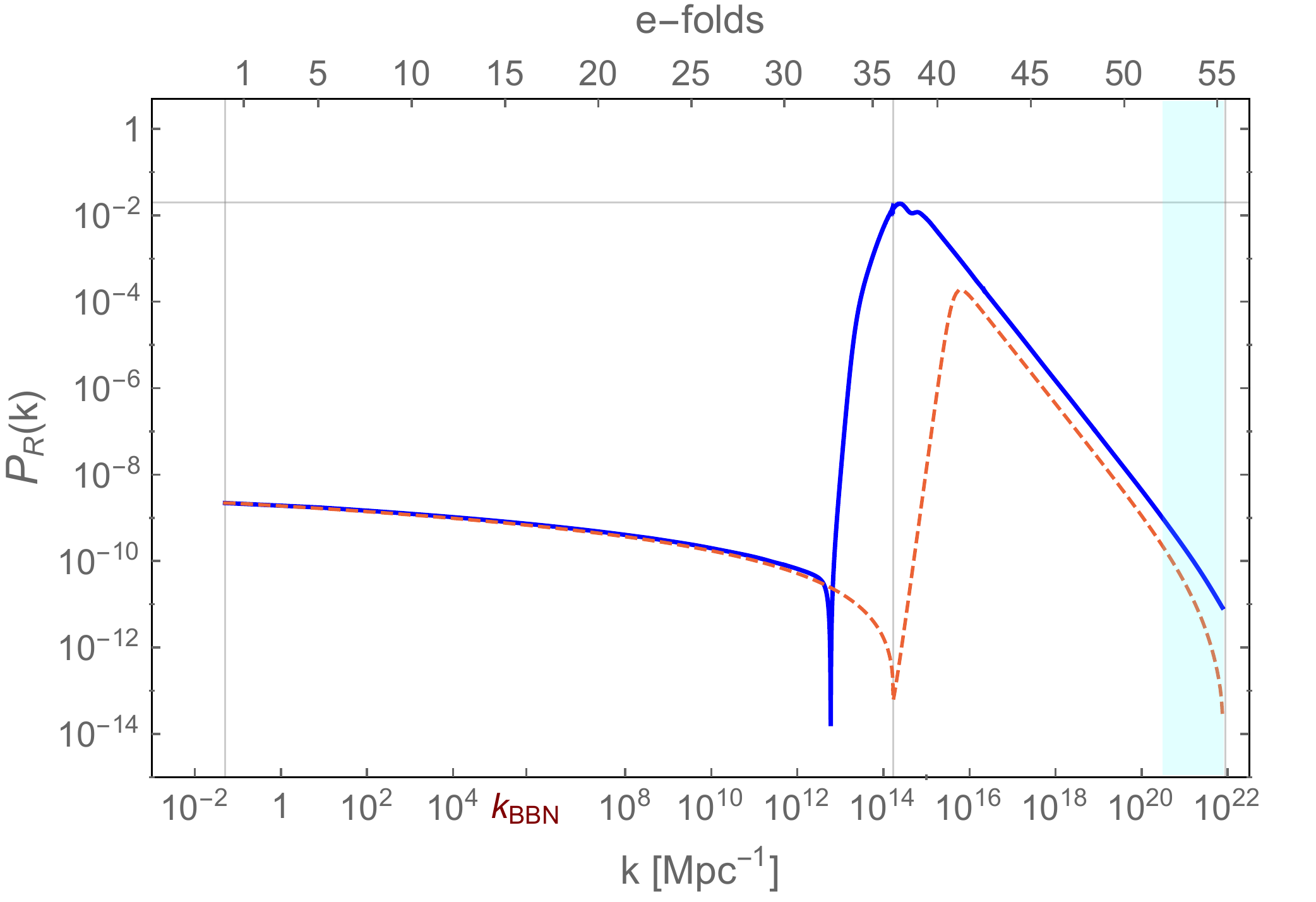}
\end{subfigure}%
\begin{subfigure}{.5\textwidth}
  \centering
  \includegraphics[width=1.\linewidth]{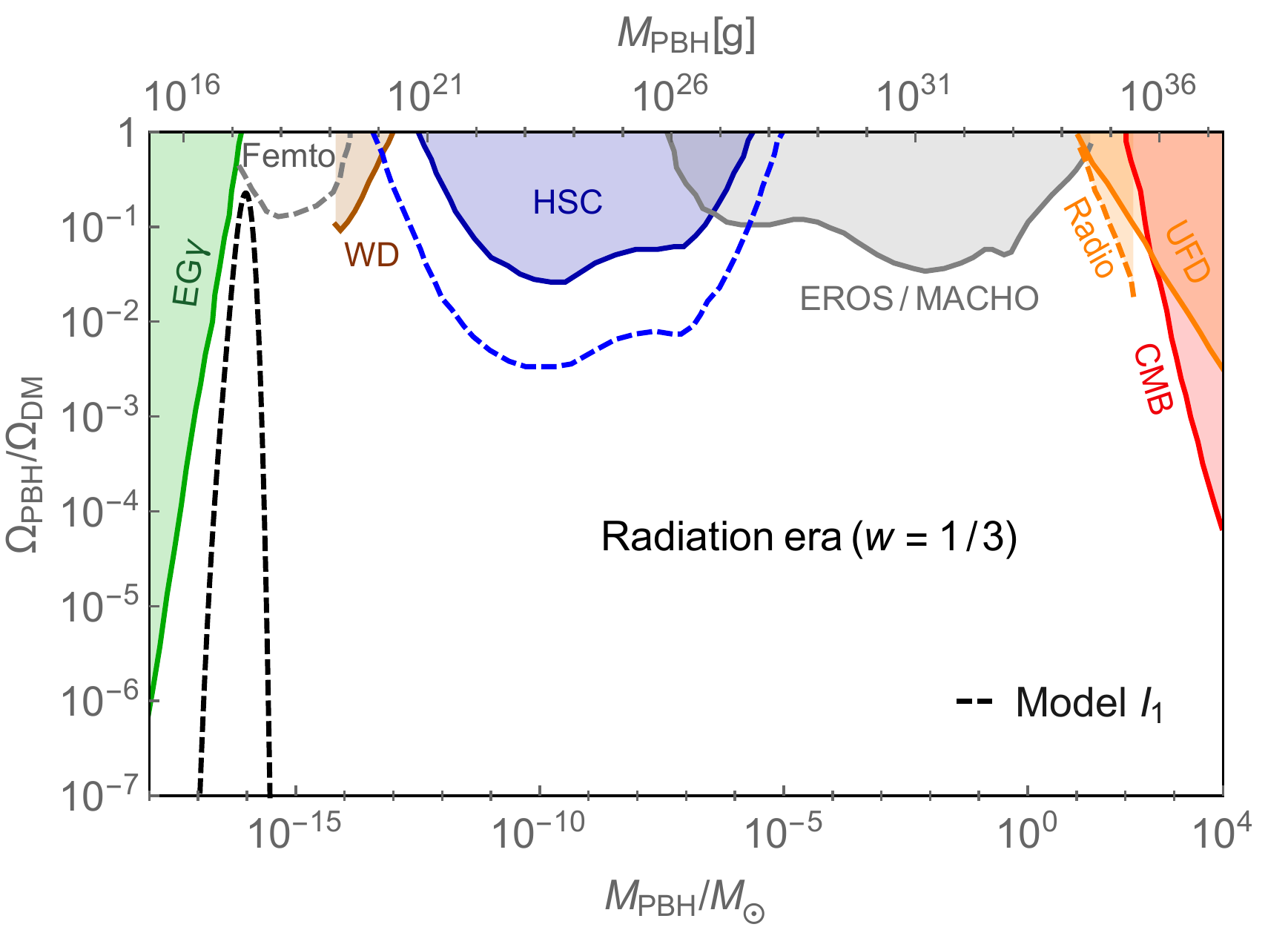}
\end{subfigure}
\caption{\label{FigRad} 
{\it Left panels}: The power spectra of the curvature perturbation for the Models $II_1$ and $I_1$ with ${\cal P_R}\sim 10^{-2}$, examples $\#$ 7 and 2, see Tables \ref{tabII}, \ref{tabI} and \ref{tabRad}.  The solid lines depict the power spectra calculated from the exact Mukhanov-Sasaki equation while the dashed lines depict the approximate expression  (\ref{formP}).  
The cyan areas show the scales that reenter during reheating. For the parameters chosen the models $II_2$ and $I_1$ have a short reheating stage with $T_\text{rh}\simeq6 \times 10^{10}$ GeV and  $T_\text{rh} \simeq 6 \times 10^{13}$ GeV respectively. {\it Right panels}: The corresponding fractional abundance of PBHs  is 0.1 and 0.15 respectively. 
In the upper right plot we have {\it added} the $II_1$ model, example $\#1$, that yields fractional PBH  abundance 0.81\ (gray dotted curve), after the release of the paper \cite{Katz:2018zrn}.
}
\end{figure}

\begin{center}
 \begin{tabular}{|c|c||c|c|c|c|c|c|c|c|} 
\hline
  $\#$ 
  & $\boldsymbol{\alpha}$ 
  & $\boldsymbol{N_{0.05}}$ 
  & $\boldsymbol{\tilde{N}_\text{rh}}$ 
  & $ \boldsymbol{T_\text{rh}}$ \textbf{(GeV)}  
  & $\boldsymbol{{\cal P_R}^\text{peak}}$ 
  & $\boldsymbol{\gamma}$ 
  & $\boldsymbol{\delta_c}$
  & $\boldsymbol{M^\text{peak}_\text{PBH}/M_{\odot}}$ 
  & $\boldsymbol{\Omega_\text{PBH}/\Omega_\text{DM}}$ 
%  & \textbf{Era}   
    \\ 
 \hline \hline
\textbf{1} & 1 & 56.6 & 0 & $2\times 10^{15}$ & $1.9 \times 10^{-2}$ & 0.2 &  0.325  &  $2.1\times 10^{-15}$ & 0.81
%& RD 
\\ 
 \hline
  \textbf{2} & 1 & 55.4 & 4 & $6\times 10^{13}$ & $1.8 \times 10^{-2}$ & 0.2 & 0.323 &  2.5$\times 10^{-16}$ & 0.15 
%  & RD
  \\ 
 \hline
 \textbf{7} & 1 & 52.5 & 6.9 & $1\times 10^{13}$ & $1.7 \times 10^{-2}$ & 1 & 1/3  &  7.5$\times 10^{-17}$ & 0.10 
  \\ 
 \hline
%& RD 
\end{tabular}
\captionof{table}{\label{tabRad}
The PBH abundance and characteristic mass for the  inflationary Model families $II_1$ and $I_1$ respectively, with parameters listed in the Tables \ref{tabII} and \ref{tabI}, examples $\# \,1, \,  2,\, 7$.  
The inflationary parameters imply a short reheating phase and large reheating temperature that the power spectrum peak enters during a {\it radiation dominated era}. The values for the $\delta_c$ parameter are the minimum allowed by observational constraints for the given power spectrum peak, see Figure \ref{FigRad}.
}
\end{center}

\begin{center}
 \begin{tabular}{|c|c||c|c|c|c|c|c|c|} 
\hline
  $\#$ 
  & $\boldsymbol{\alpha}$ 
  & $\boldsymbol{N_{0.05}}$ 
  & $\boldsymbol{\tilde{N}_\text{rh}}$ 
  & $ \boldsymbol{T_\text{rh}}$ (\textbf{GeV}) 
  & $\boldsymbol{{\cal P_R}^\text{peak}}$ 
  & $\boldsymbol{\gamma}$ 
  %& $\boldsymbol{\delta_c}$
  & $\boldsymbol{M^\text{peak}_\text{PBH}/M_{\odot}}$
  & $\boldsymbol{\Omega_\text{PBH}/\Omega_\text{DM}}$ 
%  & \textbf{Era}    
   \\ 
 \hline \hline
\textbf{3} & 3 & 50.8 &  25 & $7\times 10^6$ & $2\times 10^{-5} $   & 1 
%& - 
 & $6 \times 10^{-16}$ & 0.01 
%& MD 
\\ 
 \hline 
\textbf{4} & 3 & 50.2 & 27  &  2$\times 10^{6}$ &$ 4 \times 10^{-5}$ &  1 %&  - 
 &  5$\times10^{-16}$ & 0.14
 %& MD 
 \\ 
\hline 
\end{tabular}
\captionof{table}{\label{tabMD} The PBH abundance and characteristic mass for the  inflationary Model family $I_3$, with parameters listed in the Table \ref{tabMD}, examples  $\#\, 3,\, 4$.  
The inflationary parameters imply a prolonged reheating phase and quite low reheating temperature so that the power spectrum peak enters at the epoch of reheating completion. 
The amplitude of the power spectrum is relatively low to trigger PBH formation during radiation era and most of the PBH are formed during the pressureless {\it reheating stage}. For the example $\#\, 3$ the PBH mass peak is in slight offset with the power spectrum peak, see Figure \ref{FigReh}.
}
\end{center}

\begin{center}
 \begin{tabular}{|c|c||c|c|c|c|c|c|c|c|} 
\hline
  $\#$ 
  & $\boldsymbol{\alpha}$
  & $\boldsymbol{N_{0.05}}$ 
  & $\boldsymbol{\tilde{N}_\text{rh}}$ 
  & $\boldsymbol{T_\text{rh}}$% \textbf{(GeV)} 
  & $\boldsymbol{{\cal P_R}^\text{peak}}$  
  & $\boldsymbol{\tilde{N}_X}$ 
&  $\boldsymbol{ T^\text{dec}_X}$  % \textbf{(GeV)}   
  & $\boldsymbol{M^\text{peak}_\text{PBH}/M_{\odot}}$ 
  & $\boldsymbol{\Omega_\text{PBH}/\Omega_\text{DM}}$      \\ 
 \hline\hline 
\textbf{5} & 5 
 & 54 
 & 4 
  & $ 4 \times 10^{13} $
  & $ 1.6 \times 10^{-4}$ 
  & 8
  & $ 3 \times 10^{4}$   
  &  $ 5 \times 10^{-17}$  
  & 0.12   \\  
\hline 
\textbf{6 }& 1  
& 53 
& 6  
 & $1 \times 10^{13} $  
 & $ 1\times 10^{-2}$  
  & 10
& $9 \times 10^{3}$ 
 &  $6\times 10^{-17}$ 
 & 0.11   \\
\hline
\end{tabular}
\captionof{table}{ \label{tabX} The PBH abundance and characteristic mass for the inflationary Model families $I_5$ and $I_1$ respectively with parameters listed in the Table \ref{tabI}, examples  $\#\, 5,\,6$.  
The inflationary parameters imply a prolonged matter domination era after inflation partitioned into inflaton reheating and into a modulus $X$ {\it scalar field condensate}.  The temperatures are written in GeV units.
The PBH abundance  depends sensitively on the temperature the $X$  modulus dominates the energy density and decays.  
Also, the PBH mass peak is placed in a lower mass range, 
 see Figure \ref{FigMod}.  
}
\end{center}

\subsection{CMB observables and the postinflationary evolution}

The observational bounds,  provided by the Planck 2018 TT+lowE+lensing at the 95\% CL when the running of the running is included \cite{Akrami:2018odb}, at $k_\text{cmb}=0.05 \,\text{Mpc}^{-1}$ read
\begin{equation}
\begin{split}
\quad\quad\quad\quad\quad\quad\quad\quad     n_s&=0.9587\pm 0.0112 \\ 
\quad\quad\quad\quad\quad\quad\quad\quad    \alpha_s&=0.013\pm 0.024  \,
\end{split}
\quad \quad\quad \text{}\,( \, 95\%  \,\, \text{CL} \,)
\end{equation}
Our models predict for the scalar tilt, the running and the   tensor-to-scalar ratio 
%\begin{equation}
\begin{align}
%  0.943 \, \lesssim & \, \nonumber
&   n_s \,   \lesssim \, 0.949   \nonumber     \\
0.0014 \, \lesssim  &   \,  \alpha_s \,  \lesssim \, 0.0015 \\
0.008 \, \lesssim  &   \,  \frac{r}{\alpha} \,  \lesssim \, 0.009 \, \nonumber
\end{align}
%\end{equation}
The potentials examined are depicted in Figure \ref{6V}. Our results are obtained numerically\footnote{After the completion of the first version of our paper the Planck 2018  results \cite{Akrami:2018odb} were released.  In this version we added a representative example, the $\#$ 7, that is in accordance with the new constraints.}. 

The $n_s$ and $r$ values can be understood by the approximate analytic relations (\ref{nsr}),  $1-n_s \sim 2/N_\text{} $ and $r \sim 12/N^2_\text{} $, where $N=N_\text{peak} $ are the e-folds that separate the CMB and the PBH scales and it is   $N_{0.05}=N_\text{peak}+\Delta N_\text{peak}$.  
These relations describe quite well the actual behavior of the $n_s$ and $r$ at the plateau of the potential, though a more accurate analytic estimation has to take into account the higher  order corrections, $1/ N^2_\text{} $ and $1/N^3_\text{} $ for the $n_s$ and $r$ respectively. 
The peak in our models is generated at $N_\text{peak} \sim 38$ and hence the $n_s$ value is smaller than  0.96,  which is the value expected by a conventional superconformal attractor model where $N_{} \sim 57$. A way to understand the decrease in the scalar tilt value, $n_s(N)$,  in this sort of  inflationary  models that trigger the PBH formation,  is  to see the $\Delta N_\text{peak}$ e-folds of inflation that take place about the inflection point as a second inflationary phase. This would reduce the $N_{}$ as $N_{}\rightarrow N-\Delta N_\text{peak}-\tilde{N}_\text{rh}/4\simeq N_\text{peak}$.

We mention that, for inflation with polynomial superpotential (Model $I$), the  different values for the parameter $\alpha$ come with different values for the potential parameters which have to change accordingly, otherwise sufficient  peak at the power spectrum cannot be produced. 
The second model considered, the modulated chaotic potential (Model $II$), yields similar predictions  for the $n_s$, $\alpha_s$ and $r$.  For the Model $II$, larger values for the $\alpha$ parameter make the amplification of the power spectrum harder to be realized. 
The $n_s$ value,  mostly determined by the $ N_\text{peak}$, increases if we increase the $\gamma$ parameter, because the PBH mass range accordingly  increases. If we further let the $N_{0.05}$ vary beyond $\sim$57 e-folds then the  spectral index value can be shifted towards  larger values, even inside inside the 68\% CL region of the Planck collaboration data.  However, in this case values $w_\text{}>1/3$ should be considered and a change on the PBH formation rate should be expected,  since the matter perturbation  threshold $\delta_c$ depends on the background equation of state, Eq. (\ref{dc}). 
In this work we assume  $\delta_c \sim 1/3$ for RD which for ${\cal P_R} \sim 10^{-2}$ we attain the maximum allowed abundance; we also take $\gamma$  to be either  $0.2$ or 1. Adopting larger values for $\delta_c$ would require to accordingly increase the size of the ${\cal P_R}$ peak.

The number of e-folds $N_{0.05}$ determine whether the reheating stage after inflation is short or prolonged, Eq. (\ref{Nrh}). 
In the framework of superconformal attractors the reheating temperature can be determined only after the inflaton couplings to matter are specified. 
Phenomenologically, models with large $N_{0.05}$  imply that the inflaton field   decays fast and the thermal history of the universe starts with  a relatively large reheating temperature. 
Models that yield smaller $N_{0.05}$ imply that a prolonged non-thermal phase was realized after inflation and before BBN. 
The non-thermal phase and the details of the postinflationary cosmic evolution are also interrelated with the supersymmetry breaking scale and the sparticle mass spectrum \cite{Dalianis:2018afb}.

We note that, in each model, inflation ends in a different energy scale that may vary up to one order of magnitude. Hence, the maximum reheating temperature, determined by $\rho_\text{end}$ varies accordingly. 
The duration of the reheating stage and the position of the curvature power spectrum peak determine the PBH abundance. A prolonged matter domination era after the reentry of the peak gives a smaller abundance and smaller PBH masses. In our models the ${\cal P_R}$ peak value is placed at $k_\text{peak} \sim 10^{14}$ Mpc$^{-1}$ and the PBH abundance maximizes when the reheating temperature is close to the critical (\ref{Tcr}).  Thus, we find that  a power spectrum of $\alpha$--attractors can trigger a cosmologically significant abundance of PBH even if the amplitude is as small as 
\begin{align}
{\cal P_R} \sim 10^{-5}  \quad \quad\text{for} \quad\quad T_\text{rh}\sim 10^7-10^6\, \text{GeV}\,.
\end{align}
The exact values depend on whether spin effects are present or not\footnote{We note that we have made the approximation of instantaneous gravitational collapse during the MD era. }. If are not negligible, that is expected for  $\sigma(k)\lesssim 0.005$ \cite{Harada:2017fjm}, the power spectrum peak has to be few times larger but still less than $10^{-4}$ in order to generate a significant PBH abundance. Also, if the reheating temperature is much less than $10^7$ GeV, apart from the decrease in the abundance,  the PBH masses are shifted in a smaller mass range that is much constrained from the extra galactic gamma-ray background \cite{Carr:2016hva}, see Figure \ref{FigReh}. 

For the modulus domination scenario, the astrophysical bounds on the PBH abundance and masses, see Figure \ref{FigMod}, constrain the duration of the modulus dominated era. Assuming that modulus decays gravitationally  then we expect a decay rate, 
\begin{equation} \label{Xdec}
\Gamma^\text{}_X = \frac{c}{4\pi} \frac{M^3_X}{M^2_\text{Pl}}\,,
\end{equation} 
and the $X$ decay temperature is $T^\text{dec}_X \simeq (\pi^2 g_*/90)^{-1/4}(\Gamma^\text{}_X M_\text{Pl})^{1/2}$. For $c\sim1$ the $X$ decays gravitationally and $T^\text{dec}_X \sim 4 \,\text{MeV}\, (M_X/10^5\text{GeV})^{3/2} $. The temperatures $T_X^\text{dec}$ that we associate with a modulus decay, listed in Table \ref{tabX}, are large and imply a rather heavy modulus with mass $M_X\sim 10^9$ GeV.
By decreasing the $T_X^\text{dec}$ to MeV scale the PBH  abundance  would have fallen to negligible levels  and the masses would have shifted in smaller values. 
For $c\gg 1$ non-gravitational decay channels exist and in this case the $M_X$ can decrease. 
Even though heavy, the modulus can dilute an overabundant dark matter particle or produce it non-thermally.
In the context of ultra-TeV scale supersymmetry a heavy modulus that decays at $T_X^\text{dec}\sim 10^4$ GeV and dilutes $\Delta_X \sim 10^4$ times the thermal plasma heated after inflation at $T_\text{rh} \sim 10^{13}$  GeV, as it happens in our scenarios, is welcome. For example, for subleading  $X$ decay contribution, a gravitino LSP with mass $m_{3/2} \sim$ TeV can successfully account for the rest $\sim 90\%$ of the dark matter in the universe. 

The synopsis, regarding the CMB observables, is that the superconformal models examined here, modified to trigger a cosmological relevant abundance of PBH due to a curvature power spectrum peak ${\cal P_R} \gtrsim 10^{-5}$,  predict  a $n_s$  value that is  smaller  than the fiducial $n_s$ value of  a conventional $\alpha$--attractor models. 
On the other hand, the tensor-to-scalar ratio, $r$, and the running of the scalar tilt, $\alpha_s$, are found to be larger.
The small $n_s$  values, which are compatible with the Planck 2018 data at the 95\% CL when the running is non negligible,  place our models around the borderline of what present data allows and renders these models testable by the near future CMB probes. 

\subsection{The PBH mass range}

The potential is constructed in such a way that a significant amplification of the power spectrum is achieved at small scales, that triggers the  PBH production in the mass window
\begin{equation}\label{masswind}
 10^{-16} M_{\odot} \lesssim M_\text{PBH}  \lesssim 10^{-14} M_{\odot}\,.
\end{equation} 
   The small mass window is selected by the requirement the $n_s$ value to be compatible with the Planck data.  
We note that when the parameter $\alpha$ of the polynomial superpotential models changes similar results for the PBH abundance can be obtained for the same $\delta_c$ value. 
The main effect for  $\alpha>1$  is that values for the $\epsilon_1$ larger than one,  between the first slow-roll and the USR phase, are possible and hence a larger amplification of curvature power spectrum amplitude. 

The PBH relic density is constrained by a synergy of astrophysical observations, that we outline next. Our results are obtained numerically using the master  formuli given by Eq. (\ref{ftot}), (\ref{gen}) and (\ref{genMod}). We choose the model parameters that maximize the 
$\Omega_\text{PBH}h^2$ and 
  hence place our models in the observationally interesting  edge of the allowed ($\Omega_\text{PBH}h^2$, $M_\text{PBH}$) contour region,
\begin{align}
\frac{\Omega_\text{PBH}}{\Omega_\text{DM}} ={\cal O} (0.1)\,.
\end{align}
During RD era the PBH abundance depends sensitively on the amplitude of the ${\cal P_R}$ peak and the threshold density contrast  parameter $\delta_c$, whereas during MD era the PBH abundance is less sensitive on the amplitude of the power spectrum peak but it depends on  the reheating temperature.  For PBHs forming during MD, depending on the size of spin effects,  a moderate peak ${\cal P_R}\sim 10^{-5}-10^{-4}$ in the power spectrum is adequate to generate a cosmologically significant abundance if the largest curvature perturbation mode ${\cal R}_k$ reenters the horizon not long before the reheating of the universe. In this case the density of the PBH formed will have enough time to increase until the matter-radiation equality.  In addition,  if the pressureless reheating, or a modulus domination era continues for long after the reentry  of the power spectrum peak, the  PBH masses are shifted towards smaller values. 
Our consideration imply that  the details  of the transition between the reheating and the thermal radiation regime play an  important r\^ole for the determination of the PBH abundance and deserves a thorough investigation.   
 \subsubsection{Observational bounds on the abundance and the mass of the PBH}
Turning now to the observational constraints  on the $\Omega_\text{PBH}h^2$, there are actually several astrophysical and cosmological data that put  bounds on the abundance and the mass of the (primordial) black holes. 
 These bounds can be directly translated into constraints onto the shape of the primordial curvature power spectrum, ${\cal P_ R}$. The fact that makes it possible is that the evolution of the PBH masses via the accretion of matter is expected to be negligible.   Apart from the luminous galactic center, dark matter populates the galactic halos where the matter density is too small to accrete into black holes of small mass and hence it is legitimate to  claim that light halo BH, like those considered in this work, are indeed big-bang relics.  

\begin{figure}[!htbp]
\begin{subfigure}{.5\textwidth}
  \centering
  \includegraphics[width=1.\linewidth]{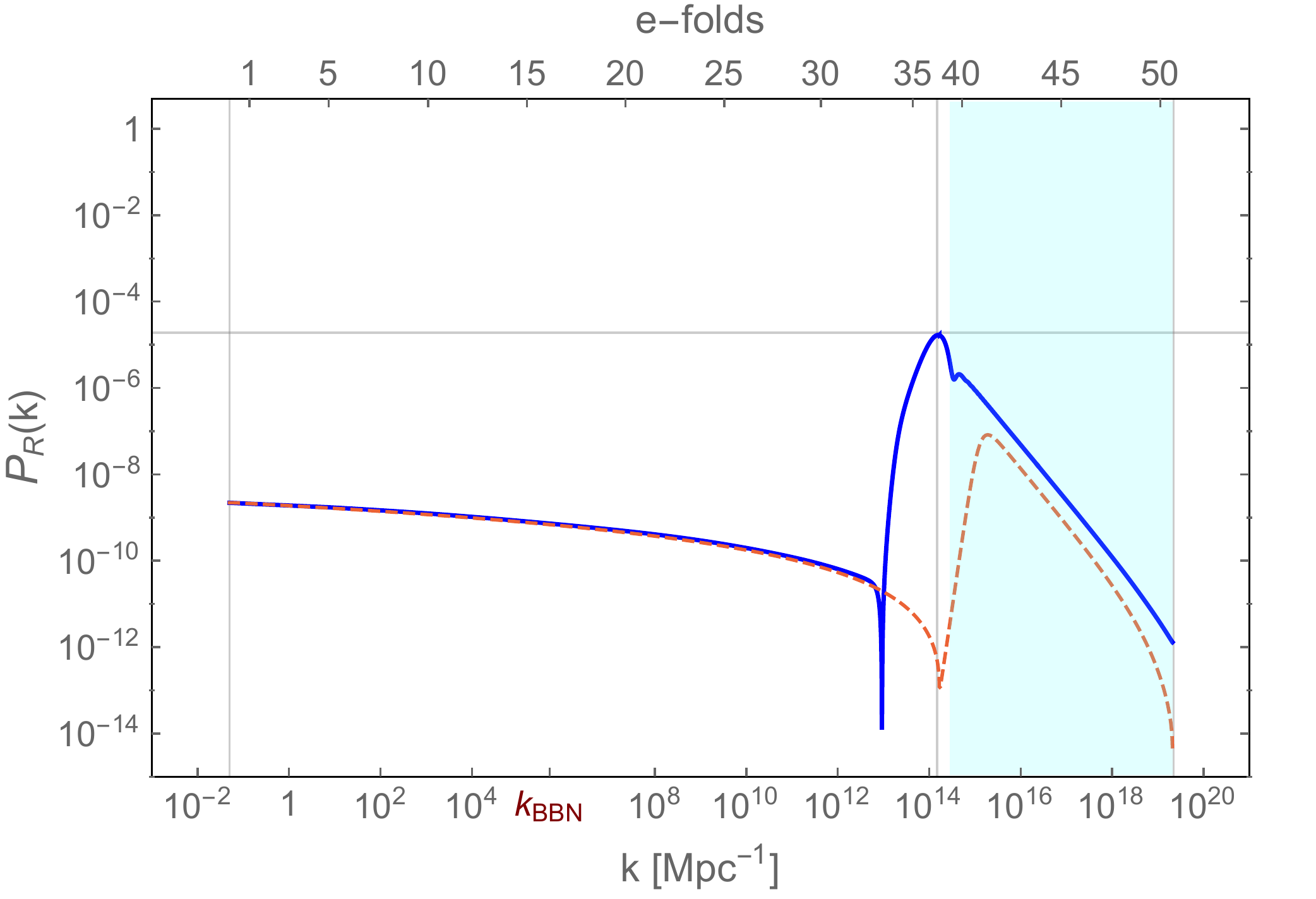}
  %\caption{1a}
  %\label{fig:sfig1}
\end{subfigure}%
\begin{subfigure}{.5\textwidth}
  \centering
  \includegraphics[width=1.\linewidth]{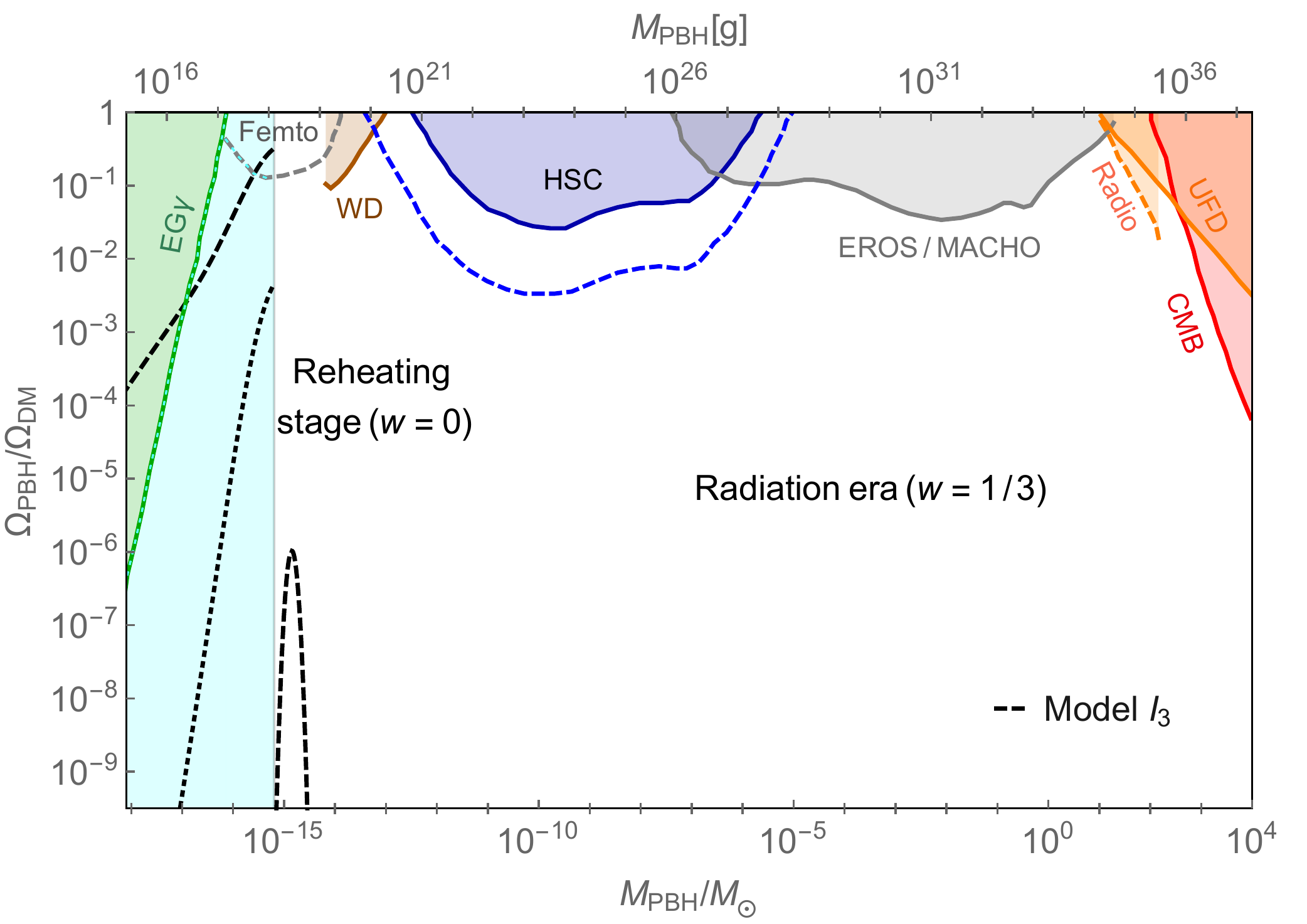}
 % \caption{1b}
  %\label{fig:sfig2}
\end{subfigure}
\\
\\
\begin{subfigure}{.5\textwidth}
  \centering
  \includegraphics[width=1.\linewidth]{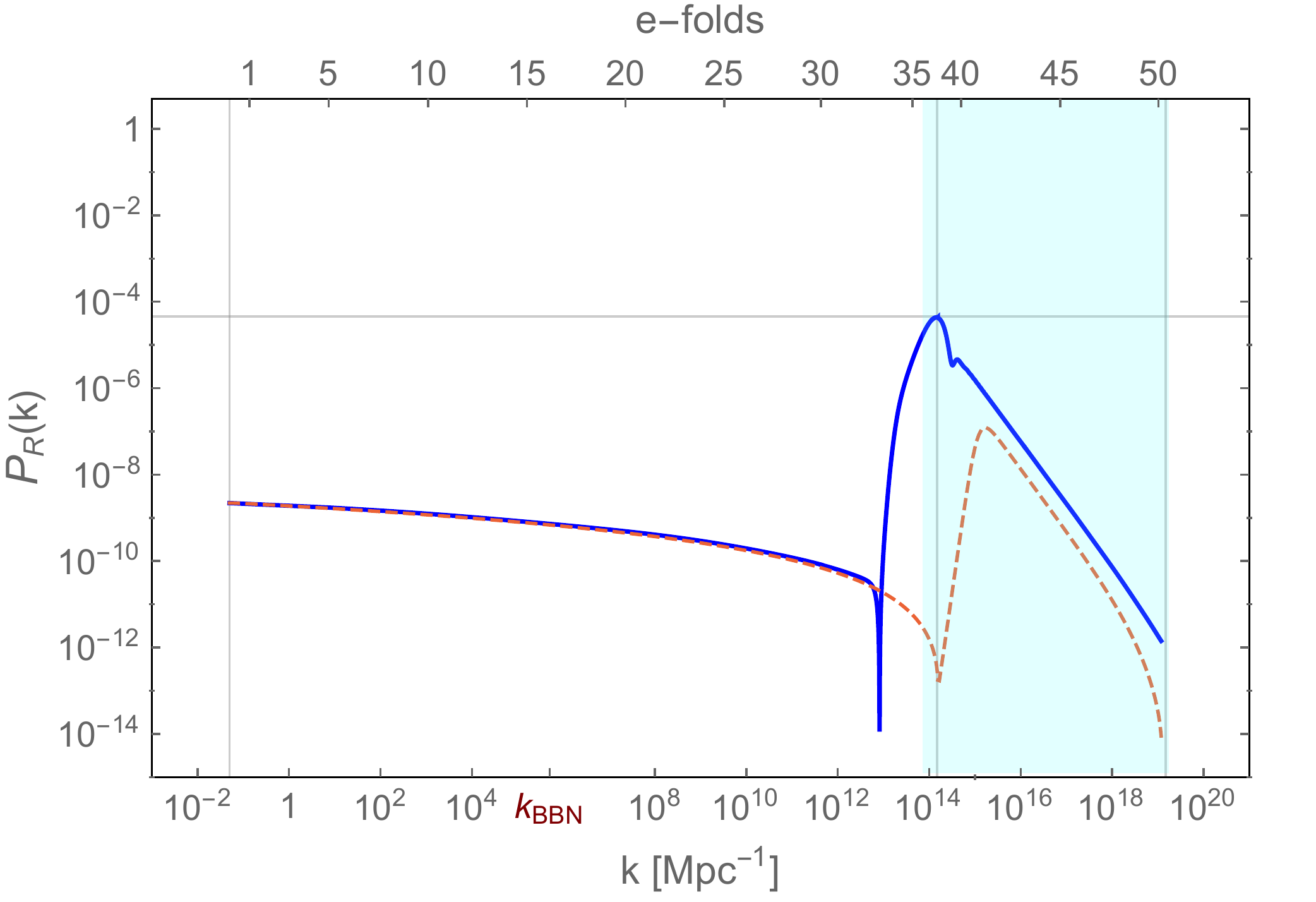}
  %\caption{1a}
  %\label{fig:sfig1}
\end{subfigure}%
\begin{subfigure}{.5\textwidth}
  \centering
  \includegraphics[width=1.\linewidth]{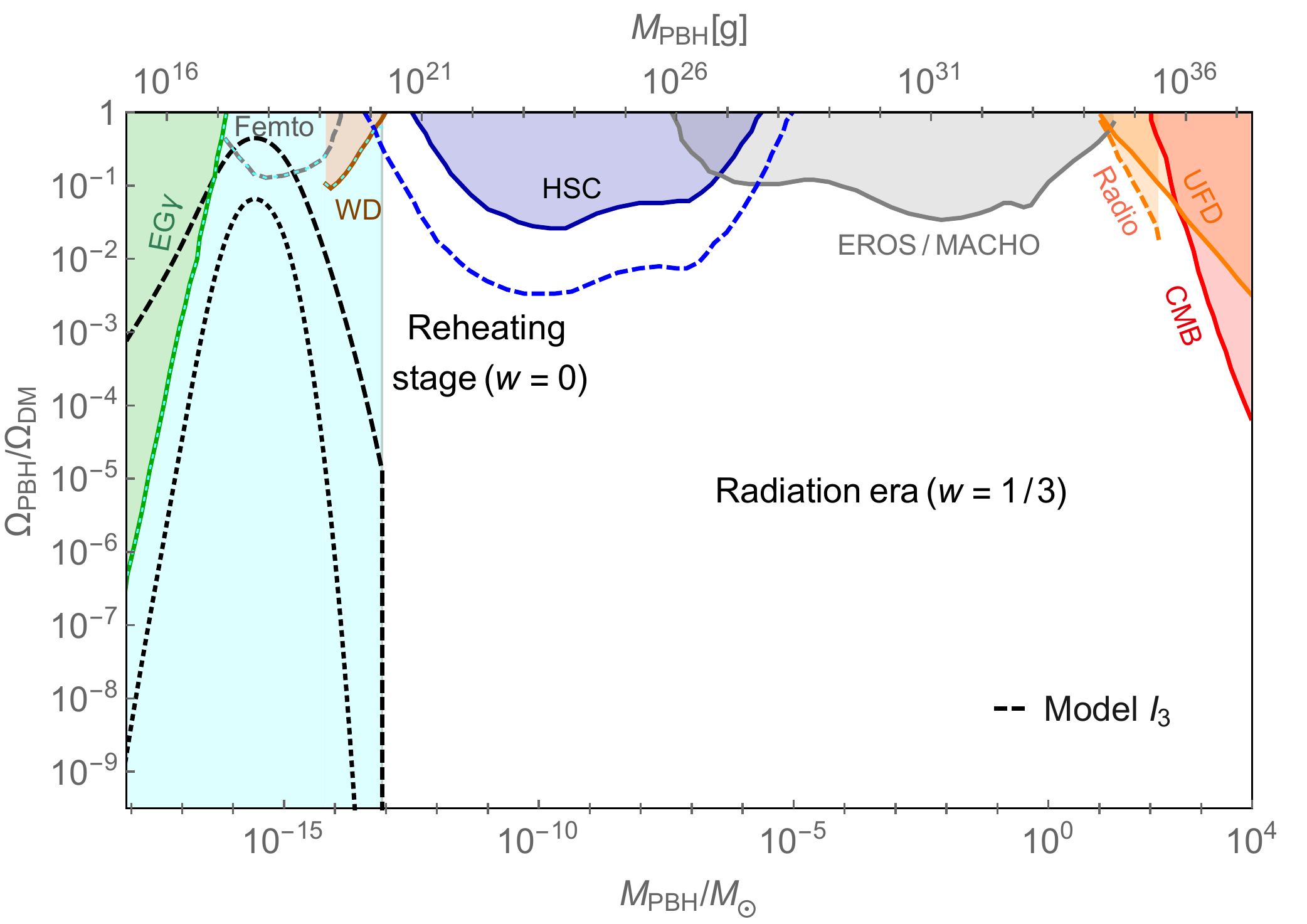}
 % \caption{1b}
  %\label{fig:sfig2}
\end{subfigure}
\caption{\label{FigReh} {\it Left panels}: The power spectra of the curvature perturbation for the Models $I_3$ with ${\cal P_R} \sim 10^{-5}$, examples $\#$ 3 and 4 respectively, see Tables {\ref{tabI}} and {\ref{tabMD}}.  
The cyan areas show the scales that reenter during reheating. For the parameters chosen the upper model has a reheating temperature  $T_\text{rh}\simeq 7\times 10^{6}$ GeV and the lower  $T_\text{rh}\simeq 2\times 10^{6}$ GeV, so that the peak of the power spectrum enters about one e-fold after and one e-fold before reheating respectively.  {\it Right panels}: The corresponding fractional abundance of PBHs, is 0.01 and 0.14 for the two models when spin effects are taken into account (dotted curves), since the variance for these models  is $\sigma(k)<0.005$ for $k$ around the peak. If there is no spin the abundance is much larger (dashed curves).  
 At the end of the reheating stage the $f_\text{PBH,tot}$ is estimated for radiation background with $w=1/3$ which  gives practically zero contribution; in the upper plot we took $\delta_c \sim 0.01$ to make the abundance peak in the RD era visible. 
}
\end{figure}
\begin{figure}[!htbp]
\begin{subfigure}{.5\textwidth}
  \centering
  \includegraphics[width=1.\linewidth]{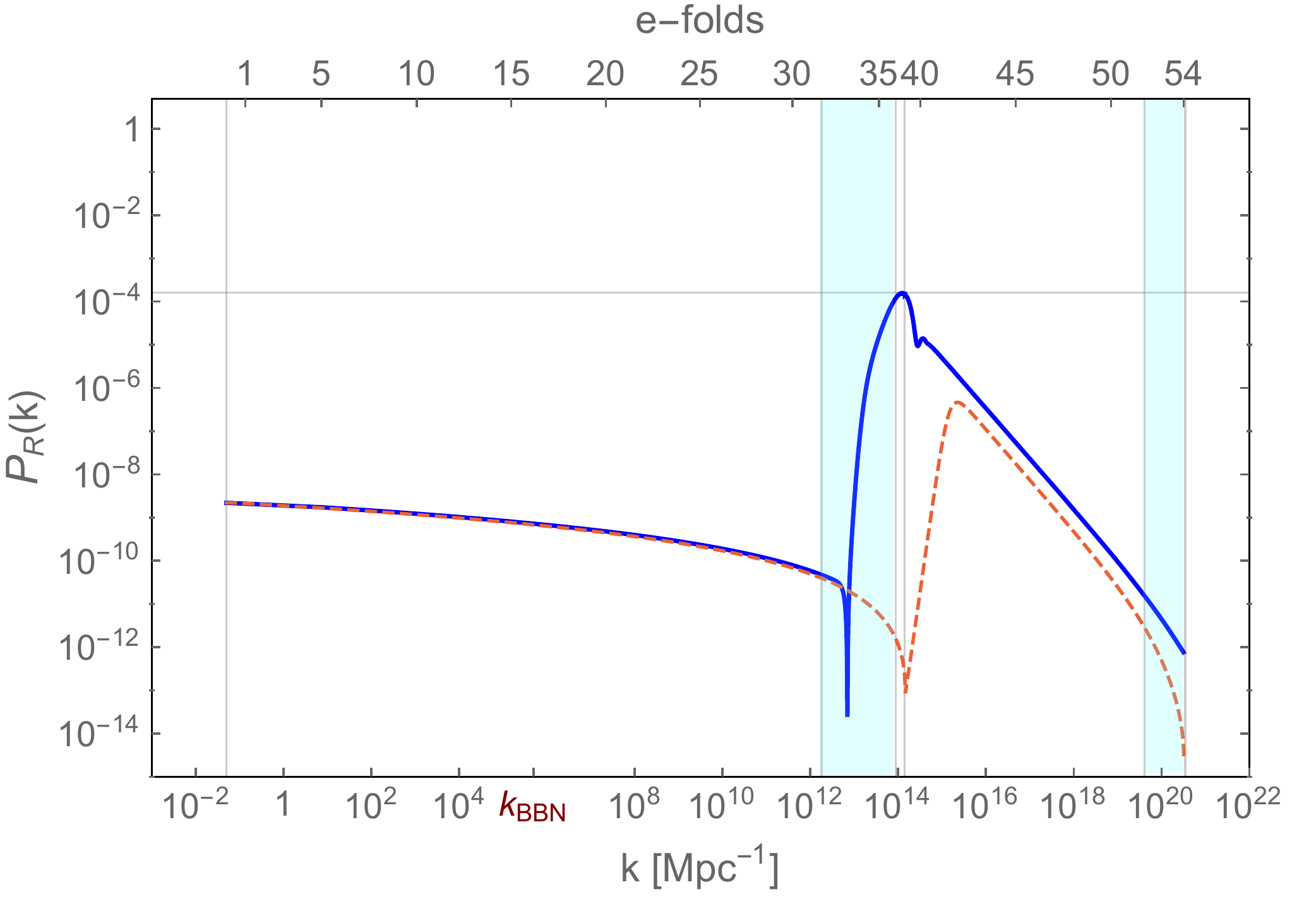}
  %\caption{1a}
  %\label{fig:sfig1}
\end{subfigure}%
\begin{subfigure}{.5\textwidth}
  \centering
  \includegraphics[width=1.\linewidth]{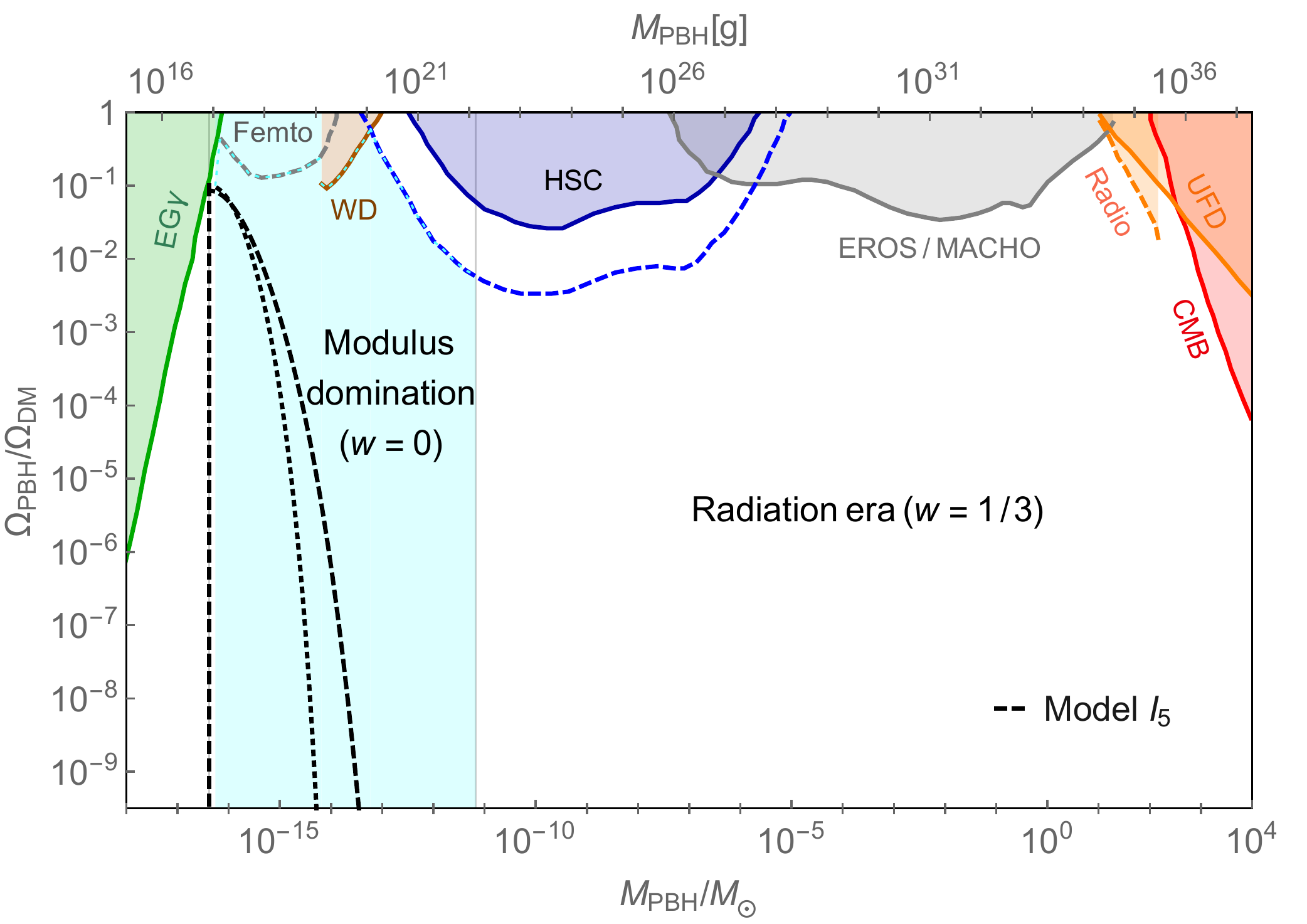}
 % \caption{1b}
  %\label{fig:sfig2}
\end{subfigure} 
\\
\\
\begin{subfigure}{.5\textwidth}
  \centering
  \includegraphics[width=1.\linewidth]{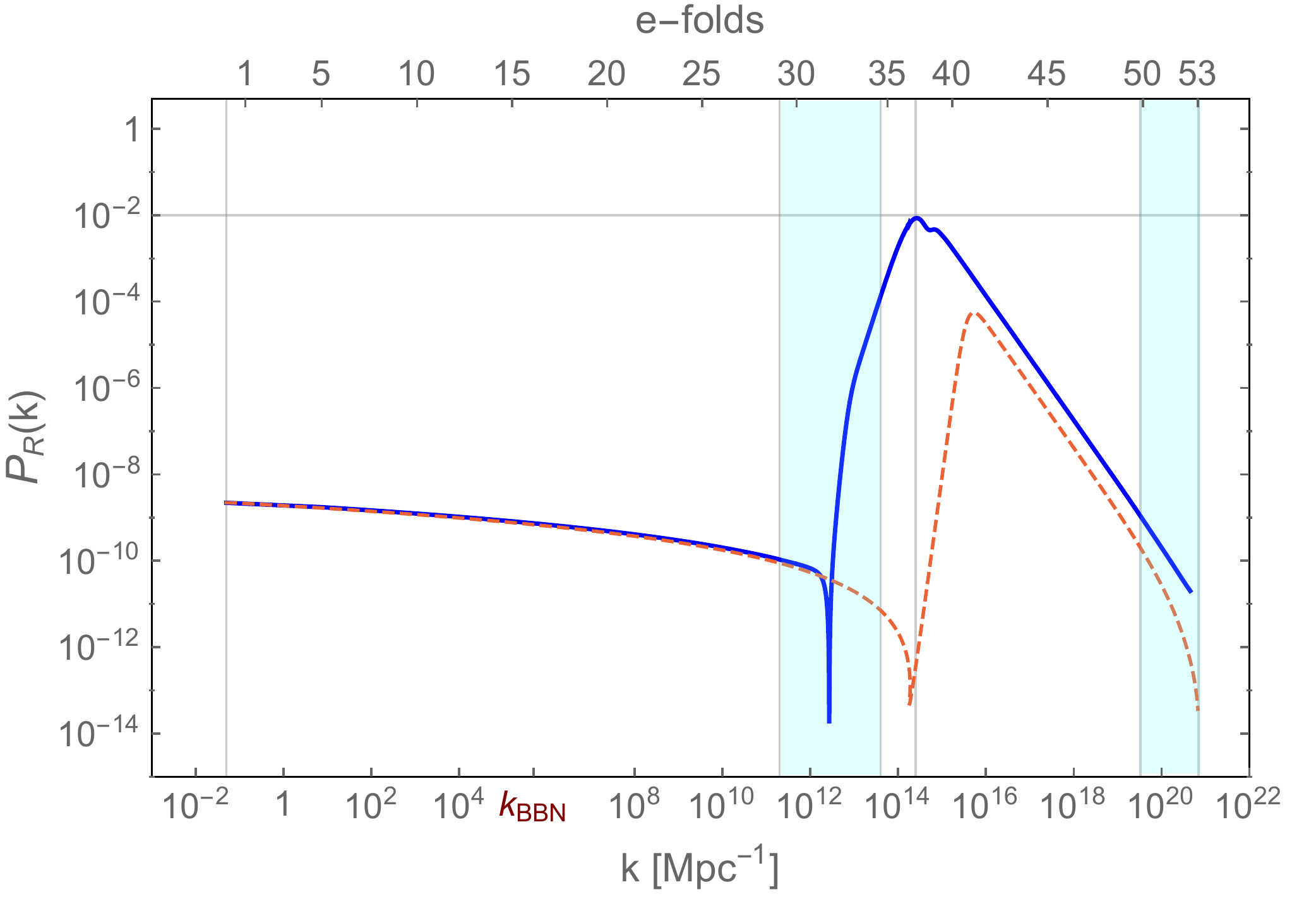}
  %\caption{1a}
  %\label{fig:sfig1}
\end{subfigure}%
\begin{subfigure}{.5\textwidth}
  \centering
  \includegraphics[width=1.\linewidth]{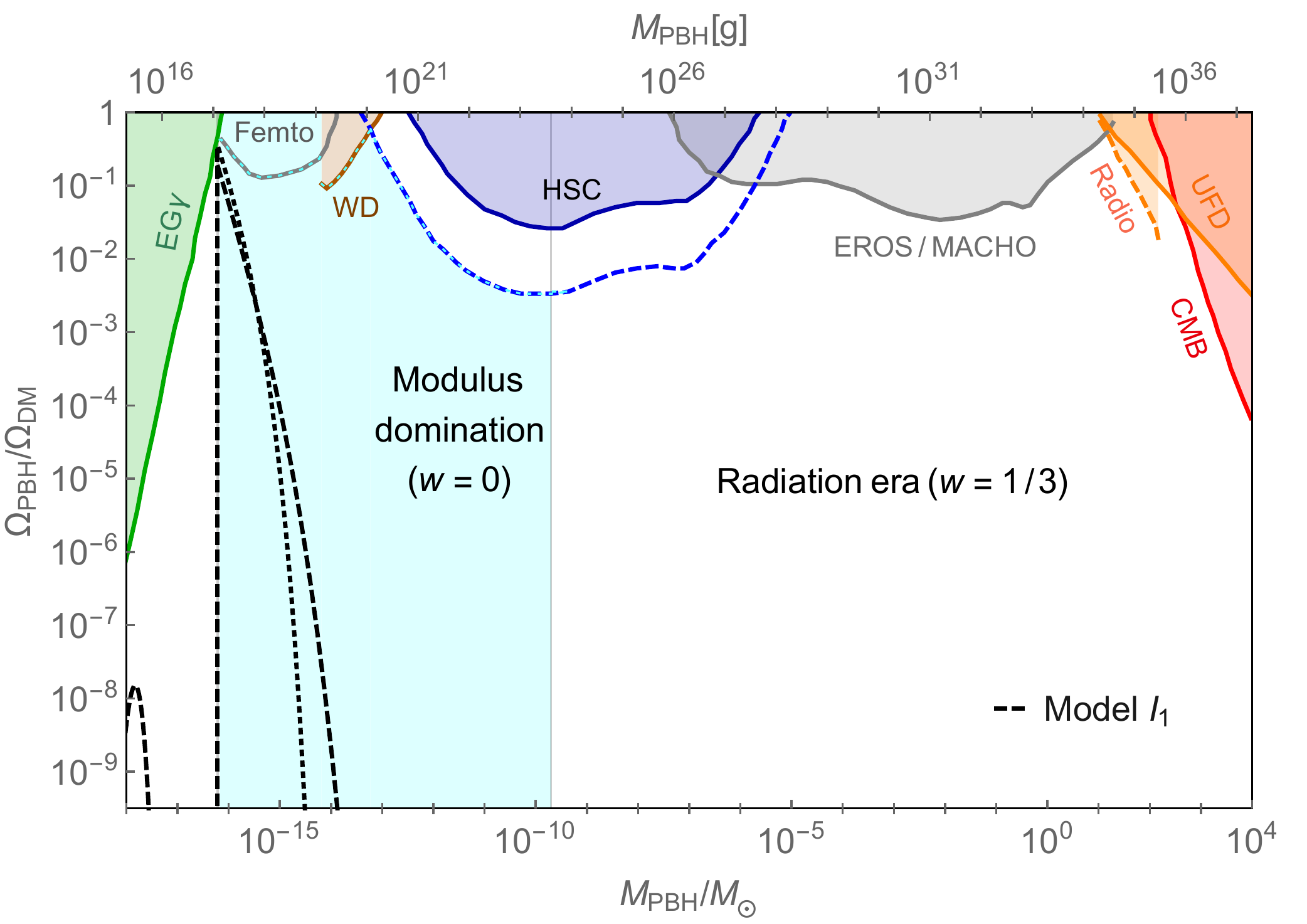}
 % \caption{1b}
  %\label{fig:sfig2}
\end{subfigure}
\caption{\label{FigMod}  
{\it Left panels}: The power spectra of the curvature perturbation for the Models $I_5$ and $I_1$ with ${\cal P_R} \sim 10^{-4}$ and ${\cal P_R} \sim 10^{-2}$ respectively,  see Tables \ref{tabI} and \ref{tabX}.
The cyan areas show the scales that reenter during a modulus dominated era. For the parameters chosen the model $I_5$ has a reheating temperature  $T_\text{rh} \simeq 4\times 10^{13}$ GeV and the model $I_1$  $T_\text{rh}= 10^{13}$ GeV. A modulus dominates the energy density for 8  and 10 e-folds  respectively that decays at about  
$T_\text{rh}\sim  10^{4}$ GeV.
  {\it Right panels}: 
The fractional PBH abundance for PBH formation during pressureless modulus domination stage (cyan area),  0.12 and 0.11 respectively.  The variance for this model for $k$ around the peak is $\sigma(k)>0.005$ and spin affects can be neglected. The PBH mass peak is placed in smaller values due to late time modulus domination. In the lower panel, the power spectrum peak, though $10^{-2}$ in size, gives a negligible PBH abundance when dilution effects are included. Here, we took $\delta_c=0.24$ to make the abundance peak during the RD era  visible.}
\end{figure}

There are four kinds of  direct observational constraints that apply on different PBH mass scales,  depicted with the colored curves and areas in Figures \ref{FigRad}, \ref{FigReh} and \ref{FigMod}. We follow the color coding of Ref.  \cite{Inomata:2017okj}. The lighter PBH  with mass less than $10^{17}$g are expected to have a cosmological relevant lifetime due to the {\it Hawking radiation}. In the late universe, the PBH evaporation rate is constrained from the extra galactic gamma-ray background \cite{Carr:2016hva} (the area in green) and in the early universe from the big-bang nucleosynthesis \cite{Carr:2009jm}.   Black holes of mass above $10^{17}$g are subject to {\it gravitational lensing} constraints \cite{Barnacka:2012bm, Tisserand:2006zx, Niikura:2017zjd} (the areas in blue and gray) \footnote{The recent results of Ref. \cite{Katz:2018zrn} remove the femtolensing constraints and we accordingly updated the plots}. Also, black holes influence the trajectory and the {\it dynamics} of other astrophysial objects such as neutron stars and white dwarfs \cite{Capela:2012jz, Capela:2013yf, Brandt:2016aco, Graham:2015apa, Gaggero:2016dpq} (the area in orange) that constrain the abundance of black holes with mass from $10^{16}$ up to $10^{25}$g. The {\it cosmic microwave background} constrains PBH with mass above $10^{33}$g because the accretion of gas and the associated emission of radiation during the recombination epoch could affect the CMB anisotropies \cite{Ricotti:2007au} (the area in orange). Recently it has been claimed that the CMB bounds on massive PBH may be relaxed due to uncertainties in the modeling of the relevant physical processes   \cite{Carr:2016drx, Clesse:2016vqa, kam}. Finally, there are indirect constraints from the \textit{ pulsar timing array} experiments on the gravitational waves associated with the formation of relatively massive PBH at the epoch of horizon entry.

We note that the PBH the mass window $10^{-16}-10^{-14} \, M_{\odot}$ that the models we studied predict is in accordance with recent searches for femtolensing effects caused by compact objects, when the presence of an extra DM component, composed of WIMP particles is considered \cite{Boucenna:2017ghj}. 
Summarizing, the synergy of different cosmological and astrophysical surveys of  large and small cosmic structures can provide us with  information about the full shape of the inflationary potential. 
An appealing feature of the models described in this work is the fact that their predictions are going to be  tested by multiple observational probes in the near future.

\section{Conclusions}

The Primordial Black Hole formation due to inflationary fluctuations has  been  attracting growing attention.  
Motivated by the substantial cosmological implications of such a hypothesis, in this paper we studied the Primordial Black Hole (PBH) production from inflationary $\alpha$--attractors considering 
different scenarios for the postinflationary evolution of the universe. 
 We analyzed two general sort of models,  labeled Models $I$ and $II$, characterized respectively by polynomial superpotential and modulated chaotic potentials that feature trigonometric functions.  The collapse of overdense region into PBH formation is triggered by a  strong peak at the power spectrum at small scales $k_\text{} \sim 10^{14}  \Mpc$.  The peak is generated in small fields values, after roughly $N=38$ e-folds of observable inflation, i.e far from the asymptotically flat regions of $\alpha$-attractors, and it is this e-folds number that determines the spectral index value, $n_s$.   
We examined the PBH formation in backgrounds dominated by a pressureless matter and by thermal radiation.  
We found that PBH in the mass window $10^{-16}-10^{-14} \, M_{\odot}$   (or $10^{17}-10^{20}$ g, which corresponds to the mass range of small asteroids) can be produced with a cosmologically significant abundance,  $\Omega_\text{PBH}/ \Omega_\text{DM}= {\cal O}(0.1)$

A pressureless matter dominated era before BBN can be attributed  to the  postinflationary  reheating stage necessary to pass to the thermal RD era.  Moreover, in the framework of superconformal attractors and generally in supergravity and string theory the presence of late decaying scalar fields, such as  moduli, is natural and can realize a postinflationary non-thermal phase. Actually, a non-thermal phase is welcome in many particle dark matter scenarios and an extra dark matter component is necessary to saturate the observed $\Omega_\text{DM}$.    Motivated by these considerations, in this work, we investigated the PBH formation during radiation,  pressureless reheating and modulus dominated eras in a consistent way and our results are presented in Tables \ref{tabRad}, \ref{tabMD} and  \ref{tabX} and in Figures \ref{FigRad}, \ref{FigReh} and \ref{FigMod}. 
We say consistent in the sense that the duration of the MD and RD eras, the reheating temperature, the PBH mass range and abundance are interrelated and explicitly determined by the inflationary stage itself.

The increase in the power spectrum is achieved by the rapid change of the Hubble flow parameters $\epsilon_1$, $\epsilon_2$ and $\epsilon_3$ in the region about the inflection point. 
The required amplitude of the power spectrum peak that yields a significant PBH abundance  depends significantly on the background pressure
that determines 
the process the primordial  inhomogeneities collapse to form black holes.  In particularly for the RD era a significant abundance can be achieved after a delicate balance between the threshold parameter $\delta_c$ and the amplitude of the power spectrum peak. During MD era  the absence or presence of angular momentum in the collapsing region is important  for PBH formation. In addition,  the reheating temperature of the universe is critical for the determination of the PBH abundance. 

 For PBHs forming during MD a moderate peak ${\cal P_R}\sim 10^{-5}-10^{-4}$ in the power spectrum is adequate to generate a cosmologically significant PBH abundance if the relevant curvature perturbation mode ${\cal R}_k$ reenters the horizon not long before the reheating of the universe. In this case the density of the PBH formed will have enough time to increase until the matter-radiation equality.  We also find  that PBH mass range is  shifted to  smaller values compared to the RD era.
 The details  of the transition between the reheating and the thermal radiation regime can play a crucial  r\^ole for the determination of the PBH abundance and deserve a further investigation.  
 
The prediction of the PBH relic density is very sensitive to the  quantity $\beta(M_k)$  \cite{Biagetti:2018pjj} .  A key parameter is the ratio of the averaged production rate in the presence of diffusion and the production rate in the absence of diffusion, $e^{\Delta_\text{qd}} \equiv  {\left\langle \beta(M_k) \right\rangle}/{\beta(M_k)}$ which has to be $|\Delta_\text{qd}|\lesssim 1$ in order the  uncertainty in  the PBH relic density estimation to be negligible.  This is an interesting issue that we leave for  a separate study.

The superconformal models lead to distinct cosmological predictions  despite the flexibility in the choice of potentials, characteristic of the $\alpha$-attractors.  
The superconformal attractors customized to generate PBH in the low-mass region    predict a relatively small $n_s$ value, and larger $r$ and $\alpha_s$ values compared to the conventional inflationary $\alpha$--attractor  models.
The $n_s$ value is 
compatible with the Planck 2015 data at the 95 \% CL. 
This improved Planck 2018 bounds further constrain our models and our predictions  are placed in the borderline of what present data allows.   
This scenario
 is possible to be tested by the next generation CMB probes that aim at pinning down the scalar tilt value with per mil accuracy. Large values for the $n_s$ will rule out this sort of models. Moreover, microlensing observational programs should be capable to search for PBH in the mass range predicted here and considerably constrain the abundance. 
 The current status is that the PBH mass window  (\ref{masswind})  is in accordance with recent searches for femtolensing effects caused by compact objects, even when the presence of an extra DM component composed of WIMP particles is considered \cite{Boucenna:2017ghj}.

The results of this work suggest that $\alpha$--attractors provide a framework %and offer new possibilities  
to implement PBH production in a mass range of cosmological and observational  interest. 
We underline, nevertheless, that our analysis and results are general and can apply to any inflationary model that generates a peak in the power spectrum of the curvature perturbation capable to trigger  PBH formation.

%Near future CMB and astrophysical probes will decisively test the predictions of this class of models.

\label{sec:Conc}

\vspace*{.5cm}

\section*{Note added}

After the completion of this work the data from Planck 2018 \cite{Akrami:2018odb} were released that increase the lower bound on the scalar tilt $n_s$. 
In addition the paper \cite{Katz:2018zrn} appeared that removes the femtolensing constraints on the $f_\text{PBH}$.  These new results have been taken into consideration in the current version of our paper and the relevant analysis, tables and figures have been updated accordingly.

\vspace*{.5cm}

\section*{Acknowledgments}

\noindent 
This work is supported by the GSRT under the EDEIL/67108600. G.T. thanks Guillermo Ballesteros and Brian Powell for correspondence as well as the Bonn-Cologne Graduate School for Physics and Astronomy for the financial support during the completion of this project.
Discussions with  K. Dimopoulos, F. Farakos,  J. Garcia-Bellido,  A. Riotto, E. Ruiz Morales and S. Sibiryakov are kindly acknowledged. The work of I.D. is supported by the IKY Scholarship Programs for Strengthening Post Doctoral Research, co-financed by the European Social Fund ESF and the Greek government.   He also thanks CERN-TH for hospitality.
A.K. thanks the Cosmology group at the D\'epartement de Physique 
  Th\'eorique  at the  Universit\'e de Gen\`eve for the kind hospitality and financial support.


\begin{thebibliography}{99}
\setlength{\itemsep}{-0.8mm}
\bibitem{ligo} B. P. Abbott et al. [LIGO Scientific and Virgo Collaborations], Phys. Rev. Lett. 116, 061102 (2016)
\arXiv{1602.03837}{gr-qc}.
  \bibitem{PBH1} B. J. Carr and S. W. Hawking, Mon. Not. Roy. Astron.
Soc. {\bf 168}, 399 (1974).  
 
  \bibitem{PBH2}  P. Meszaros, Astron. Astrophys. {\bf 37}, 225 (1974).
  
  \bibitem{PBH3} B. J. Carr, Astrophys. J. {\bf 201}, 1 (1975).

 \bibitem{PBH4} 
 %"Primordial Black Holes"
I. D. Novikov, A.G. Polnarev, A. A. Starobinsky and Ya. B. Zeldovich, Astron. Astroph. 80 , 104 (1979).
  
 
  
  \bibitem{kam} S.~Bird, I.~Cholis, J.~B.~MuÃ±oz, Y.~Ali-HaÃ¯moud, M.~Kamionkowski, E.~D.~Kovetz, A.~Raccanelli and A.~G.~Riess,
  %``Did LIGO detect dark matter?,''
  Phys.\ Rev.\ Lett.\  {\bf 116}, no. 20, 201301 (2016)
  %doi:10.1103/PhysRevLett.116.201301
  \arXiv{1603.00464}{astro-ph.CO}.
  
  %\cite{Clesse:2016vqa}
\bibitem{Clesse:2016vqa} 
  S.~Clesse and J.~García-Bellido,
  %``The clustering of massive Primordial Black Holes as Dark Matter: measuring their mass distribution with Advanced LIGO,''
  Phys.\ Dark Univ.\  {\bf 15}, 142 (2017)
%  doi:10.1016/j.dark.2016.10.002
  \arXiv{1603.05234}{astro-ph.CO}.
  %%CITATION = doi:10.1016/j.dark.2016.10.002;%%
  %98 citations counted in INSPIRE as of 24 Feb 2018
 
  %\cite{Sasaki:2016jop}
\bibitem{rep2} 
  M.~Sasaki, T.~Suyama, T.~Tanaka and S.~Yokoyama,
  %``Primordial Black Hole Scenario for the Gravitational-Wave Event GW150914,''
  Phys.\ Rev.\ Lett.\  {\bf 117}, no. 6, 061101 (2016)
  %doi:10.1103/PhysRevLett.117.061101
  \arXiv{1603.08338}{astro-ph.CO}.
  %%CITATION = doi:10.1103/PhysRevLett.117.061101;%%
  %133 citations counted in INSPIRE as of 24 Feb 2018
  
  
 
 
 \bibitem{c0} B.~Carr, M.~Raidal, T.~Tenkanen, V.~Vaskonen and H.~Veerme,
  %``Primordial black hole constraints for extended mass functions,''
  Phys.\ Rev.\ D {\bf 96}, no. 2, 023514 (2017)
   \arXiv{1705.05567}{astro-ph.CO}.
 
  \bibitem{c1} M.~Zumalacarregui and U.~Seljak,
  %``No LIGO MACHO: Primordial Black Holes, Dark Matter and Gravitational Lensing of Type Ia Supernovae,''
  \arXiv{1712.02240}{astro-ph.CO}.
  
  \bibitem{c2} 
  J.~Garcia-Bellido, S.~Clesse and P.~Fleury,
  %``LIGO Lo(g)Normal MACHO: Primordial Black Holes survive SN lensing constraints,''
 \arXiv{1712.06574}{astro-ph.CO}

 \bibitem{revPBH} 
  M.~Sasaki, T.~Suyama, T.~Tanaka and S.~Yokoyama,
  %``Primordial Black Holes - Perspectives in Gravitational Wave Astronomy -,''
  \arXiv{1801.05235}{astro-ph.CO}
  
  
  \bibitem{s1} P.~Ivanov, P.~Naselsky and I.~Novikov,
  %``Inflation and primordial black holes as dark matter,''
  Phys.\ Rev.\ D {\bf 50}, 7173 (1994).
    
  
  \bibitem{s2} J.~Garc\'{\i}a-Bellido, A.D.~Linde and D.~Wands,
  %``Density perturbations and black hole formation in hybrid inflation,''
  Phys.\ Rev.\ D {\bf 54} (1996) 6040\\
  %doi:10.1103/PhysRevD.54.6040
  \arXivold{astro-ph/9605094}.

  \bibitem{s3} %``Nonlinear metric perturbations and production of primordial black holes,''
  P.~Ivanov, Phys.\ Rev.\ D {\bf 57}, 7145 (1998)
  \arXivold{astro-ph/9708224}.
    
    
%    \bibitem{carr01}
%  B.~Carr, M.~Raidal, T.~Tenkanen, V.~Vaskonen and H.~Veermäe,
  %``Primordial black hole constraints for extended mass functions,''
%  Phys.\ Rev.\ D {\bf 96} (2017) no.2,  023514
  %doi:10.1103/PhysRevD.96.023514
%  \arXiv{1705.05567} {astro-ph.CO}.
  %%CITATION = doi:10.1103/PhysRevD.96.023514;%%
  %59 citations counted in INSPIRE as of 15 Mar 2018
    
    
%%%%  B E F O R E    B E L L I D O   2017
      
\bibitem{Alabidi:2009bk}
  L.~Alabidi and K.~Kohri,
  %``Generating Primordial Black Holes Via Hilltop-Type Inflation Models,''
  Phys.\ Rev.\ D {\bf 80} (2009) 063511
 % doi:10.1103/PhysRevD.80.063511
  \arXiv{0906.1398} {astro-ph.CO}.
  %%CITATION = doi:10.1103/PhysRevD.80.063511;%%
  %32 citations counted in INSPIRE as of 30 Mar 2018    
    
    
\bibitem{Drees:2011hb}
  M.~Drees and E.~Erfani,
  %``Running-Mass Inflation Model and Primordial Black Holes,''
  JCAP {\bf 1104} (2011) 005
  %doi:10.1088/1475-7516/2011/04/005
  \arXiv{1102.2340} {hep-ph}.
  %%CITATION = doi:10.1088/1475-7516/2011/04/005;%%
  %45 citations counted in INSPIRE as of 30 Mar 2018    
    
\bibitem{Drees:2011yz}
  M.~Drees and E.~Erfani,
  %``Running Spectral Index and Formation of Primordial Black Hole in Single Field Inflation Models,''
  JCAP {\bf 1201} (2012) 035
  %doi:10.1088/1475-7516/2012/01/035
  \arXiv{1110.6052} {astro-ph.CO}.
  %%CITATION = doi:10.1088/1475-7516/2012/01/035;%%
  %40 citations counted in INSPIRE as of 30 Mar 2018    
    
    \bibitem{sone1} J.~Garcia-Bellido and E.~Ruiz Morales,
  %``Primordial black holes from single field models of inflation,''
  Phys.\ Dark Univ.\  {\bf 18}, 47 (2017)
   \arXiv{1702.03901}{astro-ph.CO}.


\bibitem{ssm1}
  J.M.~Ezquiaga, J.~Garc\'{\i}a-Bellido and E.~Ruiz Morales,
  %``Primordial Black Hole production in Critical Higgs Inflation,''
 Phys.\ Lett.\ B {\bf 776}, 345 (2018)
  \arXiv{1705.04861}{astro-ph.CO}.
    
    
    \bibitem{sone0} K.~Kannike, L.~Marzola, M.~Raidal and H.~Veerm\"ae,
  %``Single Field Double Inflation and Primordial Black Holes,''
  JCAP {\bf 1709}, no. 09, 020 (2017)
  \arXiv{1705.06225}{astro-ph.CO}.


  
\bibitem{sone2}G.~Ballesteros and M.~Taoso,
  %``Primordial black hole dark matter from single field inflation,''
  Phys.\ Rev.\ D {\bf 97}, no. 2, 023501 (2018)
  \arXiv{1709.05565}{hep-ph}.
 
 
\bibitem{Hertzberg:2017dkh}
  M.~P.~Hertzberg and M.~Yamada,
  %``Primordial Black Holes from Polynomial Potentials in Single Field Inflation,''
  \arXiv{1712.09750} {astro-ph.CO}.
  %%CITATION = ARXIV:1712.09750;%%
  %5 citations counted in INSPIRE as of 03 Apr 2018 
 
 \bibitem{sone3} M.~Cicoli, V.~A.~Diaz and F.~G.~Pedro,
  %``Primordial Black Holes from String Inflation,''
  \arXiv{1803.02837}{hep-th}.   
    
    
   \bibitem{Ozsoy:2018flq}
  O.~\"Ozsoy, S.~Parameswaran, G.~Tasinato and I.~Zavala,
  %``Mechanisms for Primordial Black Hole Production in String Theory,''
  \arXiv{1803.07626} {hep-th}.
  %%CITATION = ARXIV:1803.07626;%%
  
\bibitem{Gao:2018pvq} 
  T.~J.~Gao and Z.~K.~Guo,
  %``Primordial Black Hole Production in Inflationary Models of Supergravity with a Single Chiral Superfield,''
  Phys.\ Rev.\ D {\bf 98}, no. 6, 063526 (2018)
 % doi:10.1103/PhysRevD.98.063526
  \arXiv{arXiv:1806.09320 [hep-ph]}.
  %%CITATION = doi:10.1103/PhysRevD.98.063526;%%
  %1 citations counted in INSPIRE as of 03 Oct 2018
  
  
  \bibitem{Kawasaki:1997ju}
  M.~Kawasaki, N.~Sugiyama and T.~Yanagida,
  %``Primordial black hole formation in a double inflation model in supergravity,''
  Phys.\ Rev.\ D {\bf 57} (1998) 6050
%  doi:10.1103/PhysRevD.57.6050
  \arXivold{hep-ph/9710259}.
  %%CITATION = doi:10.1103/PhysRevD.57.6050;%%
  %70 citations counted in INSPIRE as of 03 Apr 2018
  

  \bibitem{Kawasaki:2016pql}
  M.~Kawasaki, A.~Kusenko, Y.~Tada and T.~T.~Yanagida,
  %``Primordial black holes as dark matter in supergravity inflation models,''
  Phys.\ Rev.\ D {\bf 94} (2016) no.8,  083523
%  doi:10.1103/PhysRevD.94.083523
  \arXiv{1606.07631} {astro-ph.CO}.
  %%CITATION = doi:10.1103/PhysRevD.94.083523;%%
  %42 citations counted in INSPIRE as of 02 Apr 2018

\bibitem{Kawaguchi:2007fz}
  T.~Kawaguchi, M.~Kawasaki, T.~Takayama, M.~Yamaguchi and J.~Yokoyama,
  %``Formation of intermediate-mass black holes as primordial black holes in the inflationary cosmology with running spectral index,''
  Mon.\ Not.\ Roy.\ Astron.\ Soc.\  {\bf 388} (2008) 1426
  %doi:10.1111/j.1365-2966.2008.13523.x
  \arXiv{0711.3886} {astro-ph}.
  %%CITATION = doi:10.1111/j.1365-2966.2008.13523.x;%%
  %39 citations counted in INSPIRE as of 03 Apr 2018

\bibitem{Pi:2017gih}
  S.~Pi, Y.~l.~Zhang, Q.~G.~Huang and M.~Sasaki,
  %``Scalaron from $R^2$-gravity as a Heavy Field,''
  JCAP {\bf 1805} (2018) no.05,  042
  %doi:10.1088/1475-7516/2018/05/042
  \arXiv{1712.09896} {astro-ph.CO}.
  %%CITATION = doi:10.1088/1475-7516/2018/05/042;%%
  %4 citations counted in INSPIRE as of 01 Jun 2018
    
  
  \bibitem{stwo1} M.~Kawasaki, N.~Kitajima and T.~T.~Yanagida,
  %``Primordial black hole formation from an axionlike curvaton model,''
  Phys.\ Rev.\ D {\bf 87}, no. 6, 063519 (2013)
  \arXiv{1207.2550}{hep-ph}.

  
      \bibitem{Carr:2016drx} B.~Carr, F.~Kuhnel and M.~Sandstad,
  %``Primordial Black Holes as Dark Matter,''
  Phys.\ Rev.\ D {\bf 94}, 083504 (2016)
  %doi:10.1103/PhysRevD.94.083504
  \arXiv{1607.06077}{astro-ph.CO}.

 \bibitem{gauge} J. Garcia-Bellido, M. Peloso and C. Unal, JCAP 1612, no. {\bf 12}, 031 (2016) 
 \arXiv{1610.03763}{astro-ph.CO}.
 
 \bibitem{cs} B.~Carr, T.~Tenkanen and V.~Vaskonen,
  %``Primordial black holes from inflaton and spectator field perturbations in a matter-dominated era,''
  Phys.\ Rev.\ D {\bf 96}, no. 6, 063507 (2017)
  \arXiv{1706.03746}{astro-ph.CO}.
  
    
  \bibitem{ssm2}  J.~R.~Espinosa, D.~Racco and A.~Riotto,
  %``A Cosmological Signature of the Standard Model Higgs Vacuum Instability: Primordial Black Holes as Dark Matter,''
  Phys.\ Rev.\ Lett.\  {\bf 120}, 121301 (2018)\\
\arXiv{1710.11196}{hep-ph}.

\bibitem{ERR} 
  J.~R.~Espinosa, D.~Racco and A.~Riotto,
  %``Primordial Black Holes from Higgs Vacuum Instability: Avoiding Fine-tuning through an Ultraviolet Safe Mechanism,''
 \arXiv{1804.07731}{hep-ph}.

\bibitem{Kinney}
  W.~H.~Kinney,
  %``Horizon crossing and inflation with large eta,''
  Phys.\ Rev.\ D {\bf 72} (2005) 023515
 % doi:10.1103/PhysRevD.72.023515
  \arXivold{gr-qc/0503017}.
  %%CITATION = doi:10.1103/PhysRevD.72.023515;%%
  %107 citations counted in INSPIRE as of 02 Apr 2018

\bibitem{Biagetti:2018pjj}
  M.~Biagetti, G.~Franciolini, A.~Kehagias and A.~Riotto,
  %``Primordial Black Holes from Inflation and Quantum Diffusion,''
  \arXiv{1804.07124} {astro-ph.CO}.
  %%CITATION = ARXIV:1804.07124;%%


\bibitem{Ezquiaga:2018gbw}
  J.~M.~Ezquiaga and J.~García-Bellido,
  %``Quantum diffusion beyond slow-roll: implications for primordial black-hole production,''
  \arXiv{1805.06731} {astro-ph.CO}.
  %%CITATION = ARXIV:1805.06731;%%


\bibitem{Franciolini:2018vbk}
  G.~Franciolini, A.~Kehagias, S.~Matarrese and A.~Riotto,
  %``Primordial Black Holes from Inflation and non-Gaussianity,''
  JCAP {\bf 1803} (2018) no.03,  016
  %doi:10.1088/1475-7516/2018/03/016
  \arXiv{1801.09415} {astro-ph.CO}.
  %%CITATION = doi:10.1088/1475-7516/2018/03/016;%%
  %3 citations counted in INSPIRE as of 30 Mar 2018

% A T T R A C T O R S

\bibitem{KL1}
  R.~Kallosh and A.~Linde,
  %``Universality Class in Conformal Inflation,''
  JCAP {\bf 1307} (2013) 002
%  doi:10.1088/1475-7516/2013/07/002
  \arXiv{1306.5220}{hep-th}.
\bibitem{KL2}
  R.~Kallosh and A.~Linde,
  %``Multi-field Conformal Cosmological Attractors,''
  JCAP {\bf 1312} (2013) 006
 % doi:10.1088/1475-7516/2013/12/006
  \arXiv{1309.2015} {hep-th}.

\bibitem{CK}
  S.~Cecotti and R.~Kallosh,
  %``Cosmological Attractor Models and Higher Curvature Supergravity,''
  JHEP {\bf 1405} (2014) 114
 % doi:10.1007/JHEP05(2014)114
  \arXiv{1403.2932}{hep-th}.
  
\bibitem{KLR2} 
  R.~Kallosh, A.~Linde and D.~Roest,
  %``Superconformal Inflationary $\alpha$-Attractors,''
  JHEP {\bf 1311}, 198 (2013)
 % doi:10.1007/JHEP11(2013)198
  \arXiv{1311.0472}{hep-th}.  
  
  \bibitem{KLR0}
  R.~Kallosh, A.~Linde and D.~Roest,
  %``Universal Attractor for Inflation at Strong Coupling,''
  Phys.\ Rev.\ Lett.\  {\bf 112} (2014) no.1,  011303
 % doi:10.1103/PhysRevLett.112.011303
  \arXiv{1310.3950}{hep-th}.
  
  \bibitem{KLR}
  R.~Kallosh, A.~Linde and D.~Roest,
  %``Large field inflation and double $\alpha$-attractors,''
  JHEP {\bf 1408} (2014) 052
 % doi:10.1007/JHEP08(2014)052
  \arXiv{1405.3646}{hep-th}.

\bibitem{Kal}
  R.~Kallosh,
  %``Planck 2013 and Superconformal Symmetry,''
%  doi:10.1093/acprof:oso/9780198728856.003.0012
  \arXiv{1402.0527}{hep-th}.
  %%CITATION = doi:10.1093/acprof:oso/9780198728856.003.0012;%%
  %12 citations counted in INSPIRE as of 16 Mar 2018



\bibitem{FKS} 
  S.~Ferrara and R.~Kallosh,
  %``Seven-disk manifold, $\alpha$-attractors, and $B$ modes,''
  Phys.\ Rev.\ D {\bf 94}, no. 12, 126015 (2016)
%  doi:10.1103/PhysRevD.94.126015
  \arXiv{1610.04163}{hep-th}.



\bibitem{Kallosh:2010ug}
  R.~Kallosh and A.~Linde,
  %``New models of chaotic inflation in supergravity,''
  JCAP {\bf 1011} (2010) 011
%  doi:10.1088/1475-7516/2010/11/011
  \arXiv{1008.3375}{hep-th}.
  %%CITATION = doi:10.1088/1475-7516/2010/11/011;%%
  %196 citations counted in INSPIRE as of 16 Mar 2018


\bibitem{FvP}
  D.~Z.~Freedman and A.~Van Proeyen,
  ``Supergravity,'' Cambridge Univ. Pr. 2012, 607 p.

\bibitem{Stewart:1994ts}
  E.~D.~Stewart,
  %``Inflation, supergravity and superstrings,''
  Phys.\ Rev.\ D {\bf 51} (1995) 6847
 % doi:10.1103/PhysRevD.51.6847
\arXiv {hep-ph/9405389}.
  %%CITATION = doi:10.1103/PhysRevD.51.6847;%%
  %229 citations counted in INSPIRE as of 03 Oct 2018

\bibitem{Roest:2015qya}
  D.~Roest and M.~Scalisi,
  %``Cosmological attractors from α-scale supergravity,''
  Phys.\ Rev.\ D {\bf 92} (2015) 043525
%  doi:10.1103/PhysRevD.92.043525
  \arXiv {1503.07909} {hep-th}.
  %%CITATION = doi:10.1103/PhysRevD.92.043525;%%
  %47 citations counted in INSPIRE as of 01 Jun 2018


\bibitem{FKLP1}
  S.~Ferrara, R.~Kallosh, A.~Linde and M.~Porrati,
  %``Minimal Supergravity Models of Inflation,''
  Phys.\ Rev.\ D {\bf 88} (2013) no.8,  085038
 % doi:10.1103/PhysRevD.88.085038
  \arXiv{1307.7696}{hep-th}.
  
 \bibitem{FK}
  S.~Ferrara and A.~Kehagias,
  %``Higher Curvature Supergravity, Supersymmetry Breaking and Inflation,''
  Subnucl.\ Ser.\  {\bf 52} (2017) 119
%  doi:10.1142/9789813148680_0003
  \arXiv{1407.5187}{hep-th}.
  
  
\bibitem{Starobinsky:1980te}
  A.~A.~Starobinsky,
  %``A New Type of Isotropic Cosmological Models Without Singularity,''
  Phys.\ Lett.\ B {\bf 91} (1980) 99
   [Phys.\ Lett.\  {\bf 91B} (1980) 99]
   [Adv.\ Ser.\ Astrophys.\ Cosmol.\  {\bf 3} (1987) 130].
  %doi:10.1016/0370-2693(80)90670-X
  %%CITATION = doi:10.1016/0370-2693(80)90670-X;%%
  %3895 citations counted in INSPIRE as of 04 Jan 2019  
  
 
\bibitem{Planck1} 
  P.~A.~R.~Ade {\it et al.} [Planck Collaboration],
  %``Planck 2015 results. XX. Constraints on inflation,''
  Astron.\ Astrophys.\  {\bf 594}, A20 (2016)
  %doi:10.1051/0004-6361/201525898
  \arXiv{1502.02114} {astro-ph.CO}.
  %%CITATION = doi:10.1051/0004-6361/201525898;%%

\bibitem{Planck2} 
  P.~A.~R.~Ade {\it et al.} [Planck Collaboration],
  %``Planck 2015 results. XIII. Cosmological parameters,''
  Astron.\ Astrophys.\  {\bf 594}, A13 (2016)
  %doi:10.1051/0004-6361/201525830
  \arXiv{1502.01589} {astro-ph.CO}.
  %%CITATION = doi:10.1051/0004-6361/201525830;%%


% \bibitem{Kinneyz} 
%  W.~H.~Kinney,
  %``Horizon crossing and inflation with large eta,''
%  Phys.\ Rev.\ D {\bf 72}, 023515 (2005)
 % doi:10.1103/PhysRevD.72.023515
%  \arXivold{gr-qc/0503017}.
  %%CITATION = doi:10.1103/PhysRevD.72.023515;%%


\bibitem{Leach:2001zf}
  S.~M.~Leach, M.~Sasaki, D.~Wands and A.~R.~Liddle,
  %``Enhancement of superhorizon scale inflationary curvature perturbations,''
  Phys.\ Rev.\ D {\bf 64} (2001) 023512
  %doi:10.1103/PhysRevD.64.023512
  \arXivold{astro-ph/0101406}.
  %%CITATION = doi:10.1103/PhysRevD.64.023512;%%
  %126 citations counted in INSPIRE as of 23 Apr 2018

\bibitem{Martin:2012pe}
  J.~Martin, H.~Motohashi and T.~Suyama,
  %``Ultra Slow-Roll Inflation and the non-Gaussianity Consistency Relation,''
  Phys.\ Rev.\ D {\bf 87} (2013) no.2,  023514
 % doi:10.1103/PhysRevD.87.023514
 \arXiv{1211.0083} {astro-ph.CO}
  %%CITATION = doi:10.1103/PhysRevD.87.023514;%%
  %80 citations counted in INSPIRE as of 01 Jun 2018

\bibitem{Motohashi:2014ppa} 
  H.~Motohashi, A.~A.~Starobinsky and J.~Yokoyama,
  %``Inflation with a constant rate of roll,''
  JCAP {\bf 1509}, no. 09, 018 (2015)
%  doi:10.1088/1475-7516/2015/09/018
  \arXiv{1411.5021} {astro-ph.CO}.
  %%CITATION = doi:10.1088/1475-7516/2015/09/018;%%
  %48 citations counted in INSPIRE as of 23 Apr 2018


  \bibitem{Germani} 
  C.~Germani and T.~Prokopec,
  %``On primordial black holes from an inflection point,''
  Phys.\ Dark Univ.\  {\bf 18}, 6 (2017)
  %doi:10.1016/j.dark.2017.09.001
  \arXiv{1706.04226}{astro-ph.CO}.
  %%CITATION = doi:10.1016/j.dark.2017.09.001;%%


  \bibitem{Dimopoulos} 
  K.~Dimopoulos,
  %``Ultra slow-roll inflation demystified,''
  Phys.\ Lett.\ B {\bf 775}, 262 (2017)
  %doi:10.1016/j.physletb.2017.10.066
  \arXiv{1707.05644}{hep-ph}.
  %%CITATION = doi:10.1016/j.physletb.2017.10.066;%%

\bibitem{Motohashi:2017kbs}
  H.~Motohashi and W.~Hu,
  %``Primordial Black Holes and Slow-Roll Violation,''
  Phys.\ Rev.\ D {\bf 96} (2017) no.6,  063503
  %doi:10.1103/PhysRevD.96.063503
  \arXiv{1706.06784} {astro-ph.CO}.
  %%CITATION = doi:10.1103/PhysRevD.96.063503;%%
  %24 citations counted in INSPIRE as of 03 Apr 2018


\bibitem{Kallosh:2014vja}
  R.~Kallosh, A.~Linde and B.~Vercnocke,
  %``Natural Inflation in Supergravity and Beyond,''
  Phys.\ Rev.\ D {\bf 90} (2014) no.4,  041303
%  doi:10.1103/PhysRevD.90.041303
  \arXiv{1404.6244}{hep-th}.


\bibitem{SW1}
  E.~Silverstein and A.~Westphal,
  %``Monodromy in the CMB: Gravity Waves and String Inflation,''
  Phys.\ Rev.\ D {\bf 78} (2008) 106003
%  doi:10.1103/PhysRevD.78.106003
  \arXiv{0803.3085}{hep-th}.
  
\bibitem{SW2}
  L.~McAllister, E.~Silverstein and A.~Westphal,
  %``Gravity Waves and Linear Inflation from Axion Monodromy,''
  Phys.\ Rev.\ D {\bf 82} (2010) 046003
 % doi:10.1103/PhysRevD.82.046003
  \arXiv{0808.0706}{hep-th}.  
  

\bibitem{Easther:2013kla}
  R.~Easther and R.~Flauger,
  %``Planck Constraints on Monodromy Inflation,''
  JCAP {\bf 1402} (2014) 037
%  doi:10.1088/1475-7516/2014/02/037
  \arXiv{1308.3736}{astro-ph.CO}.
  
  \bibitem{Flauger:2009ab}
  R.~Flauger, L.~McAllister, E.~Pajer, A.~Westphal and G.~Xu,
  %``Oscillations in the CMB from Axion Monodromy Inflation,''
  JCAP {\bf 1006} (2010) 009
%  doi:10.1088/1475-7516/2010/06/009
  \arXiv{0907.2916}{hep-th}.
  %%CITATION = doi:10.1088/1475-7516/2010/06/009;%%
  
\bibitem{Kobayashi:2014ooa}
  T.~Kobayashi, O.~Seto and Y.~Yamaguchi,
  %``Axion monodromy inflation with sinusoidal corrections,''
  PTEP {\bf 2014} (2014) no.10,  103E01
%  doi:10.1093/ptep/ptu145
  \arXiv{1404.5518}{hep-ph}.  
  
  
%  \bibitem{KLV}
%  R.~Kallosh, A.~Linde and B.~Vercnocke,
  %``Natural Inflation in Supergravity and Beyond,''
%  Phys.\ Rev.\ D {\bf 90} (2014) no.4,  041303
 % doi:10.1103/PhysRevD.90.041303
%  \arXiv{1404.6244}{hep-th}.
  %%CITATION = doi:10.1103/PhysRevD.90.041303;%%


%\cite{Starobinsky:1979ty}
\bibitem{Starobinsky:1979ty}
  A.~A.~Starobinsky,
  %``Spectrum of relict gravitational radiation and the early state of the universe,''
  JETP Lett.\  {\bf 30} (1979) 682
   [Pisma Zh.\ Eksp.\ Teor.\ Fiz.\  {\bf 30} (1979) 719].
  %%CITATION = JTPLA,30,682;%%
  %1341 citations counted in INSPIRE as of 04 Jan 2019


\bibitem{Mukhanov:1981xt}
  V.~F.~Mukhanov and G.~V.~Chibisov,
  %``Quantum Fluctuations and a Nonsingular Universe,''
  JETP Lett.\  {\bf 33} (1981) 532
   [Pisma Zh.\ Eksp.\ Teor.\ Fiz.\  {\bf 33} (1981) 549].
  %%CITATION = JTPLA,33,532;%%
  %1193 citations counted in INSPIRE as of 08 Nov 2017



\bibitem{Press} 
  W.~H.~Press and P.~Schechter,
  %``Formation of galaxies and clusters of galaxies by selfsimilar gravitational condensation,''
  Astrophys.\ J.\  {\bf 187}, 425 (1974).
  %doi:10.1086/152650
  %%CITATION = doi:10.1086/152650;%%


\bibitem{Niemeyer:1997mt}
  J.~C.~Niemeyer and K.~Jedamzik,
  %``Near-critical gravitational collapse and the initial mass function of primordial black holes,''
  Phys.\ Rev.\ Lett.\  {\bf 80} (1998) 5481
  %doi:10.1103/PhysRevLett.80.5481
  \arXivold{astro-ph/9709072}.
  %%CITATION = doi:10.1103/PhysRevLett.80.5481;%%
  %145 citations counted in INSPIRE as of 03 Apr 2018

\bibitem{Shibata:1999zs}
  M.~Shibata and M.~Sasaki,
  %``Black hole formation in the Friedmann universe: Formulation and computation in numerical relativity,''
  Phys.\ Rev.\ D {\bf 60} (1999) 084002
  %doi:10.1103/PhysRevD.60.084002
  [gr-qc/9905064].
  %%CITATION = doi:10.1103/PhysRevD.60.084002;%%
  %122 citations counted in INSPIRE as of 23 May 2018

\bibitem{Musco:2008hv} 
  I.~Musco, J.~C.~Miller and A.~G.~Polnarev,
  %``Primordial black hole formation in the radiative era: Investigation of the critical nature of the collapse,''
  Class.\ Quant.\ Grav.\  {\bf 26}, 235001 (2009)
 % doi:10.1088/0264-9381/26/23/235001
  \arXiv{0811.1452} {gr-qc}.
  %%CITATION = doi:10.1088/0264-9381/26/23/235001;%%
  %59 citations counted in INSPIRE as of 23 May 2018



\bibitem{Musco:2012au}
  I.~Musco and J.~C.~Miller,
  %``Primordial black hole formation in the early universe: critical behaviour and self-similarity,''
  Class.\ Quant.\ Grav.\  {\bf 30} (2013) 145009
  %doi:10.1088/0264-9381/30/14/145009
  \arXiv{1201.2379} {gr-qc}.
  %%CITATION = doi:10.1088/0264-9381/30/14/145009;%%
  %55 citations counted in INSPIRE as of 23 May 2018


\bibitem{Harada:2013epa}
  T.~Harada, C.~M.~Yoo and K.~Kohri,
  %``Threshold of primordial black hole formation,''
  Phys.\ Rev.\ D {\bf 88} (2013) no.8,  084051
   Erratum: [Phys.\ Rev.\ D {\bf 89} (2014) no.2,  029903]
  %doi:10.1103/PhysRevD.88.084051, 10.1103/PhysRevD.89.029903
  \arXiv{1309.4201} {astro-ph.CO}.
  %%CITATION = doi:10.1103/PhysRevD.88.084051, 10.1103/PhysRevD.89.029903;%%
  %60 citations counted in INSPIRE as of 03 Apr 2018

\bibitem{Germani:2018jgr}
  C.~Germani and I.~Musco,
  %``The abundance of primordial black holes depends on the shape of the inflationary power spectrum,''
  \arXiv{1805.04087} {astro-ph.CO}.
  %%CITATION = ARXIV:1805.04087;%%
  %1 citations counted in INSPIRE as of 23 May 2018

\bibitem{Byrnes:2018clq}
  C.~T.~Byrnes, M.~Hindmarsh, S.~Young and M.~R.~S.~Hawkins,
  %``Primordial black holes with an accurate QCD equation of state,''
  \arXiv{1801.06138} {astro-ph.CO}.
  %%CITATION = ARXIV:1801.06138;%%
  %2 citations counted in INSPIRE as of 23 May 2018


  \bibitem{Young:2014ana} 
  S.~Young, C.~T.~Byrnes and M.~Sasaki,
  %``Calculating the mass fraction of primordial black holes,''
  JCAP {\bf 1407}, 045 (2014)
  %doi:10.1088/1475-7516/2014/07/045
  \arXiv{1405.7023}{gr-qc}.
  %%CITATION = doi:10.1088/1475-7516/2014/07/045;%%

\bibitem{Khlopov:1980mg}
  M.~Y.~Khlopov and A.~G.~Polnarev,
  %``Primordial Black Holes As A Cosmological Test Of Grand Unification,''
  Phys.\ Lett.\  {\bf 97B} (1980) 383.
%  doi:10.1016/0370-2693(80)90624-3
  %%CITATION = doi:10.1016/0370-2693(80)90624-3;%%
  %120 citations counted in INSPIRE as of 01 Jun 2018

\bibitem{Polnarev:1986bi}
  A.~G.~Polnarev and M.~Y.~Khlopov,
  %``Cosmology, Primordial Black Holes, And Supermassive Particles,''
  Sov.\ Phys.\ Usp.\  {\bf 28} (1985) 213
   [Usp.\ Fiz.\ Nauk {\bf 145} (1985) 369].
  %doi:10.1070/PU1985v028n03ABEH003858
  %%CITATION = doi:10.1070/PU1985v028n03ABEH003858;%%
  %67 citations counted in INSPIRE as of 01 Jun 2018

\bibitem{Helfer:2016ljl}
  T.~Helfer, D.~J.~E.~Marsh, K.~Clough, M.~Fairbairn, E.~A.~Lim and R.~Becerril,
  %``Black hole formation from axion stars,''
  JCAP {\bf 1703} (2017) no.03,  055
 % doi:10.1088/1475-7516/2017/03/055
  \arXiv{1609.04724} {astro-ph.CO}.
  %%CITATION = doi:10.1088/1475-7516/2017/03/055;%%
  %29 citations counted in INSPIRE as of 21 May 2018

\bibitem{Harada:2016mhb}
  T.~Harada, C.~M.~Yoo, K.~Kohri, K.~i.~Nakao and S.~Jhingan,
  %``Primordial black hole formation in the matter-dominated phase of the Universe,''
  Astrophys.\ J.\  {\bf 833} (2016) no.1,  61
%  doi:10.3847/1538-4357/833/1/61
  \arXiv{1609.01588} {astro-ph.CO}.
  %%CITATION = doi:10.3847/1538-4357/833/1/61;%%
  %19 citations counted in INSPIRE as of 27 Apr 2018

\bibitem{Harada:2017fjm}
  T.~Harada, C.~M.~Yoo, K.~Kohri and K.~I.~Nakao,
 % ``Spins of primordial black holes formed in the matter-dominated phase of the Universe,''
  Phys.\ Rev.\ D {\bf 96} (2017) no.8,  083517
 %  doi:10.1103/PhysRevD.96.083517
  \arXiv{1707.03595} {gr-qc}.
  %%CITATION = doi:10.1103/PhysRevD.96.083517;%%
  %5 citations counted in INSPIRE as of 27 Apr 2018



\bibitem{Liddle:2003as}
  A.~R.~Liddle and S.~M.~Leach,
  %``How long before the end of inflation were observable perturbations produced?,''
  Phys.\ Rev.\ D {\bf 68} (2003) 103503
  %doi:10.1103/PhysRevD.68.103503
  \arXivold{astro-ph/0305263}.
  %%CITATION = doi:10.1103/PhysRevD.68.103503;%%
  %335 citations counted in INSPIRE as of 30 Mar 2018




\bibitem{Dalianis:2018afb}
  I.~Dalianis and Y.~Watanabe,
  %``Probing the BSM physics with CMB precision cosmology: an application to supersymmetry,''
  JHEP {\bf 1802} (2018) 118
  %doi:10.1007/JHEP02(2018)118
  \arXiv{1801.05736} {hep-ph}.
  %%CITATION = doi:10.1007/JHEP02(2018)118;%%


\bibitem{Carr:2018nkm}
  B.~Carr, K.~Dimopoulos, C.~Owen and T.~Tenkanen,
  %``Primordial Black Hole Formation During Slow Reheating After Inflation,''
  \arXiv{1804.08639} {astro-ph.CO}.
  %%CITATION = ARXIV:1804.08639;%%

\bibitem{Georg:2017mqk}
  J.~Georg and S.~Watson,
  %``A Preferred Mass Range for Primordial Black Hole Formation and Black Holes as Dark Matter Revisited,''
  JHEP {\bf 1709} (2017) 138
   [JHEP {\bf 1709} (2017) 138]
  %doi:10.1007/JHEP09(2017)138
  \arXiv{arXiv:1703.04825 [astro-ph.CO]}.
  %%CITATION = doi:10.1007/JHEP09(2017)138;%%
  %23 citations counted in INSPIRE as of 03 Oct 2018


\bibitem{Akrami:2018odb}
  Y.~Akrami {\it et al.} [Planck Collaboration],
  %``Planck 2018 results. X. Constraints on inflation,''
  arXiv:1807.06211 [astro-ph.CO].
  %%CITATION = ARXIV:1807.06211;%%
  %50 citations counted in INSPIRE as of 03 Oct 2018



\bibitem{Carr:2016hva}
  B.~J.~Carr, K.~Kohri, Y.~Sendouda and J.~Yokoyama,
  %``Constraints on primordial black holes from the Galactic gamma-ray background,''
  Phys.\ Rev.\ D {\bf 94} (2016) no.4,  044029
  %doi:10.1103/PhysRevD.94.044029
  \arXiv{1604.05349} {astro-ph.CO}.
  %%CITATION = doi:10.1103/PhysRevD.94.044029;%%
  %16 citations counted in INSPIRE as of 23 Mar 2018


\bibitem{Inomata:2017okj}
  K.~Inomata, M.~Kawasaki, K.~Mukaida, Y.~Tada and T.~T.~Yanagida,
  %``Inflationary Primordial Black Holes as All Dark Matter,''
  Phys.\ Rev.\ D {\bf 96} (2017) no.4,  043504
  %doi:10.1103/PhysRevD.96.043504
  \arXiv{1701.02544} {astro-ph.CO}.
  %%CITATION = doi:10.1103/PhysRevD.96.043504;%%
  %29 citations counted in INSPIRE as of 23 Mar 2018




\bibitem{Carr:2009jm}
  B.~J.~Carr, K.~Kohri, Y.~Sendouda and J.~Yokoyama,
  %``New cosmological constraints on primordial black holes,''
  Phys.\ Rev.\ D {\bf 81} (2010) 104019
 %  doi:10.1103/PhysRevD.81.104019
  \arXiv{0912.5297} {astro-ph.CO}.
  %%CITATION = doi:10.1103/PhysRevD.81.104019;%%
  %354 citations counted in INSPIRE as of 23 Mar 2018

\bibitem{Barnacka:2012bm}
  A.~Barnacka, J.~F.~Glicenstein and R.~Moderski,
  %``New constraints on primordial black holes abundance from femtolensing of gamma-ray bursts,''
  Phys.\ Rev.\ D {\bf 86} (2012) 043001
%  doi:10.1103/PhysRevD.86.043001
  \arXiv{1204.2056 }{astro-ph.CO}.
  %%CITATION = doi:10.1103/PhysRevD.86.043001;%%
  %74 citations counted in INSPIRE as of 23 Mar 2018
  
\bibitem{Niikura:2017zjd}
  H.~Niikura {\it et al.},
  %``Microlensing constraints on primordial black holes with the Subaru/HSC Andromeda observation,''
 \arXiv{1701.02151} {astro-ph.CO}.
  %%CITATION = ARXIV:1701.02151;%%
  %45 citations counted in INSPIRE as of 23 Mar 2018
   
   %\cite{Katz:2018zrn}
\bibitem{Katz:2018zrn}
  A.~Katz, J.~Kopp, S.~Sibiryakov and W.~Xue,
  ``Femtolensing by Dark Matter Revisited,''
  %Submitted to: JCAP
  [arXiv:1807.11495 [astro-ph.CO]].
  %%CITATION = ARXIV:1807.11495;%%
  %2 citations counted in INSPIRE as of 03 Oct 2018 
    
  \bibitem{Tisserand:2006zx}
  P.~Tisserand {\it et al.} [EROS-2 Collaboration],
  %``Limits on the Macho Content of the Galactic Halo from the EROS-2 Survey of the Magellanic Clouds,''
  Astron.\ Astrophys.\  {\bf 469} (2007) 387
 % doi:10.1051/0004-6361:20066017
  \arXivold{astro-ph/0607207}.
  %%CITATION = doi:10.1051/0004-6361:20066017;%%
  %338 citations counted in INSPIRE as of 23 Mar 2018

\bibitem{Capela:2012jz}
  F.~Capela, M.~Pshirkov and P.~Tinyakov,
  %``Constraints on Primordial Black Holes as Dark Matter Candidates from Star Formation,''
  Phys.\ Rev.\ D {\bf 87} (2013) no.2,  023507
  % doi:10.1103/PhysRevD.87.023507
  \arXiv{1209.6021} {astro-ph.CO}.
  %%CITATION = doi:10.1103/PhysRevD.87.023507;%%
  %35 citations counted in INSPIRE as of 23 Mar 2018

\bibitem{Capela:2013yf}
  F.~Capela, M.~Pshirkov and P.~Tinyakov,
  %``Constraints on primordial black holes as dark matter candidates from capture by neutron stars,''
  Phys.\ Rev.\ D {\bf 87} (2013) no.12,  123524
 % doi:10.1103/PhysRevD.87.123524
  \arXiv{1301.4984} {astro-ph.CO}.
  %%CITATION = doi:10.1103/PhysRevD.87.123524;%%
  %69 citations counted in INSPIRE as of 23 Mar 2018

\bibitem{Brandt:2016aco}
  T.~D.~Brandt,
  %``Constraints on MACHO Dark Matter from Compact Stellar Systems in Ultra-Faint Dwarf Galaxies,''
  Astrophys.\ J.\  {\bf 824} (2016) no.2,  L31
 % doi:10.3847/2041-8205/824/2/L31
  \arXiv{1605.03665} {astro-ph.GA}.
  %%CITATION = doi:10.3847/2041-8205/824/2/L31;%%
  %71 citations counted in INSPIRE as of 23 Mar 2018

\bibitem{Graham:2015apa}
  P.~W.~Graham, S.~Rajendran and J.~Varela,
  %``Dark Matter Triggers of Supernovae,''
  Phys.\ Rev.\ D {\bf 92} (2015) no.6,  063007
 % doi:10.1103/PhysRevD.92.063007
  \arXiv{1505.04444} {hep-ph}.
  %%CITATION = doi:10.1103/PhysRevD.92.063007;%%
  %40 citations counted in INSPIRE as of 23 Mar 2018

\bibitem{Gaggero:2016dpq}
  D.~Gaggero, G.~Bertone, F.~Calore, R.~M.~T.~Connors, M.~Lovell, S.~Markoff and E.~Storm,
  %``Searching for Primordial Black Holes in the radio and X-ray sky,''
  Phys.\ Rev.\ Lett.\  {\bf 118} (2017) no.24,  241101
  %doi:10.1103/PhysRevLett.118.241101
  \arXiv{1612.00457} {astro-ph.HE}.
  %%CITATION = doi:10.1103/PhysRevLett.118.241101;%%
  %33 citations counted in INSPIRE as of 23 Mar 2018

\bibitem{Ricotti:2007au}
  M.~Ricotti, J.~P.~Ostriker and K.~J.~Mack,
  %``Effect of Primordial Black Holes on the Cosmic Microwave Background and Cosmological Parameter Estimates,''
  Astrophys.\ J.\  {\bf 680} (2008) 829
 % doi:10.1086/587831
  \arXiv{0709.0524} {astro-ph}.
  %%CITATION = doi:10.1086/587831;%%
  %151 citations counted in INSPIRE as of 23 Mar 2018

%\bibitem{Carr:2016drx}
%  B.~Carr, F.~Kuhnel and M.~Sandstad,
  %``Primordial Black Holes as Dark Matter,''
%  Phys.\ Rev.\ D {\bf 94} (2016) no.8,  083504
  %doi:10.1103/PhysRevD.94.083504
 % \arXiv{1607.06077} {astro-ph.CO}.
  %%CITATION = doi:10.1103/PhysRevD.94.083504;%%
  %163 citations counted in INSPIRE as of 23 Mar 2018

%\bibitem{Clesse:2016vqa}
%  S.~Clesse and J.~García-Bellido,
  %``The clustering of massive Primordial Black Holes as Dark Matter: measuring their mass distribution with Advanced LIGO,''
%  Phys.\ Dark Univ.\  {\bf 15} (2017) 142
  %doi:10.1016/j.dark.2016.10.002
%  \arXiv{1603.05234} {astro-ph.CO}.
  %%CITATION = doi:10.1016/j.dark.2016.10.002;%%
  %100 citations counted in INSPIRE as of 23 Mar 2018

\bibitem{Boucenna:2017ghj}
  S.~M.~Boucenna, F.~Kuhnel, T.~Ohlsson and L.~Visinelli,
  ``Novel Constraints on Mixed Dark-Matter Scenarios of Primordial Black Holes and WIMPs,''
  \arXiv{1712.06383} {hep-ph}.
  %%CITATION = ARXIV:1712.06383;%%

\end{thebibliography}
\end{document}